\documentclass[useAMS,usenatbib]{mn2e}
\usepackage{graphicx}
\usepackage{epsfig}
\usepackage{longtable}

\newcommand{\HI}{H\,{\sc i}}
\newcommand{\HII}{H\,{\sc ii}}
\newcommand{\SII}{[S\,{\sc ii}]}

\newcommand{\OIII}{[O\,{\sc iii}]}
\newcommand{\OII}{[O\,{\sc ii}]}
\newcommand{\OI}{[O\,{\sc i}]}

\newcommand{\NII}{[N\,{\sc ii}]}
\newcommand{\HeII}{He\,{\sc ii}}
\newcommand{\HeI}{He\,{\sc i}}
\newcommand{\NeII}{[Ne\,{\sc ii}]}
\newcommand{\NeIII}{[Ne\,{\sc iii}]}
\newcommand{\ArII}{[Ar\,{\sc ii}]}
\newcommand{\ArIII}{[Ar\,{\sc iii}]}

\newcommand{\NeV}{[Ne\,{\sc v}]}
\newcommand{\NeVI}{[Ne\,{\sc vi}]}
\newcommand{\FeII}{[Fe\,{\sc ii}]}
\newcommand{\OIV}{[O\,{\sc iv}]}
\def\p0{\phantom{0}}

\title[A multiwavelength analysis of LMC PNe]{A multiwavelength analysis of planetary nebulae in the Large Magellanic Cloud}
\author[Warren A. Reid]{Warren A. Reid$^{1,2}$\thanks{e-Mail:
warren.reid@mq.edu.au; war@aao.gov.au } \\
$^{1}$Department of Physics and Astronomy, Macquarie University, Sydney, NSW 2109, Australia\\
$^{2}$Centre for Astronomy, Astrophysics and Astrophotonics, Macquarie University, Sydney, NSW 2109, Australia\\
}
\begin{document}

\date{}

\pagerange{\pageref{firstpage}--\pageref{lastpage}} \pubyear{2002}

\maketitle

\label{firstpage}

\begin{abstract}
This paper examines, compares and plots optical, near- and mid-infrared (MIR) photometric data for 605 planetary nebulae (PNe) in the Large Magellanic Cloud (LMC). With the aid of multi-wavelength surveys such as the $Spitzer$ legacy programme Surveying the Agents of a Galaxy's Evolution, the Two Micron All Sky Survey and the Magellanic Cloud Photometric Survey, plots have been constructed to expose the relative contributions from molecular hydrogen, polycyclic aromatic hydrocarbons, forbidden emission lines, warm dust continuum and stellar emission at various bands. Besides identifying trends, these plots have helped to reveal PN mimics including six previously known PNe in the outer LMC which are re-classified as other object types. Together with continuing follow-up optical observations, the data have enabled a substantial reduction in the number of PNe previously tagged as `likely' and `possible'. The total number of LMC PNe is adjusted to 715 but with a greater degree of confidence in regard to classification.

In each colour--colour plot, the more highly evolved LMC PNe are highlighted for comparison with younger, brighter PNe. The faintest and most evolved PNe typically cluster in areas of colour--colour space occupied by ordinary stars. Possible reasons for the wide disparity in infrared colour--colour ratios, such as evolution and dust composition, are presented for evaluation. A correlation is found between the optical luminosity of PNe, emission-line ratios and the MIR dust luminosity at various bands. Luminosity functions using the four Infrared Array Camera and Multiband Imaging Photometer of $Spitzer$ (MIPS) [24] bands are directly compared, revealing an increasing accumulation of PNe within the brightest two magnitudes at longer wavelengths. A correlation is also found between the MIPS [24] band and the \OIII\,5007 and H$\beta$ fluxes.

\end{abstract}


\section{Introduction}

The Large Magellanic Cloud (LMC) has proven to be an excellent laboratory for the study of numerous astrophysical objects. Its known $\sim$50kpc distance, useful angle of inclination (35$^{o}$: van der Marel \& Cioni 2001) and location above the plane of the Milky Way allow the properties of individual objects and their kinematic interaction to be studied in detail. In particular, LMC planetary nebulae (PNe) have the advantage of being located in an environment of low and relatively consistent reddening of about $\textsl{E}$($\textsl{V-I}$) = 0.09 $\pm$ 0.07 mag (Haschker et al. 2011) where valuable kinematic and abundance data can be gained both from the main bar and the outer regions of the galaxy. Since the LMC has an average metallicity half that of the Milky Way (Caputo et al. 1999) and a correspondingly lower dust-to-gas ratio (Weingartner \& Draine 2001), LMC PNe afford the opportunity to study how these factors affect late stage stellar evolution. The LMC is located close enough for PNe to be isolated and in some cases resolved with earth-based telescopes. The detailed examination of LMC PNe, confirmed through optical spectroscopy and multi-wavelength observations, is of enormous value for population studies, luminosity functions, metallicity and evolutionary studies. It also provides a unique insight into the kinematic motion of the older LMC population enabling comparison with young populations and the \HI~disk (eg. Reid \& Parker, 2006b).

The United Kingdom Schmidt Telescope (UKST) H$\alpha$ deep stack survey of the central 25deg$^{2}$ of the LMC was initially used by Reid \& Parker to identify faint emission-line candidates for follow-up. From the resulting spectroscopy, 462 PN candidates were identified and 170 previously known PNe confirmed (Reid \& Parker, 2006a,b). Subsequent spectroscopic confirmation and preliminary multi-wavelength comparison resulted in these numbers being modified to 411 new and 164 previously known PNe (Reid \& Parker 2010b).

Their search was recently extended to include the outer 64deg$^{2}$ of the LMC using both the Magellanic Cloud Emission-Line Survey (MCELS~\footnote{C. Smith, S. Points, the MCELS Team and NOAO/AURA/NSF} Smith et al. 1998) and the UKST H$\alpha$/short red Magellanic Cloud survey as the source maps for candidate identification (Reid \& Parker 2013). Candidates were followed up spectroscopically using AAOmega on the Anglo-Australian Telescope (AAT) and 6dF on the UKST. The spectra were examined in conjunction with photometry and false colour images from the $\textit{Spitzer}$ SAGE (Surveying the Agents of a Galaxy's Evolution) survey (Meixner et al. 2006) and Two Micron All Sky Survey (2MASS; Cutri et al., 2003, 2012) in order to assess and classify new PN candidates. Additional photometry was gained from the Magellanic Cloud Photometric Survey (MCPS; Zaritsky et al. 2004) depending on coverage and magnitude limits. The new PN discoveries in the outer LMC were published in Reid \& Parker (2013).

This paper investigates the optical, near-intrared (NIR) and mid-infrared (MIR) properties of all LMC PNe, where data are available, including those discovered in the outer regions. Comparison plots based on the NIR $J$, $H$ and $K_{s}$ bands, MIR 3.6, 4.5, 5.8, 8 and 24$\mu$m bands plus the optical $U$, $B$, $V$ and $I$ bands are provided. The results are evaluated within the context of evolutionary theories, the low metallicity of the LMC, and the properties of nebulae and their central stars. Unlike several previous studies that have relied on a limited number of \OIII-bright and mainly young PNe (eg, Stanghellini et al. 2007; Hora et al. 2008), or were focused on confined regions of the LMC (Miszalski et al. 2011b), the scope of this analysis covers the entire galaxy and includes a large proportion of low and very low excitation PNe with their characteristic weak \OIII~and high (\NII/H$\alpha$) ratios (Reid \& Parker 2006a; 2013). In addition, this work explores the role of metallicity in dust production and its effect on the formation and evolution of PNe. Combining all the multi-wavelength data has allowed the initial Reid \& Parker (RP) object classifications to be firmed up.

In agreement with the findings of Hora et al. (2008), a comparison of 2MASS and SAGE photometry shows that many faint and highly evolved PNe have the appearance of ordinary stars. Judging by IR plots alone, faint PNe can be almost indistinguishable from evolved stars or young stellar objects (YSOs). In such cases, spectral energy distributions (SEDs), wide spectroscopic coverage in the optical and MIR along with high-resolution imaging together provide greater confidence in classification.

The body of the paper is divided into four sections. Section~\ref{section2} describes the NIR and MIR data used for object confirmation and SED comparisons. Section~\ref{section3} contains a series of multiwavelength colour--colour plots comparing LMC PNe to a sample of 3,000 primarily main-sequence stars where magnitudes are available. In Section~\ref{section4} luminosity functions for all four Infrared Array Camera (IRAC) bands and the [24] Multiband Imaging Photometer of $Spitzer$ (MIPS) band are shown. Combining and comparing the shape of the functions reveals trends that are then compared against \OIII\,5007~and H$\beta$ fluxes. Section~\ref{section5} describes the spectroscopic and multiwavelength object follow-up and includes a table of objects that have been re-classified. It also provides notes on RP objects of particular interest (Section~\ref{subsection5.2}) and a brief discussion of objects in the 30 Doradus region (Section~\ref{subsection5.3}).



\section{NIR and MIR data}
\label{section2}

\subsection{2MASS photometry}

Using the 2MASS 6x catalogue for the LMC (Cutri et al., 2003, 2012) magnitudes were obtained for 274 PNe in $J$, 269 in $H$ and 263 in $K_{s}$. A search radius of 1.3 arcsec was allowed as the majority of sources are extended and can span up to 5 arcsec diameter in the 2MASS images. Where a number of sources were detected in close proximity, the extent of the emission was checked using the deep H$\alpha$ stacked maps and images at each of the 2MASS bands. Where there were two objects that were difficult to separate, or if there was only one detectable source despite 2MASS returning two separate point sources within it, the data were excluded from the comparisons. In all there were 206 2MASS $J$ sources that had corresponding SAGE 3.6 mag. That number decreases slightly for $H$ and $K_{s}$ magnitudes as it does for the IRAC 4.5, 5.6 and 8$\mu$m channels. A number of previous studies (eg. Gruendl \& Chu 2009) have linked 2MASS magnitudes together with SAGE data. Before using these data, coordinates and magnitudes were checked in order to verify the astrometry and the photometry being extracted from each catalogue.

One issue for 2MASS data in the LMC is its sensitivity cut-off. Although the 2MASS sensitivity at S/N = 10 is reached at 15.8, 15.1 and 14.3 mag for $\textsl{J}$, $\textsl{H}$ and $\textsl{K$_{s}$}$ respectively, some strong emission lines which often dominate these bands can still be detected below this limit in the PN sample. The brightest line in the $\textsl{J}$ band is Paschen $\beta$ although a spectrum would also reveal the presence of \HeI, \FeII~and \OI. The $\textsl{H}$ band can be dominated by the Brackett series with \HeI\,1.7002$\mu$m and occasionally \FeII\,1.6440$\mu$m. The brightest line in the $\textsl{K$_{s}$}$ band is Brackett $\gamma$ with \HeI 2.058, 2.112$\mu$m. Although it is useful to know that these lines may be present, they may be missing in young PNe where a strong warm dust continuum is likely to dominate (Hora et al, 1999).

\subsection{IRSF photometry}

To increase the number of detections available for comparison magnitudes were also obtained from the InfraRed~Survey~Facility (IRSF) Magellanic Clouds Point Source Catalogue (Kato et al., 2007) which covered 40deg$^{2}$ of the LMC. This catalogue provided 259 $J$ magnitudes, 297 $H$ and 232 $K_{s}$ magnitudes. A total of 158 IRSF $H$-band sources were found to correspond to 2MASS $H$-band detections. Comparison of these 158 measurements produced a standard deviation of 0.25 mag and a correlation coefficient of 0.978. Furthermore, more than 60\% of 2MASS and IRSF magnitudes agreed to within 0.1 mag allowing both of these data sets to be used, thereby increasing the number of data points in the plots. Where there were two or more published values for the same object or variations in their error estimates, error estimates were allowed to increase proportionately.

\begin{figure*}
\begin{center}
  \includegraphics[width=0.85\textwidth]{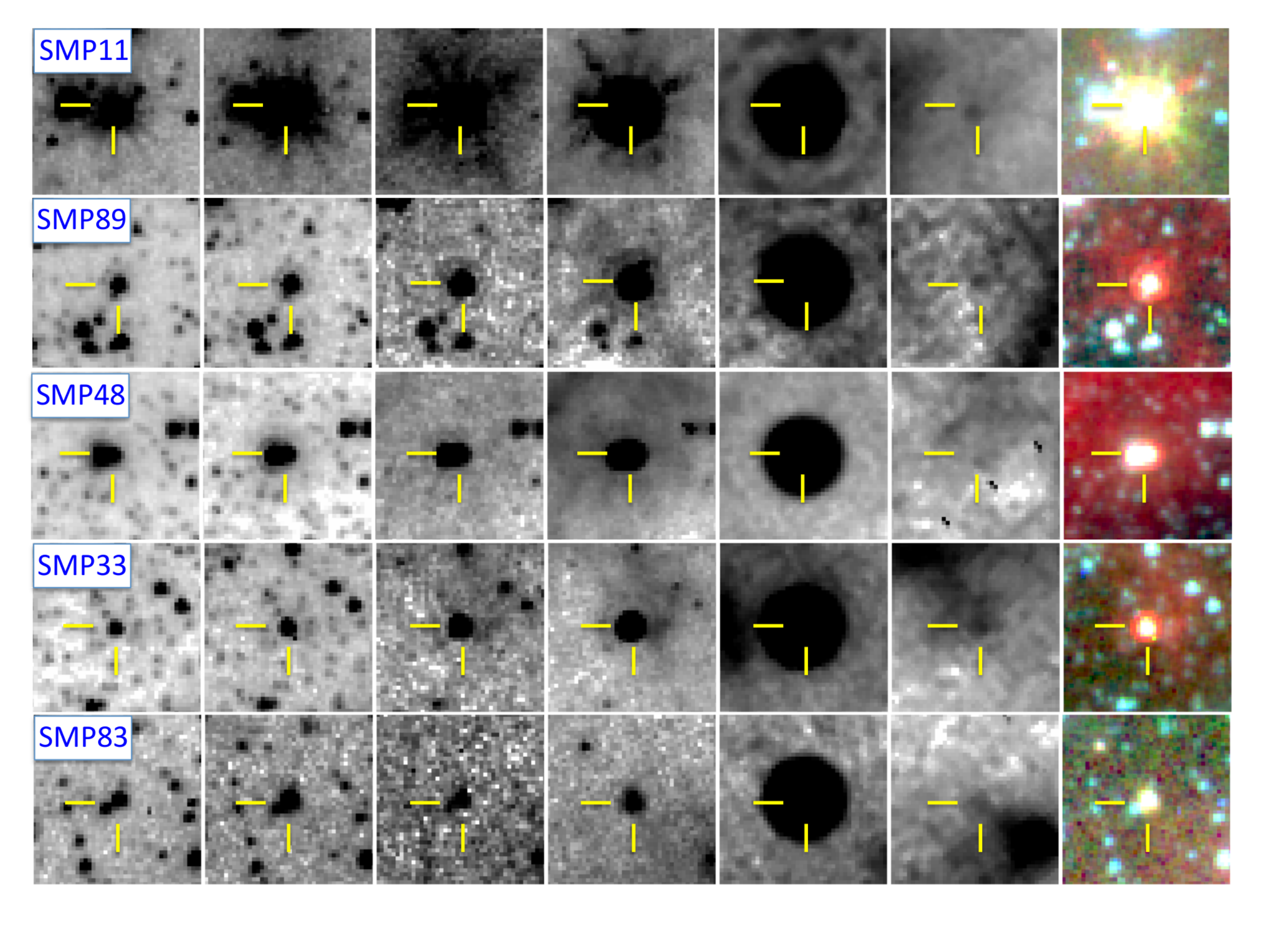}\\
  \includegraphics[width=0.85\textwidth]{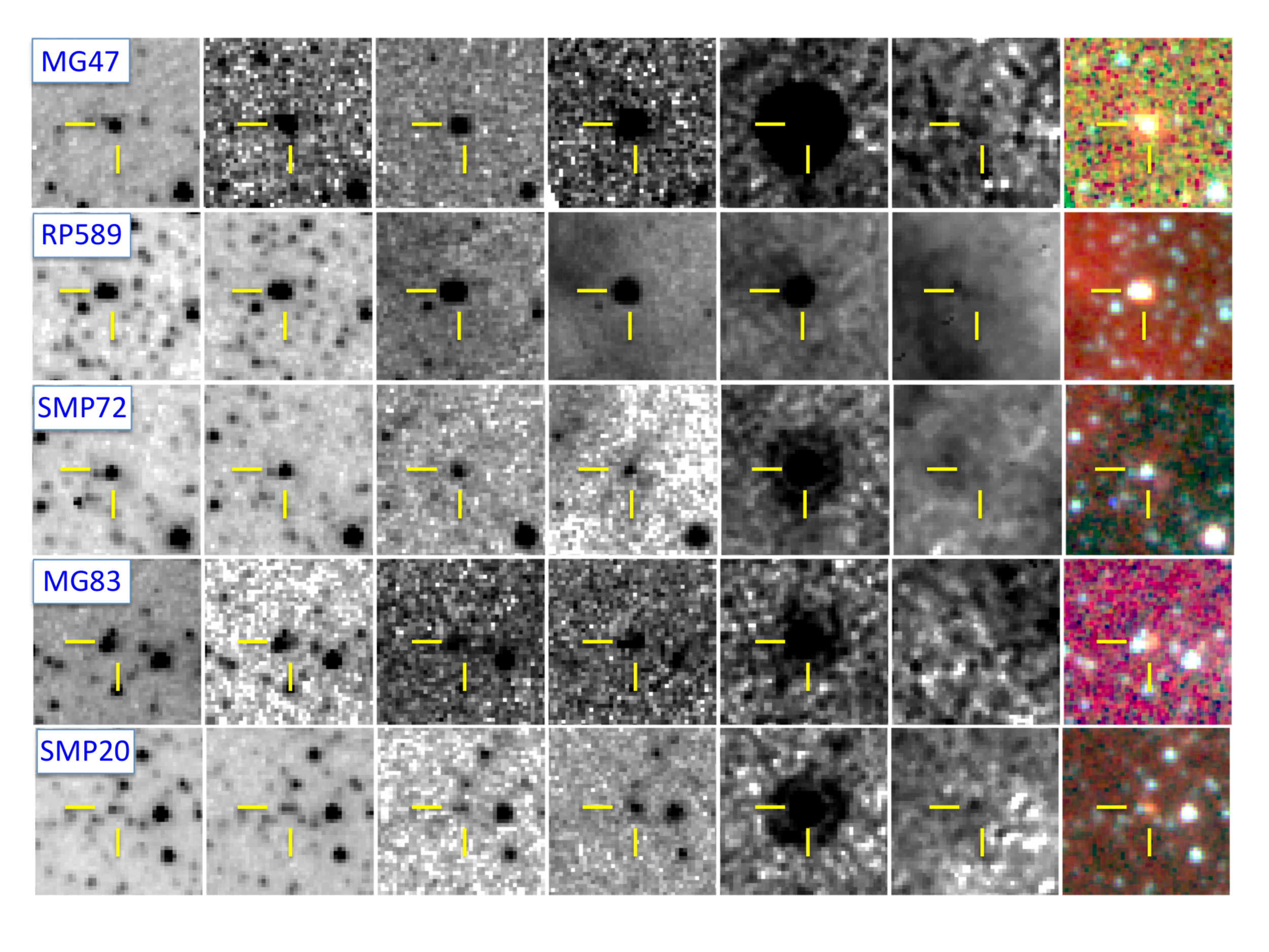}\\
  \caption{A sample of the brightest to medium luminosity LMC PNe in the MIR bands. Columns from left to right indicate IRAC [3.6], [4.5], [5.8], [8], [24], followed by MIPS [24] and [70]. The final column shows a false colour mosaic of IRAC channels [4.5] (blue), [5.8] (green) and [8] (red). Each image is centred in a 1 arcmin $\times$ 1 arcmin box and shown at the same linear scale.}
  \label{Figure1}
  \end{center}
  \end{figure*}

  \begin{figure*}
\begin{center}
  \includegraphics[width=0.85\textwidth]{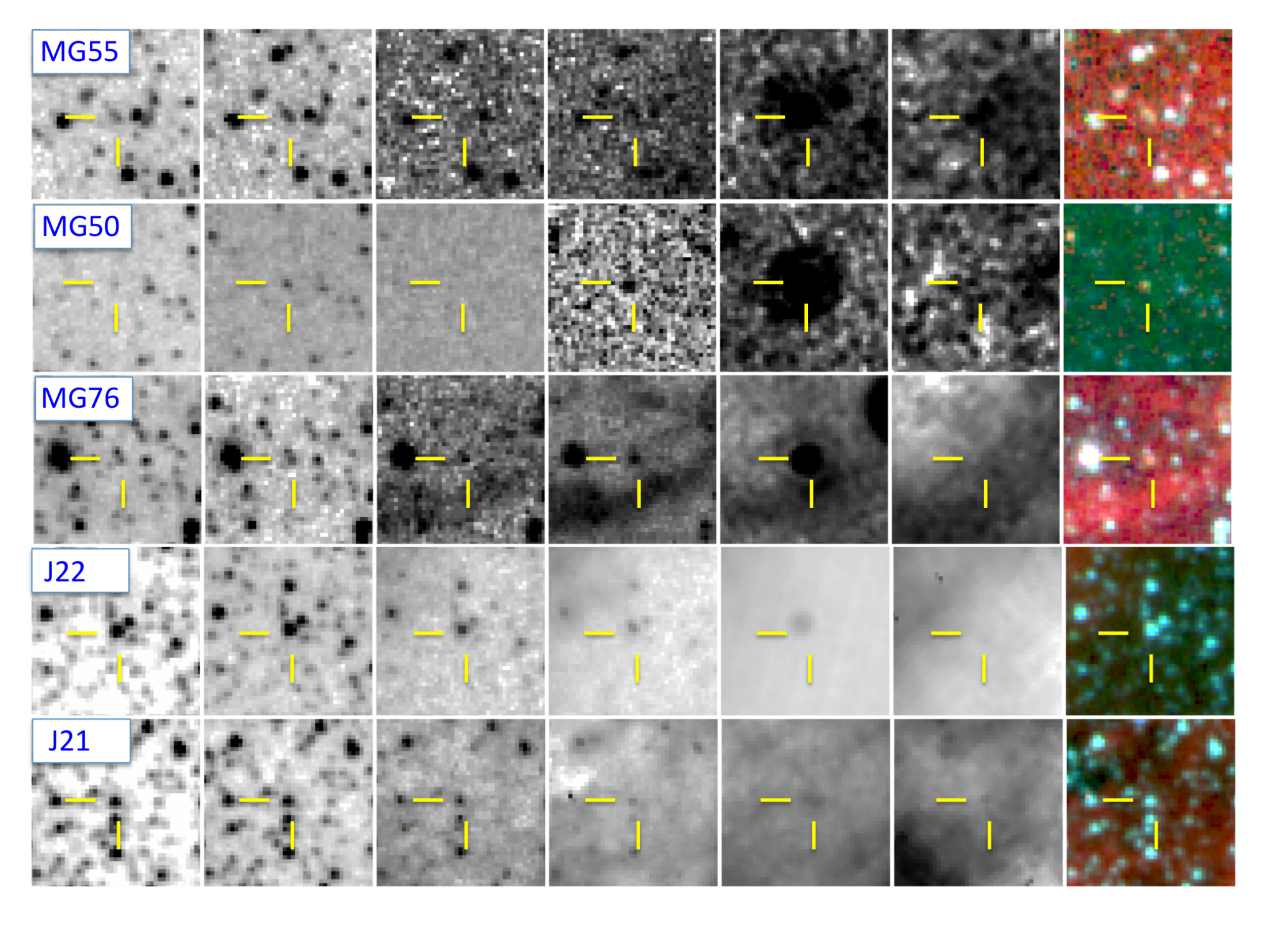}\\
  \caption{A sample of the faintest LMC PNe in the MIR bands. Columns from left to right indicate IRAC [3.6], [4.5], [5.8], [8], [24], followed by MIPS [24] and [70]. The final column shows a false colour mosaic of IRAC channels [4.5] (blue), [5.8] (green) and [8] (red). Each image is centred in a 1 arcmin $\times$ 1 arcmin box and shown at the same linear scale.}
  \label{Figure2}
  \end{center}
  \end{figure*}

\subsection{IRAC photometry}

The $Spitzer~Space~Telescope$ ($Spitzer$) has the necessary angular resolution and sensitivity to provide singular and discrete data for individual PNe. The 3.6, 4.5, 5.8 and 8\,$\mu$m bands were obtained with the IRAC (Fazio et al. 2004) on board $Spitzer$. This study used the archival data from the $Spitzer$ legacy programme SAGE (Meixner et al. 2006) which mapped the central 7 $\times$ 7 deg$^{2}$ area of the LMC.

Magnitudes were taken from the IRAC data of Hora et al. (2008) and Gruendl and Chu (2009). Both of these studies used basic calibrated data rather than drawing their magnitudes from the SAGE catalogue since the latter comprises point sources which may lead to non-detections or inaccurate results for extended nebulae. By incorporating their own photometry, each study achieved higher sensitivity while making the data less susceptible to instrumental artefacts and cosmic rays. The main difference between the two data sets is that Hora et al. used a 2.8 arcsec diameter aperture size compared to a 3.6 arcsec radius aperture centered on each source by Gruendl and Chu. In the highly populated regions of the LMC, the smaller 2.8 arcsec diameter aperture seems to be a very good choice, especially where PNe are small, faint and/or in crowded environments. In the case of several bright PNe, I preferred to use the measurements from Gruendl and Chu (2009) since the combined optical and MIR imaging showed that the larger aperture was better at accounting for all the emission (especially at 8$\mu$m) including the point spread function (PSF) seen in IRAC images. Hora et al. (2008) nonetheless tested their magnitudes against the standard IRAC calibration, which used a 12.2 arcsec aperture, and checked images for extended objects, giving them confidence that their magnitudes were accounting for the total emission from each extended source.

A total of 185 sources were found to be common to both catalogues. Comparing these two sets of magnitudes gave a mean difference of $\pm$0.0129 mag with a standard deviation of 0.212 mag and a correlation coefficient of 0.993. Such close agreement gave me confidence to use magnitudes from both data sets in the plots without differentiating them. Gruendl and Chu (private communication) provided images and magnitudes for many of the candidate PNe (Reid \& Parker, 2006b) and so their data set is larger by 193 objects.

 \subsection{MIPS photometry}

 The MIPS data were also obtained from both the Hora et al. (2008) and Gruendl and Chu (2009) studies. The resulting large number of sources at 24$\mu$m is indicative of the high sensitivity, low noise, diffraction-limited performance afforded by the three MIPS arrays and the peak of the SEDs for young PNe where dust corresponds to a temperature of $\sim$100K. There were 209 PNe detected at 24$\mu$m, 173 of which had detections across all four IRAC bands.

 Figs.~\ref{Figure1} and~\ref{Figure2} show the IRAC 3.6, 4.5, 5.8 and 8.0\,$\mu$m and MIPS 24 and 70\,$\mu$m single band images followed by a false colour mosaic of IRAC channels [4.5] (blue), [5.8] (green) and [8] (red) for sample LMC PNe. The accepted PN name is shown in the first image of each row. Each object is indicated by the cross-hairs in each 1 arcmin $\times$ 1 arcmin image. The PNe shown in these figures are often the outliers in the comparison plots. They are displayed in decreasing order of magnitude with Fig.~\ref{Figure1} containing the brightest and Fig.~\ref{Figure2} containing the faintest. SMP11 is the brightest and J21 is one of the faintest PNe to be detected in the SAGE survey. Figs~\ref{Figure1} and~\ref{Figure2} show that detections at 70$\mu$m are extremely faint and with only $\sim$20 published magnitudes available, this band had to be omitted from the analysis.

 \subsection{MCPS photometry}

 The Magellanic Cloud Photometric Survey (MCPS: Zaritsky et al. 2004) provided $U$, $B$, $V$ and $I$ stellar photometry for comparison with MIR magnitudes. The survey covers 64 deg$^{2}$ of the LMC, closely matching the area encompassed by MCELS. Data were obtained using the Las Campanas Swope 1\,m telescope and the Great Circle Camera with a 2K CCD. Drift-scan image exposures between 3.8 and 5.2 min were obtained in Johnson $U$, $B$, and $V$ and Gunn $\textit{i}$~with a pixel scale of 0.7 arcsec pixel$^{-1}$. The magnitude limits at each band vary as a function of wavelength and stellar crowding. Zaritsky et al. (2004) report sensitivity incompleteness becoming evident for stars with 20 $<$ $V$ $<$ 21. Severe incompleteness was found at $B$ = 23.5, $V$ = 23, $U$ = 21.5 and $I$ = 22. Please see Zaritsky et al. (2004) and references therein for further information such as data reduction, extinction, photometric and astrometric accuracy.

\section{Multi-wavelength comparisons}
\label{section3}

\begin{figure*}
\begin{center}
  \includegraphics[width=0.9\textwidth]{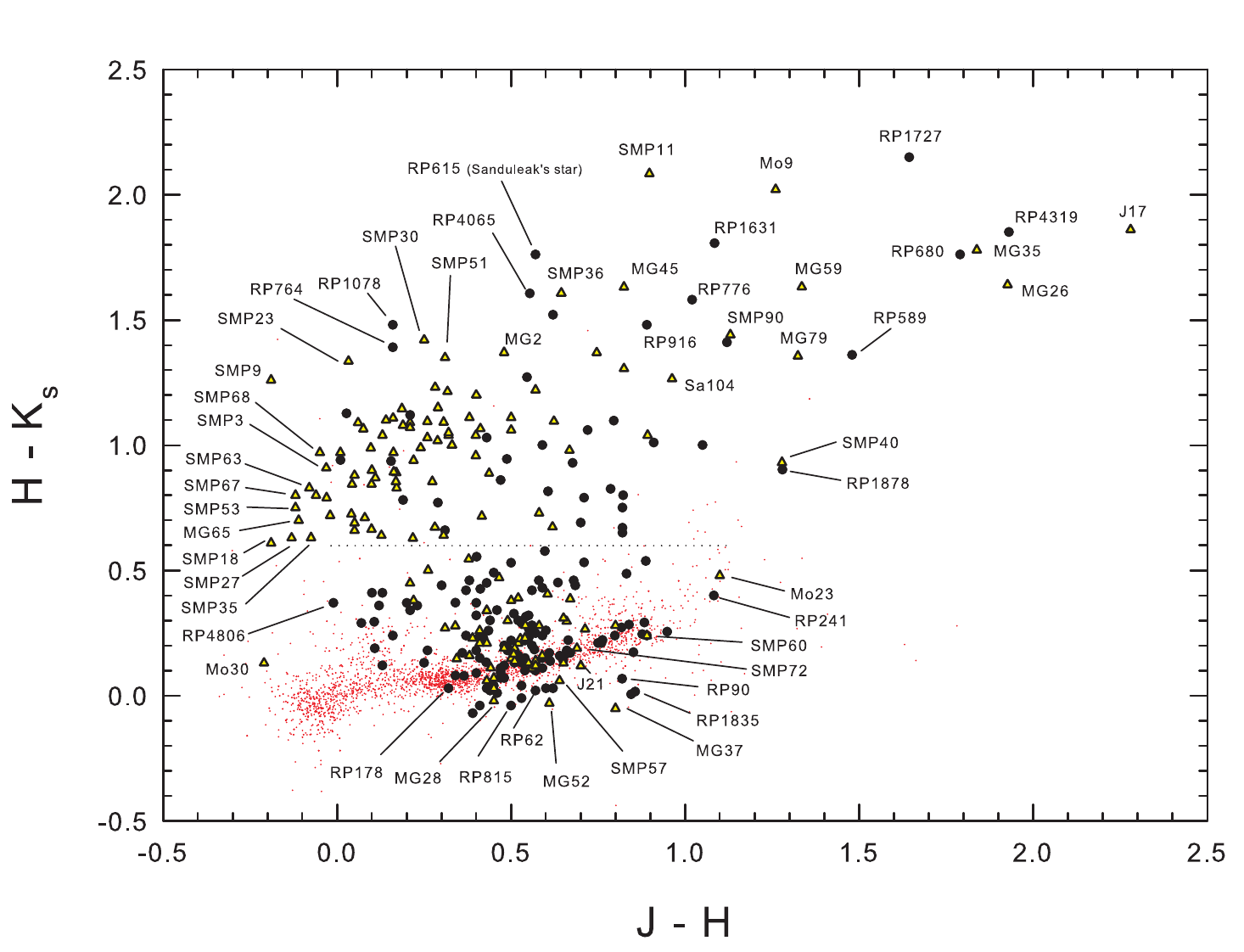}\\
  \caption{The $\textsl{J -- H}$ versus $\textsl{H -- K}$$_{s}$ colour-colour diagram for LMC PNe. The objects are from the complete catalogue of LMC PNe and include PNe known prior to 2006 (shown as yellow-filled triangles) and those uncovered after that time (Reid \& Parker 2006a,b, 2013) shown as black dots. PNe at the extremities of the plot have been named in order to facilitate comparisons in other plots. For comparison, LMC stars are shown as red dots. An arbitrary line placed at $\textsl{H -- K}$$_{s}$=0.6 divides the PNe in order to compare those close to and far from the stellar population (see Figure~\ref{Figure4}~and the text for details). Also included is RP615, a probable symbiotic star with extensive jets.}
  \label{Figure3}
  \end{center}
  \end{figure*}

The 2MASS, IRAC, MIPS and MCPS data for the LMC PNe are shown in Figs 3--13 while a summary of detections in each band is given in Table~\ref{table1}. The full set of magnitudes in each band and from each survey are provided in the appendix. Despite the large sample of PNe now discovered in the LMC, none of the plots contain all of the identified PNe to date. The reasons for this are twofold. First, detection limits vary at different bands, particularly affecting the fainter sources. Secondly, detections in 2MASS do not always astrometrically coincide with detections in SAGE and vice versa. Positional uncertainty due to object density was still found to be a major problem in the LMC where even a very small aperture search ($\sim$1.5 arcsec radius) could return magnitudes for two or three separate sources. Since this is such a potentially serious problem, where uncertainty due to object density occurs, the individual PNe images in H$\alpha$, 2MASS and SAGE were checked and overlaid for positional accuracy. Further checks comparing ratios of 2MASS and SAGE magnitudes to the ratios of 2MASS and SAGE image data counts occasionally affirmed which magnitude detection was associated with the PN. In a number of cases, blending with stellar sources and high background noise from extended dust and emission created too much uncertainty. This meant that it was safer not to associate magnitude data with a PN even though the position was accurately known.
\begin{table}
\begin{center}
\begin{tabular}{|l|c|c|c|r|}
  \hline
  Band & No. of PNe & Median & Std dev.  & 5$\sigma$  \\
     &     detected   &   mag  &  median   &   sensitivity   \\
  \hline\hline
  $J$ & 466 & 17.08 & 1.83 & 17.2 \\
  $H$ & 408 & 16.35 & 1.83 & 16.2 \\
  $K$$_{s}$ & 354 & 15.45 & 1.76 & 15.6 \\
  3.6 & 440 & 15.47 & 1.83 & 19.3 \\
  4.5 & 444 & 14.90 & 1.83 & 18.5 \\
  5.8 & 267 & 13.01 & 1.84 & 16.1 \\
  8 & 294 & 11.70 & 2.01 & 15.4\\
  24 & 209 & 7.35 & 2.11 & 10.4 \\
  70 & 18 & 2.68 & 0.67 & 3.5 \\
  \hline
\end{tabular}
\caption{Detection limits and other statistics for NIR and MIR data used in this survey. As in Hora et al. (2008), the 5$\sigma$ point source sensitivity limits for the IRAC and MIPS data are from Meixner et al. (2006) and the 2MASS limits are from Cutri et al. (2003, 2012).}\label{table1}
\end{center}
\end{table}

 Within Figs. 3--13, the PNe discovered prior to 2006 are plotted as yellow filled triangles, those found after 2006 are plotted as black circles, while stars (the majority of which are main sequence) are plotted as red dots. Most of the stars have optical $V$ magnitudes between 10 and 13 and therefore qualify as guide stars. Judging by their MIR colours they are unlikely to be Be or B[e] stars, asymptotic giant branch (AGB) stars or YSOs which may support an emission shell. As with the PNe themselves, only stars with the required magnitudes were included in any figure. In each plot, PNe at the outskirts or extremities have been named in order to compare their placement from one plot to another, and thereby provide clues about the nature of unusual PNe. The names also facilitate referral to specific objects and areas within the plot. Errors, which are typically less than 0.25\,mag, are smaller than the symbols used for data points. These errors, where derived for individual magnitudes, are also provided in the appendix.

 \subsection{\textsl{J -- H} versus \textsl{H -- K}$_{s}$}

   \begin{figure}
\begin{center}
  \includegraphics[width=0.49\textwidth]{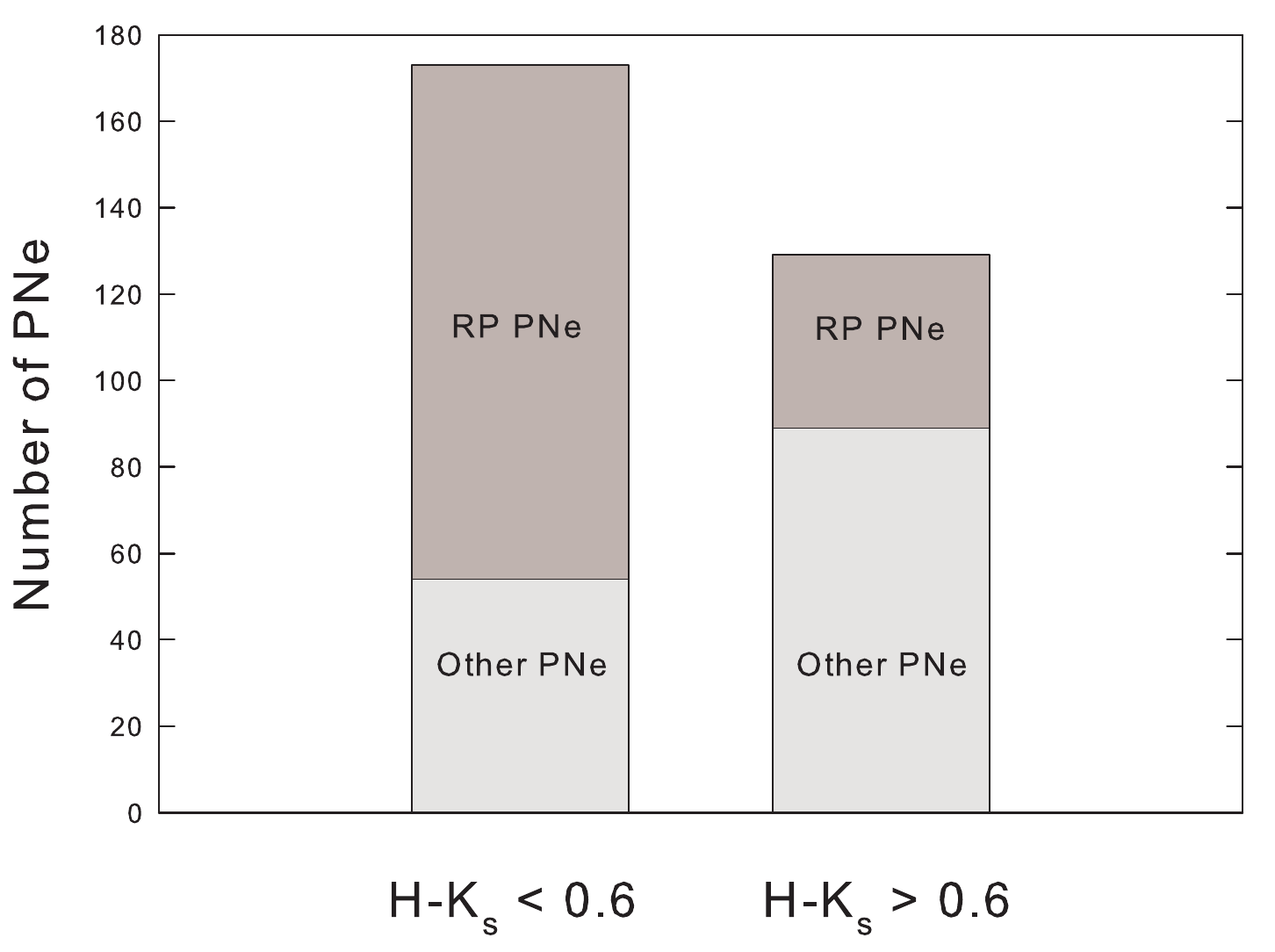}\\
  \caption{The number of previously known and newly identified LMC PNe with \textsl{H-K$_{s}$} $<$\,0.6 and $>$\,0.6. The separation between the two groups is shown as a dotted line on the plot in Fig.~\ref{Figure1}. PNe in the \textsl{H-K$_{s}$} $<$\,0.6 bin are likely to be more evolved, carbon poor and lacking in solid-state features.}
  \label{Figure4}
  \end{center}
  \end{figure}

 \begin{figure*}
\begin{center}
  \includegraphics[width=0.9\textwidth]{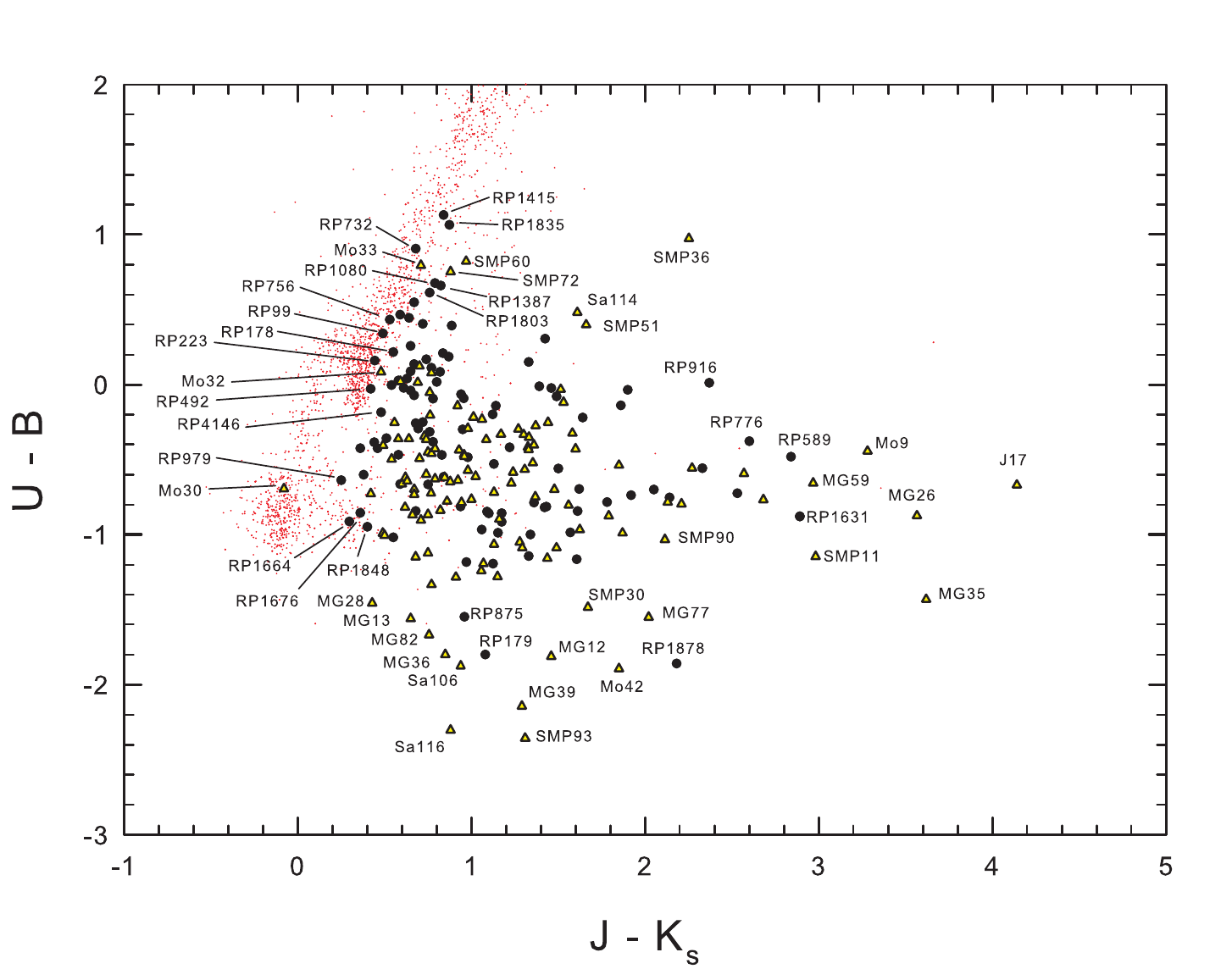}\\
  \caption{Same as Fig.~\ref{Figure3} but for the \textsl{J -- K} versus \textsl{U -- B} colour--colour diagram.}
  \label{Figure5}
  \end{center}
  \end{figure*}

  \begin{figure*}
\begin{center}
  \includegraphics[width=0.9\textwidth]{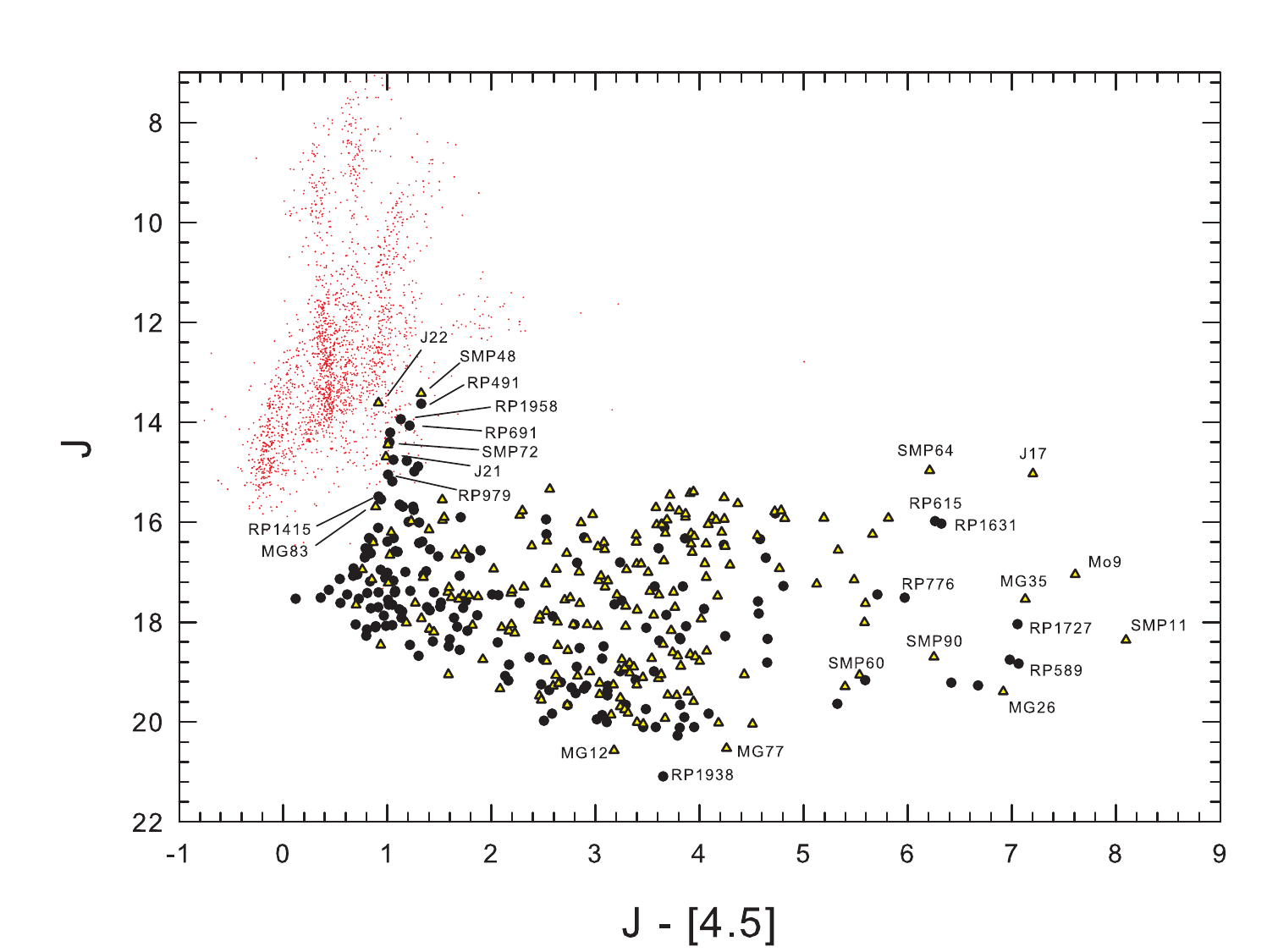}\\
  \caption{Same as Fig.~\ref{Figure3} but for the \textsl{J} -- [4.5] versus \textsl{J} colour--magnitude diagram.}
  \label{Figure6}
  \end{center}
  \end{figure*}

  \begin{figure*}
\begin{center}
  \includegraphics[width=0.9\textwidth]{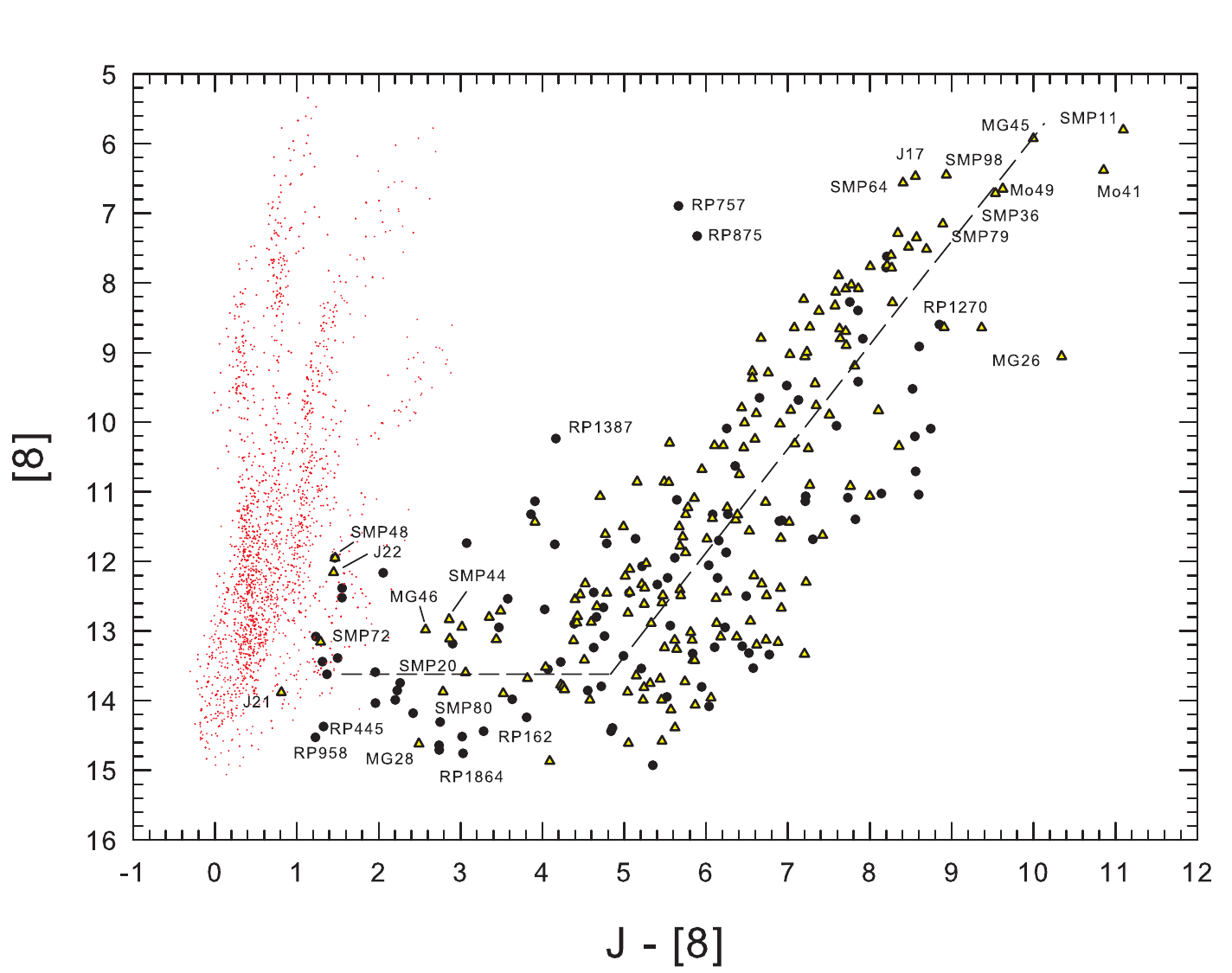}\\
  \caption{Same as Fig.~\ref{Figure3} but for the \textsl{J} -- [8] versus [8] colour--magnitude diagram. A line of best fit, derived either side of \textsl{J} -- [8] $\sim$ 5, has been included in order to show the changing trend in faint PNe.}
  \label{Figure7}
  \end{center}
  \end{figure*}

  \begin{figure*}
\begin{center}
  \includegraphics[width=0.9\textwidth]{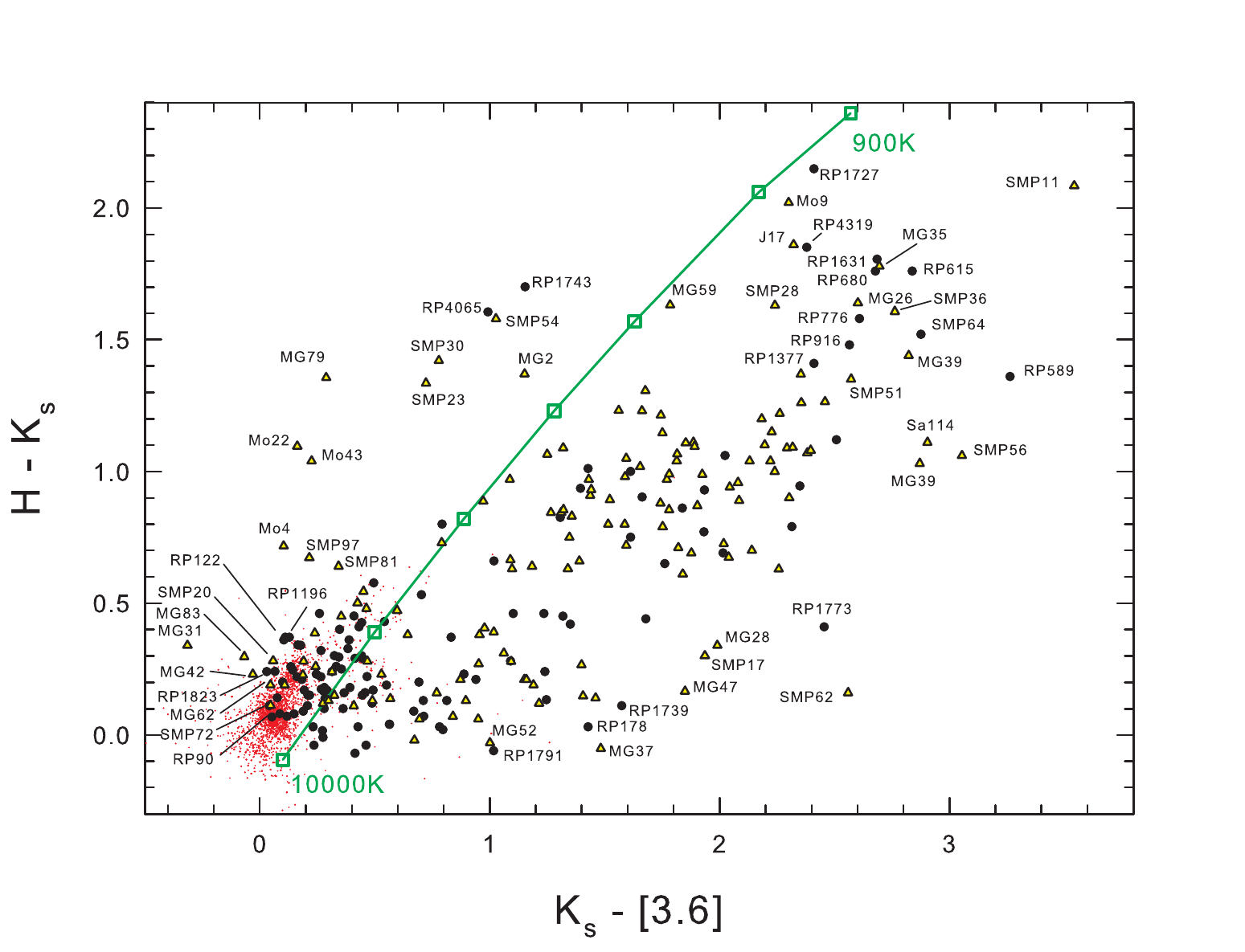}\\
  \caption{Same as Fig.~\ref{Figure3} but for the \textsl{K} -- [3.6] versus \textsl{H -- K} colour--colour diagram. Blackbody temperatures are plotted and shown as connected green squares. From the top down the squares indicate blackbody temperatures of 900, 1000, 1250, 1500, 2000, 3000 and 10\,000K.}
  \label{Figure8}
  \end{center}
  \end{figure*}

  \begin{figure*}
\begin{center}
  \includegraphics[width=0.9\textwidth]{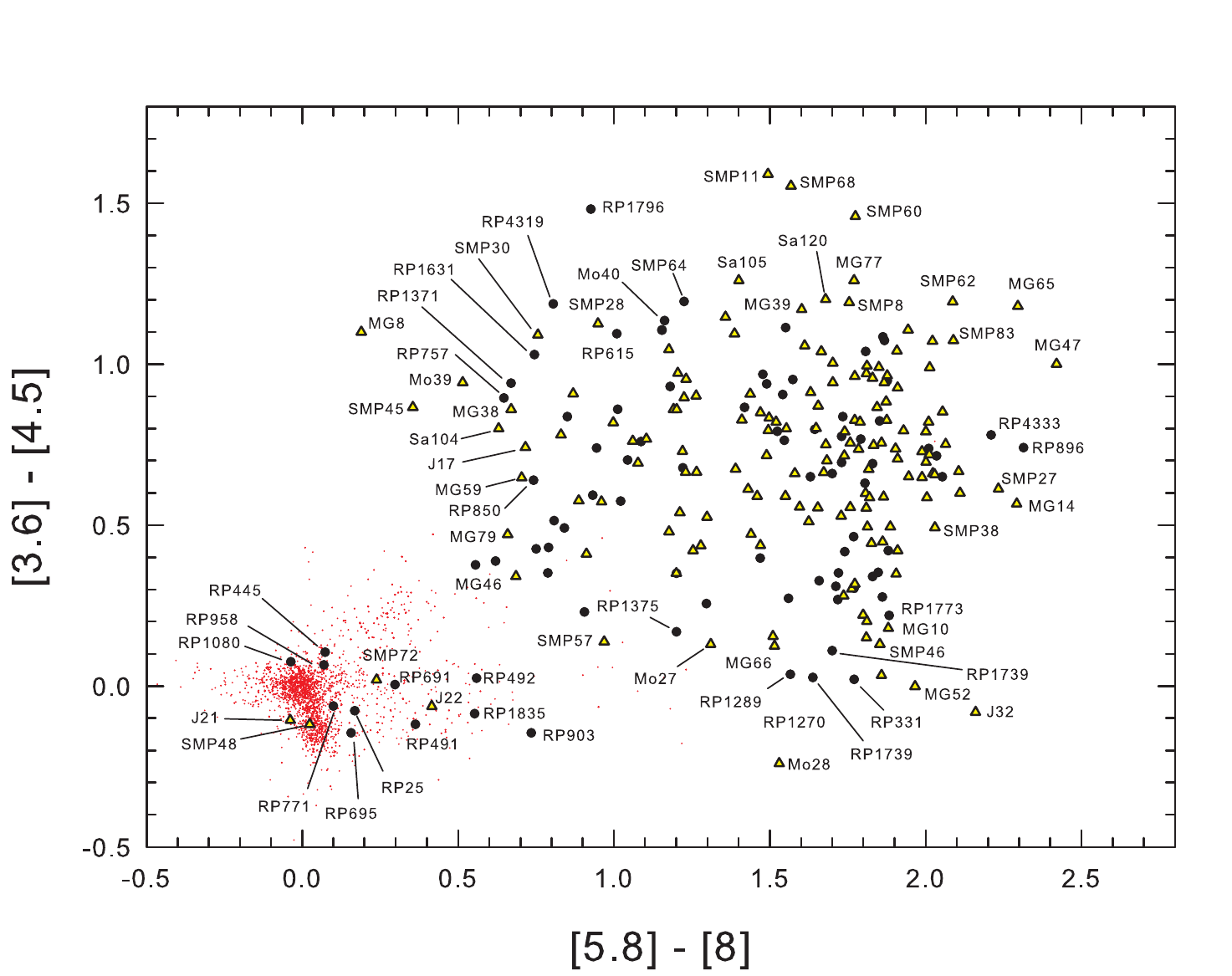}\\
  \caption{Same as Fig.~\ref{Figure3} but for the [5.8] -- [8] versus [3.6] -- [4.5] colour--colour diagram.}
  \label{Figure9}
  \end{center}
  \end{figure*}

  \begin{figure*}
\begin{center}
  \includegraphics[width=0.9\textwidth]{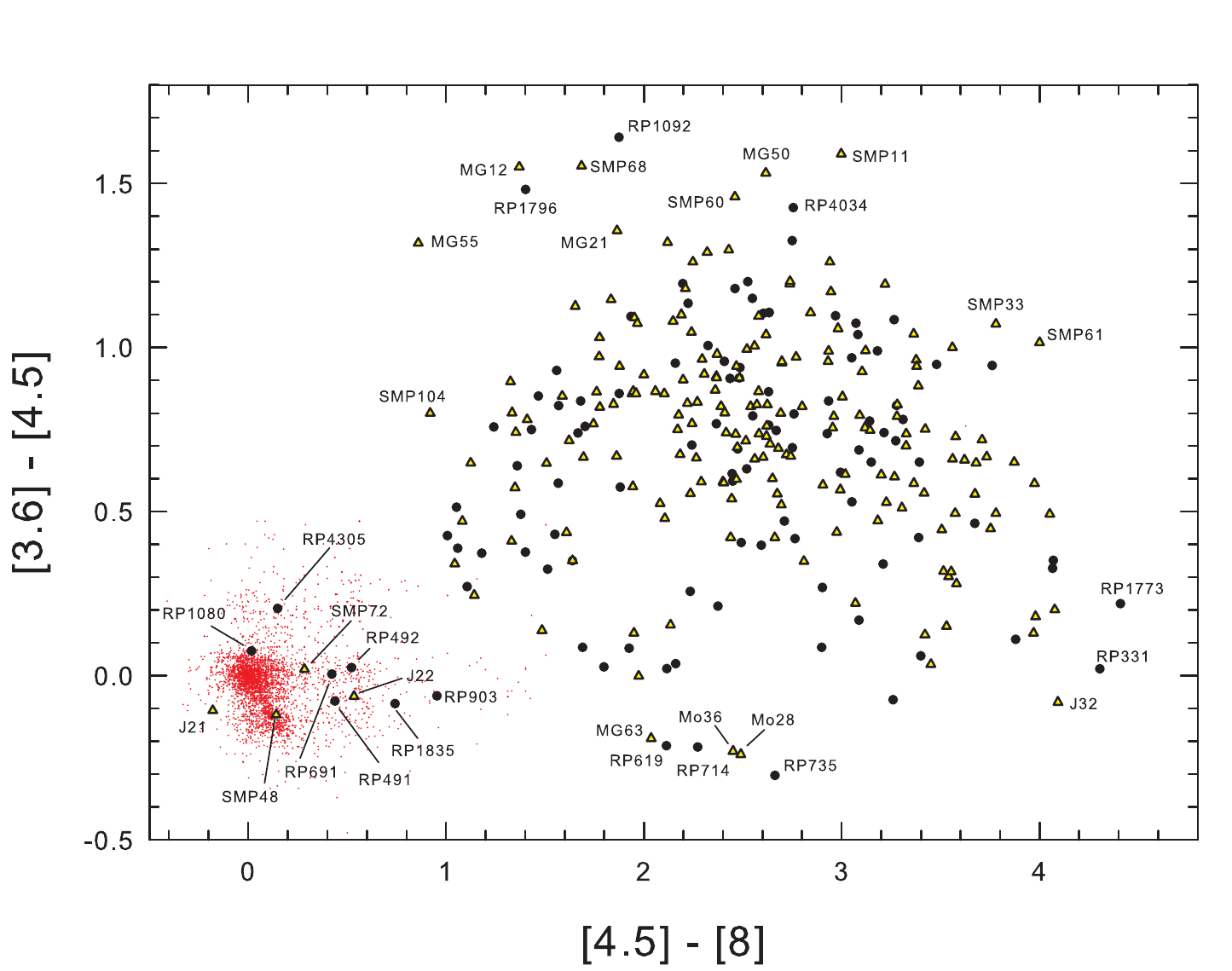}\\
  \caption{Same as Fig.~\ref{Figure3} but for the [4.5] -- [8] versus [3.6] -- [4.5] colour--colour diagram.}
  \label{Figure10}
  \end{center}
  \end{figure*}

   \begin{figure*}
\begin{center}
  \includegraphics[width=0.9\textwidth]{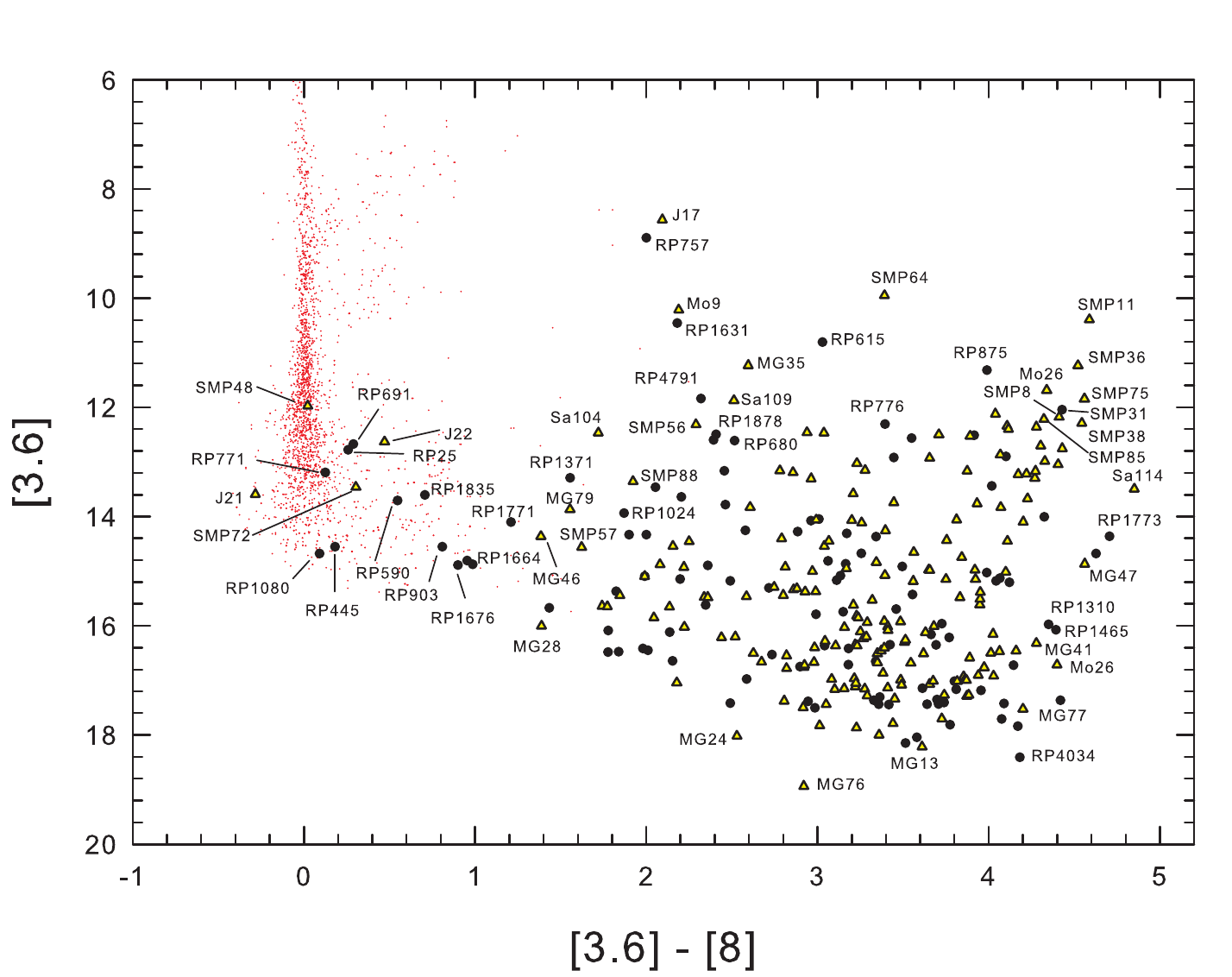}\\
  \caption{Same as Fig.~\ref{Figure3} but for the [3.6] -- [8] versus [3.6] colour--magnitude diagram.}
  \label{Figure11}
  \end{center}
  \end{figure*}

   \begin{figure*}
\begin{center}
  \includegraphics[width=0.9\textwidth]{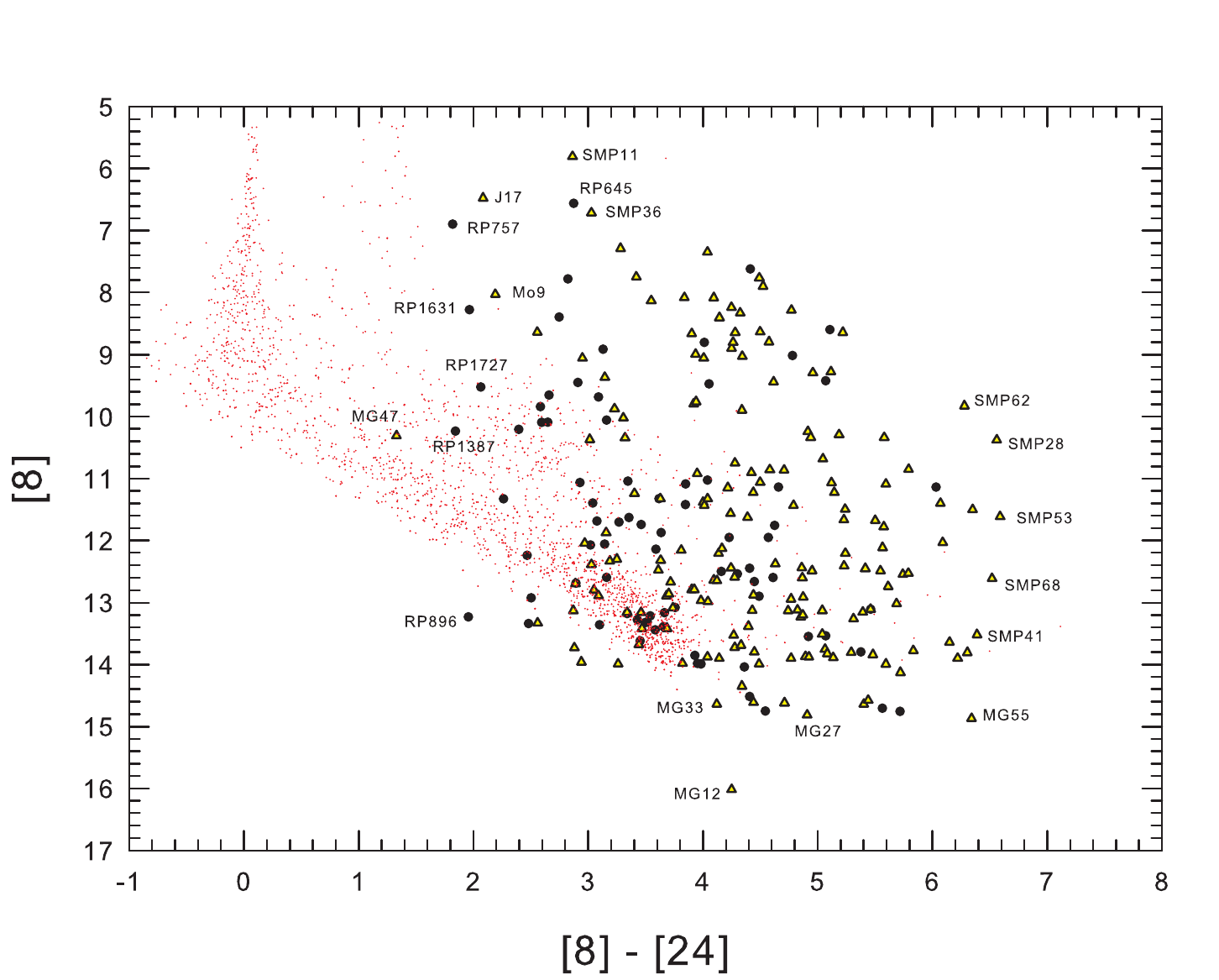}\\
  \caption{The [8] -- [24] versus [8] colour--magnitude diagram. }
  \label{Figure12}
  \end{center}
  \end{figure*}

 Fig.~\ref{Figure3} shows the three NIR $\textsl{J}$, $\textsl{H}$ and $\textsl{K}$$_{s}$ bands compared using $\textsl{J -- H}$ and $\textsl{H -- K}$$_{s}$. This plot gives the NIR ratios for LMC PNe and compares them to 3000 primarily main-sequence stars shown as red dots. A large percentage of stars clump around $y$-axis = 0 : $x$-axis = 0 (0 : 0) which indicates a relatively flat spectral energy across these bands. Nonetheless, with increasing separation between $\textsl{J}$ and $\textsl{H}$, the stars form an elongated group extending from the (0.0 : -0.2) position to the (0.4 : 1.0) position. Clearly, there is an increase in $\textsl{H -- K}$$_{s}$ for main sequence stars as we move towards the red in $\textsl{J -- H}$. The majority of stars around (0 : 0), including those with negative values on both axes, are likely to be very hot B-type stars. Moving to the right, the spectral types cool until reaching class M, late-type stars, which congregate around (0.3 : 0.9).

 The PNe form two groups. The first is clustered with the field stars. The other is a more discordant group, all of which have $\textsl{H}$ -- $\textsl{K}$$_{s}$ $>$ 0.6 but range in $\textsl{J}$ -- $\textsl{H}$ from 0.015 to 2.3. This separation of PNe into two groups was first seen in the colour--colour plots of Allen \& Glass (1974), where one large group of PNe was tightly centred on the cool M-type stars (where we find SMP60) and the other PNe were more dispersed but centred near where we find Sa104 in Fig.~\ref{Figure3}. Since highly evolved PNe have flatter SEDs, I believe that the shared position of PNe with ordinary stars in $\textsl{H -- K}$$_{s}$ is not in itself a reason to dismiss these objects as PNe. Multi-wavelength tools, however, can be used to search for other characteristics that may assist our understanding of these PNe.

 A dotted line is placed at the $\textsl{H -- K}$$_{s}$~position of $\sim$0.6 in order to neatly separate the two groups. The dotted line does not represent a real division but is arbitrarily placed for the sole purpose of comparison. The position was determined on the basis of a deficit of PNe in this region and a drop in stellar density above this line. Since the Reid \& Parker (2006a,b) objects are the fainter and presumably more evolved PNe according to their mainly medium- to low-excitation class classifications (Reid \& Parker 2010a) and central star temperatures (Reid 2013), there is a value in comparing the ratios of RP PNe to other PNe within these two bins. Using $\textsl{H}$--$\textsl{K}$$_{s}$ = 0.6 as the dividing line, the RP versus other PN ratios are shown in Fig.~\ref{Figure4}. With 173 PNe, the $\textsl{H}$ -- $\textsl{K}$$_{s}$ $<$\,0.6 bin contains the largest sample, representing 57\% of the total. Of these, 119 or 68.7\% are RP PNe. The $\textsl{H}$ -- $\textsl{K}$$_{s}$ $>$ 0.6 bin contains 129 or 42\% of the sample with 41 or 31.7\% RP PNe.

Since the distribution of RP to other PNe shown in Fig.~\ref{Figure4} is clearly disproportionate to the size of the bins, and PNe with $\textsl{H}$ -- $\textsl{K}$$_{s}$ $<$\,0.6 share colour space with the LMC stellar population, it raises the question of whether there are other quantifiable characteristics that may be attributed to PNe with $\textsl{H}$--$\textsl{K}$$_{s}$ $<$\,0.6. The data show that such PNe do not appear to be influenced by the intrinsic magnitude of the emission in the $\textsl{H}$ or $\textsl{K}$$_{s}$ band. PNe above $\textsl{H}$--$\textsl{K}$$_{s}$ = 0.6 have a mean $\textsl{H}$ magnitude of 16.62$\pm$1.2 whereas those below have a mean $\textsl{H}$ magnitude of 16.35$\pm$1.24. Since PNe below the 0.6 line have an average $H$ magnitude only 0.27mag lower than those above, the NIR magnitude itself can safely be ruled out as a contributing factor.

The answer may lie in characteristics that are peculiar to LMC PNe. It has previously been found that, in many cases, LMC PNe brighten only gradually through the NIR due to low dust continuum levels -- a feature not seen in Galactic PNe (Stanghellini et al. 2007). The low levels of grain emission appear to extend to the MIR as does the absence of dust features. It is in the MIR that the first major differences between those PNe above and below $\textsl{H}$ -- $\textsl{K}$$_{s}$ = 0.6 begin to emerge. PNe with $\textsl{H}$ -- $\textsl{K}$$_{s}$ $>$ 0.6 have a greater magnitude excess in the 8$\mu$m band. For $\textsl{H}$ -- $\textsl{K}$$_{s}$ $>$ 0.6, the mean 3.6$\mu$m -- 8$\mu$m excess is 3.27 $\pm$1.38mag suggesting a strong contribution from polycyclic aromatic hydrocarbons (PAHs) and/or warm dust
continuum. This in turn suggests that they are more likely to be carbon rich rather than evolved PNe that are weak in their solid-state features. The same 3.6$\mu$m -- 8$\mu$m mean excess for PNe with $\textsl{H}$ -- $\textsl{K}$$_{s}$ $<$\,0.6 is only 1.8 $\pm$2.44mag. The relative flux deficiency and higher
standard deviation in the 8$\mu$m band for PNe $\textsl{H}$ -- $\textsl{K}$$_{s}$ $<$\,0.6 suggest that the majority of these PNe are likely to be carbon poor although some may have oxygen-rich dust, which may increase the magnitude at 8$\mu$m depending on the strength of other lines such as \ArII, \NeVI~and \ArIII~within that band. Since the $\textsl{H}$ -- $\textsl{K}$$_{s}$ ratio appears to be strongly influenced by dust composition, there is likely to be an overall effect on forbidden emission in PNe where $\textsl{H}$ -- $\textsl{K}$$_{s}$ $<$\,0.6. Using optical emission lines from every measured PN in the LMC, the following values are derived in order to highlight the relation between the dust composition and forbidden emission.\\
 \\
 The mean for PNe with $\textsl{H}$--$\textsl{K}$$_{s}$$>$\,0.6:\\
 $\bullet$ flux at \OIII\,5007\AA = 1.05E--12\,$\pm$1.86E--12\\
 $\bullet$ \OIII\,5007\AA/\,H$\beta$ ratio = 7.48\,$\pm$5.3\\
 $\bullet$ \NII\,6583/\,H$\alpha$ ratio = 0.57\,$\pm$1.35.\\
 \\
 The mean for PNe with $\textsl{H}$--$\textsl{K}$$_{s}$$<$\,0.6:\\
 $\bullet$ flux at \OIII\,5007\AA = 1.29E--13\,$\pm$4.46E--13\\
 $\bullet$ \OIII\,5007\AA/\,H$\beta$ ratio = 6.22\,$\pm$3.9\\
 $\bullet$ \NII\,6583/\,H$\alpha$ ratio = 1.26\,$\pm$1.59.\\


  The higher \OIII\,5007\AA~flux level and higher \OIII\,5007\AA/H$\beta$ ratio seen in PNe with $\textsl{H}$ -- $\textsl{K}$$_{s}$$>$\,0.6 are consistent with carbon-rich dust in the nebula. In particular, it is the \NII\,6583/H$\alpha$ ratio that aids in separating PNe with carbon-rich dust from PNe with oxygen-rich dust. PNe with $\textsl{H}$ -- $\textsl{K}$$_{s}$ $>$ 0.6 show a steady increase in $\textsl{J -- H}$ with increasing $\textsl{H -- K}$$_{s}$. PNe at the top right of the plot show the steepest and most consistent energy rise across $\textsl{J}$, $\textsl{H}$ and $\textsl{K}$$_{s}$. These PNe, including RP1727, Mo9, MG35, MG26 and J17, are likely to contain warm carbon-rich dust and probably contain some combination of PAHs and Pa$\beta$ nebula emission. In particular, the six PNe with $\textsl{J--H}$ $>$ 1.5 have a mean \NII\,6563/H$\alpha$ ratio of only 0.05 with the highest ratio in the group only 0.11. This allows us to predict the NIR SED simply by measuring key optical emission-line ratios. It also provides a reason for the flat NIR SEDs seen in a large number of LMC PNe.

  LMC PNe with $\textsl{H}$ -- $\textsl{K}$$_{s}$ $<$\,0.6 have a mean \NII\,6583\AA/H$\alpha$ ratio which is more than twice that found for \OIII\,5007\AA~bright PNe with $\textsl{H}$ -- $\textsl{K}$$_{s}$ $>$ 0.6. This high \NII\,6583\AA/H$\alpha$ ratio is a strong indicator of carbon-poor dust in the nebula with no significant contribution from PAHs. The high mass PNe have experienced the hot bottom burning (HBB) nucleosynthesis process which depletes carbon while increasing nitrogen in the nebula (eg., Marigo 2001). Hence, their spectra may show oxygen-rich emission features such as crystalline silicates but we should expect that a large proportion may be termed `featureless' or lacking in solid-state MIR features. This absence of solid-state emission features may be attributed to either the highly evolved state of the PN or the object falling below the detection limit (see Stanghellini et al. 2007).

  The low metallicity of the LMC ($\sim$50\% Galactic; e.g. Caputo et al. 1999) also plays an important role in dust production as well as its composition, grain size and the period of time it is warmed by UV radiation from the central star. The correlation between the decreasing metallicity and increasing carbon- to oxygen-rich ratio for Local Group galaxies is well documented (eg. Cioni et al 2003; Schulthesis et al 2004). The amount of dust per gas unit mass in the LMC is $\sim$3.3 times smaller than that for the Milky Way. The amount of dust per H atom in CO-traced molecular gas is $\sim$3.7 times less than that in atomic H gas in the LMC (Bernard et al. 2008) whereas the ratio is the same in the Milky Way. Because AGB stars in the LMC are more opaque than those in the Galaxy (e.g., Zhukovska \& Henning 2013) it follows that less dust will be formed around AGB stars in the LMC's low-metallicity environment. Since dust plays an important role in the mass-loss efficiency of AGB stars, it has been suggested that AGB stars in the Magellanic Clouds may spend a longer period dredging up and synthesizing nuclear-processed material prior to the PN stage (Stanghellini et al. 2007). That being the case, we should expect to find LMC PNe with lower dust content than those in the Galaxy. Likewise there should also be a higher ratio of PNe with oxygen-rich dust and higher \NII/H$\alpha$ emission ratios. The high proportion of LMC PNe with $\textsl{H}$ -- $\textsl{K}$$_{s}$$<$\,0.6 is a strong indication that this is correct. Since the most massive PN progenitors experience HBB, which depletes gaseous carbon (Leisy \& Dennefeld, 2006), it follows that many PNe within $\textsl{H}$ -- $\textsl{K}$$_{s}$$<$\,0.6 had high-mass progenitors.


\subsection{\textsl{J -- K}$_{s}$ versus \textsl{U -- B}}

The $\textsl{J -- K}$$_{s}$ versus $\textsl{U -- B}$ plot is more successful at isolating a large percentage of PNe from main sequence stars (see Fig.~\ref{Figure5}). The majority of PNe are encased in a triangular region between points (1 : 0.8), (-1.4 : 0.2) and (-0.6 : 4.1) converging on J17. To the right of the plot, where $\textsl{J -- K}$$_{s}$ $>$ 3, there are a number of PNe, such as Mo9, MG26, SMP11, MG35 and J17, with steep NIR SEDs. This is an area of the plot where we would also expect to find symbiotic stars (eg. Miszalski et al. 2011b). At the top end of the plot we find a number of PNe with $\textsl{U--B}$ $>$ 0.2 which follow the red boundary of main stellar population. These PNe either have very low dust reddening or have evolved to the point where the central star is unable to fully ionize the nebulae. In the latter case, the nebulae are ionization-bound, meaning that the radius of the ionized region around the central star is determined by the absorption of the ionizing photons coming from the inner nebula. As such, even if larger amounts of gas exist in the outer regions of the PN, they will remain neutral. One method suggested for assessing whether a nebula is ionization-bound is to detect the presence of \OI$\lambda$6300 (Bieging et al. 2008) since this line originates in the region of transition between the ionized and neutral hydrogen. All of these PNe do exhibit the \OI$\lambda$6300 line but the magnitudes and \OI$\lambda$6300/H$\alpha$ ratios are not significantly higher than those found where $\textsl{U -- B}$ $>$ -1.0. The same applies to the presence of neutral or molecular hydrogen possibly together with other molecules (Rodr\'{\i}guez et al. 2009). Without a detailed spectrum for each object in the MIR this association is not able to be determined at the present time.

In total, there are 34 PNe with $\textsl{U -- B}$ $>$ 0.2, but due to the absence of published $\textsl{J}$ or $\textsl{K}$$_{s}$ NIR measurements, 11 PNe including Mo10 ($\textsl{U--B}$=1.137), SMP10 ($\textsl{U -- B}$=0.986), J31 ($\textsl{U -- B}$=0.481) and MG18 ($\textsl{U -- B}$=0.47) were unable to be plotted. Of the 23 PNe with $\textsl{U -- B}$ $>$ 0.2 that were able to be plotted in Fig.~\ref{Figure5}, 18 are RP PNe, 11 of which (in order of $U$ -- $B$ strength: RP1415, RP1835, RP1080, RP1354, RP1387, RP771, RP679, RP756, RP900, RP102 and RP735) are low excitation PNe with \NII $>$~H$\alpha$. The remaining seven RP PNe and SMP36, SMP51, SMP60, Mo33, SMP72 and Sa114 have a small measure of continuum (generally $\leq$1 the level of H$\beta$) evident in their low-resolution spectra. Of these PNe, the $U$ and $B$ photometry for RP1267, RP178, RP90, RP903, RP1803, Mo33, SMP51, SMP60 and SMP36 may include light from bright, nearby stars where coincidental or very close objects can be seen optically and spectroscopically. SAGE images in particular show that SMP36 is partially covered by the large radial extent from its neighbour: the infrared source 2MASS J05104124--6836066, also known as MSX LMC45. This object is not visible in the H$\alpha$/SR stacked images (composed of 12 exposures made between 1997 and 2000; see Reid \& Parker 2006a,b) but is much brighter than SMP36 in SAGE bands. Only RP99 (very faint), RP732 (very faint) and Sa114 are well separated from nearby sources yet show some level of stellar continuum in their spectrum, which accounts for their positive $U$ -- $B$ position in Fig.~\ref{Figure5}. A false colour mosaic of the SAGE [4.5], [5.6] and [8] bands (see Fig.~ref{Figure1}) hints at the same possible eclipsing of SMP72 by a field star that was first seen in the H$\alpha$/SR stacked images and subsequently confirmed by five independent spectroscopic observations (Reid \& Parker 2006b, 2010b).

Based on these findings it follows that objects located above $\sim$$\textsl{U -- B}$=2 are either faint, low-excitation PNe where the central star is beginning to dominate the system or the $U$ and $B$ photometry is collecting additional light from nearby stellar sources. In some cases, such as RP735, RP900, RP756 and RP679, both scenarios are likely at the same time.

  \subsection{\textsl{J} -- [4.5] versus \textsl{J}}

  In the $\textsl{J}$ -- [4.5] versus $\textsl{J}$ plot (Fig.~\ref{Figure6}), there is a distinct blue-end cut-off in the data within 17.8 $<$ $\textsl{J}$ $<$ 14 as $\textsl{J}$ -- [4.5] moves from 0.2 to 1.0. This cutoff, not seen before, shows that as the $\textsl{J}$ flux of the central star decreases, the minimum $J$ -- [4.5] value decreases exponentially towards a flat $\textsl{J}$ -- [4.5] SED at $J$\,=\,17.5. The most conspicuous outlying object at the blue end of the plot is J22. This object has a strong continuum (eight times the strength of H$\beta$) and probably suffers with contamination from a bright star 2 arcsec to the NW. If such a star is dominating the photometry, we should not be surprised by the position of this object in the plot. The PNe at the blue end of the plot ( $\textsl{J}$ -- [4.5] $<$ 1) are less influenced by cool dust, allowing the central star to contribute most of the measured flux at [4.5] thereby creating an interesting correlation between what appears to be the minimum physically allowable $\textsl{J}$ -- [4.5] colours with decreasing $\textsl{J}$ flux.

  PNe brighter than $\textsl{J}$ mag 15 with low $\textsl{J}$ -- [4.5] colours ($\sim$1) reside in an area of the plot also occupied by main-sequence field stars in the LMC. The best method to explain the position of these objects is to examine each one with respect to its location (in the case of crowding), optical spectrum and combined MIR colours. In the case of RP1958 the occulted PN is extremely difficult to separate from the bright star only 1 arcsec to the north. Despite the measured high \NII/H$\alpha$ ratio of 2.9, the \OIII\,5007\AA~emission is only 3$\sigma$ the noise and there is little else in its spectrum to suggest it is a PN. The MIR SED is quite flat and there is some continuum in the most recent optical spectrum. This object has therefore been removed from the PN catalogue.

  Despite its position on the plot, RP491 has \OIII/H$\beta$ = 5.9 and \HeII\,4686/H$\beta$ = 0.94, strongly indicating a PN. Although there is little evidence of continuum, the optical image suggests that RP491 may be partially obscured by another star in the line of sight. SAGE images show some excess at 8$\mu$m but the SED is quite flat. Because of the strong spectral confirmation this object should continue to be accepted as a PN which may have at least one intervening star. The same may also apply to J22, where \NII $>$ H$\alpha$$\sim$1.6 and \OIII\,4959/H$\beta$$\sim$3 but with a strong stellar continuum and evidence of two nearby stars. With the H$\alpha$ excess visible only as a small arc to the NE of the stellar source recorded as J22, it is possible that most of the PN has actually been obscured from view. On the other hand, it is still possible that J22 is a PN mimic since there appears to be a decreasing MIR excess, positioning it together with other field stars.

  The H$\alpha$~images of RP691 and SMP72 show emission to the north of the main object. The spectrum of RP691 has \NII\,6583\,$/$\,H$\alpha$\,=\,1.4 and \OIII\,4959\,/\,H$\beta$\,=\,5 while SMP72 has strong \OIII~and no \NII. Neither of these objects have any MIR excess. In the case of RP691, there is evidence of coincidence with a bright star 1.5 arcsec to the south obscuring most of the H$\alpha$ flux. With a convincing PN spectrum, this object is left in the PN catalogue at present.

  RP979 has the characteristic 3-to-1 \OIII/H$\beta$~ratios but has a relatively strong continuum. Since the main body of emission is to the immediate north of this object it is highly possible that the star in question is not associated with the emission. In either case, it is safer to change this object from a `true' PN to a `possible' PN. RP1415 shows \NII/H$\alpha$= 4.8 but there is some continuum. It resides in a very dense dust cloud and this may have an impact on dust that was expelled during the post-AGB stage.

  The small amount of H$\alpha$ emission seen in the H$\alpha$/SR image to the right of the object known as J21 suggests that the main object is a star rather than a PN. With a strong continuum seen in the spectrum, either the star is foreground to the very small and highly evolved PN, possibly detected to the right, or it is not a PN at all. In short, all the PNe in this part of the plot can be summarized as showing some degree of continuum in their spectrum. Possible reasons for the continuum are quite varied and may need to be confirmed using very deep, space-based imaging.

  The other main feature of this plot is the linear cut-off within 0.2 $<$ $\textsl{J}$ -- [4.5] $<$ 4 with decreasing $\textsl{J}$ mag from 17.8 to 20.5. This is a strong indication of the sensitivity cut-off for [4.5] in IRAC. Interestingly, there appears to be a faint nadir for $\textsl{J}$ magnitudes at about $\textsl{J}$ -- [4.5] = 3.7. The PNe in this region (MG12, RP1938 and MG77) all have very faint detections at $\textsl{J}$ but are approximately 4 times brighter in [4.5]. Because their emission line ratios are varied, there appear to be no optical characteristics to link them.

  For values of $\textsl{J}$ -- [4.5] higher than $\sim$4, the increasing brightness of the [4.5] flux sees a small rise in the minimum $\textsl{J}$ flux, due largely to a steepening SED for these, mainly young, PNe. At this far end of the plot, we find very bright MIR colours so the outlying examples have been specifically named. The MIR detections show a substantial rise through [3.6] and [4.5] to be extremely bright in [5.6] and [8] making them some of the steepest PN SEDs in the LMC sample. I note that Miszalski et al. (2011b) classify all objects beyond $\textsl{J}$ -- [4.5] = 5 as symbiotic stars since they occupy an area of the plot where symbiotic stars are to be found. After examining spectral features and NIR and MIR colours, I have decided it is inadvisable to classify objects purely on the basis of their position in this plot.

  \subsection{$\textsl{J}$ -- [8] versus [8]}

  Fig.~\ref{Figure7} may be compared to similar plots in Hora et al. (2008) and Blum et al. (2006). The field stars separate into four distinct spurs as they increase in 8$\mu$m mag. The two spurs on the left are most likely to consist of main-sequence stars while the next to the right is likely to include M supergiants and luminous oxygen-rich stars. The fourth spur, extending within 8.5 $<$ [8] $<$ 10.5, is likely to consist of carbon-rich AGB stars. If extreme AGB stars were present in this plot, they could be expected to form an extension of this spur leading to SMP64. In between this spur and the M supergiants, there is a small group around (10 : 1.5) that is likely to contain some oxygen-rich AGB stars (eg. Hora et al. 2008; Whitney et al. 2008).

  The PNe form a group gradually reddening through 1 $<$ $\textsl{J}$ -- [8] $<$\,4.5, after which they substantially increase in 8$\mu$m mag with a smaller increase in $\textsl{J}$ mag. Although they overlap with background galaxy candidates, as shown in Blum et al. (2006), the main body of LMC PNe in this plot extend to somewhat fainter magnitudes in [8] than previously seen. To the left end of the plot, PNe bluer than ($\textsl{J}$ -- [8]) = 4.8 go no fainter than mag 15 in [8]. Although this cut-off is mainly due to the sensitivity limit of IRAC in [8], these PNe ($\textsl{J}$ -- [8]) $<$ 4.8 begin to move towards the main clump of stars. A line of best fit, which shows this changing trend, is included in Fig.~\ref{Figure7}. Such a change is attributable to an increase in $\textsl{J}$ magnitude which will also result in a flatter SED.  This was hinted at in Fig. 7 of Hora et al. (2008) but they found only a few LMC PNe in this area. The reason for the flatter SED and more prominent $\textsl{J}$ mag in many PNe with [8] mag less than 12 could be their evolved state. For this type of PNe the H recombination lines, such as Pa$\beta$ at 1.28$\mu$m, dominate the NIR emission (Hora et al. 1999), making evolved objects much brighter than they would be if flux were coming from stellar continuum alone. Correspondingly, the dust temperature is expected to decrease with time, aided by the gradual destruction of molecules and sputtering of material off the surface of dust grains. This process has been shown to correlate with the radius of the nebula as it expands with age, suggesting a power-law decline of $T$$_{dust}$\,$\propto$\,$R$$^{-0.25}$ (Stanghellini et al. 2007). This increasingly affects the objects named at the left of the plot. Two PNe, SMP48 and J22, lie above the main group with brighter flux at 8$\mu$m. Like J22, mentioned earlier, J21 lies at the edge of the stellar group in most of the figures, indicating a strong stellar component.

  At the other end of the plot, objects with very strong 8$\mu$m emission (9 $\leq$ 8$\mu$m $\leq$ 5.5) increase substantially in $\textsl{J}$ -- [8], yet the data show that the range of $\textsl{J}$ magnitudes only increases about five times from the extreme blue to red sides of the plot. A high [8] luminosity therefore loosely correlates with a steepening SED contributed by dust and PAHs, over and above the stellar component. High-excitation PNe, such as SMP11, RP1337, SMP36, J17 and SMP64, lie in the area of the plot where extreme AGB stars are to be found. PNe in this part of the plot are expected to have strong PAH bands at [8] and continuum dust emission, typical of very young PNe. The red-ward cut-off stretching from MG26 down to fainter fluxes in [8] (15 : 6) is likely due to the detection limits of the 2MASS and IRSF surveys. Although new data have extended the plot for LMC PNe further redward than was previously achievable, the inclusion of Galactic PNe by Hora et al. (2008) in a similar plot indicates that extremely faint, low excitation LMC PNe would extend the data at least 1 mag further to the red if $Spitzer$ sensitivity limits were increased.

  \subsection{ $\textsl{K}$$_{s}$ -- [3.6] versus $\textsl{H -- K}$$_{s}$}

  By plotting $\textsl{K}$$_{s}$ -- [3.6] versus $\textsl{H -- K}$$_{s}$ we are mainly looking at stellar continuum and the starlight that is scattered from surrounding dust, although dust of various kinds will make some contribution to the observed levels, particularly at $\textsl{K}$$_{s}$ -- [3.6]. A larger percentage of PNe in Fig.~\ref{Figure8}~can therefore be expected to share the area of the plot occupied by stars. In their version of this plot, Hora et al. (2008) found that the PNe are divided into two main groups, one clustering at the locus of the stars and the other at (1.0 : 2.2) with a large gap within 0.8 $<$ $\textsl{K}$$_{s}$ -- [3.6] $<$ 1.6. Although they found several PNe, such as MG52, that are slightly redder than the stellar group, the larger number of PNe show that these two groups meet up. Although about a third of the PNe cluster tightly with the field stars, the broad cluster centred at (1.0 : 2.2) is still a key feature. The current multiwavelength data show that most of the PNe gathering close to the field stars have medium to low levels of 8$\mu$m emission, indicating low PAH levels and probably low level dust continuum.

  Following Fig. 5 of Hora et al. (2008), the blackbody emission at various temperatures has been included. With dust grains formed into clusters and PAHs, UV radiation from the central star continually evaporates the dust until there is little to be observed at late evolutionary stages. This is a process that continues to affect the nebula throughout the PN stage. While thermal dust emission with temperatures $\sim$800K may be characteristic at the central locus for PNe in the plot (1.0 : 2.2; Allen \& Glass 1974), higher temperature grains in the order of 1000K around (0.5 : 1.0) can still exist around the PNe. At condensation levels above about 1000K, however, it is likely that the dust grains have evaporated.

  There are three PNe that lie close to (1.3 : 0.3). These objects have SEDs that rise steeply through $\textsl{H -- K}$$_{s}$ but plateau from $\textsl{K}$$_{s}$ -- [3.6]. Although Mo43 is clearly a PN, MG79 is an example where the magnitudes are almost certainly contaminated by those of adjacent stars. In addition, the spectrum is not consistent with it being a PN. Mo22 has a PN-like \OIII\,5007/H$\beta$ ratio but extremely low H$\alpha$ levels. There is no evident overlap with other sources but, as confirmed by spectra, no H$\alpha$ excess can be seen in the UKST H$\alpha$/short red map. Both MG79 and Mo22 are unlikely to be PNe and are discussed in Section~\ref{subsection5.2}.

  \subsection{[5.8] -- [8] versus [3.6] -- [4.5]}

  Fig.~\ref{Figure9} provides a full comparison of the four IRAC colour--colour bands in order to assess the SED for each object through the MIR. In this plot the stars cluster very strongly around the (0 : 0) point with a prominent `tail' towards negative values in [3.6] -- [4.5]. A sparse scattering of stars is also found to the right of the main group, encompassing objects RP1664, SMP72, J22 and RP691. Because the stars have been individually selected, I have avoided the inclusion of background galaxies and active galactic nuclei (AGNs) which centre on positions (1 : 1) and (0$\rightarrow$0.5 : 0$\rightarrow$3) as shown in the similar plot of Hora et al. (2008). Plotting these bands has largely proven to be insufficient to separate YSOs which also populate the same positions as PNe. Only the optical spectrum, combined with all the MIR data, can effectively separate YSOs from highly evolved PNe.

  Although the PNe are widely scattered (-0.3$\rightarrow$1.6 : 0$\rightarrow$2.4), there is some linear clumping from (0$\rightarrow$1.2 : 1.6$\rightarrow$2). In other words, from the position of MG77 to that of MG52 there is a high density of PNe marking the line at which band [8] is twice the strength of band [5.8]. Much of this will be due to PAHs which are twice as strong in [8] as they are in [5.8]. Those PNe to the right of [5.8] -- [8] = 2 probably have other ionized gas lines and carbon-rich dust (Bernard-Salas et al. 2006; Stanghellini et al. 2007), adding to the [8] excess. Those less than [5.8] -- [8] $\sim$1.6 are more likely to have lower levels of PAHs compared to the contribution from continuum emission and molecular lines.

  At (0 : 0) we encounter PNe with very few MIR features. These objects were classed as `F' for featureless by Stanghellini et al. (2007), since they show no evidence of solid-state features. The absence of grain emission is to be expected in these highly evolved PNe (Stanghellini et al. 2007) but since the dust is neither carbon nor oxygen rich, it may even fall just below the SAGE detection limit. PNe at the (0 : 0) area, such as RP1835, J21, J22, SMP48, RP695 and SMP72, are either severely occulted by intervening stars or they are highly evolved PNe. If the latter is true, then recombination of the nebula gas is probably preventing other emission features from forming.

\subsection{ [4.5] -- [8] versus [3.6] -- [4.5]}

\begin{figure}
\begin{center}
  \includegraphics[width=0.49\textwidth]{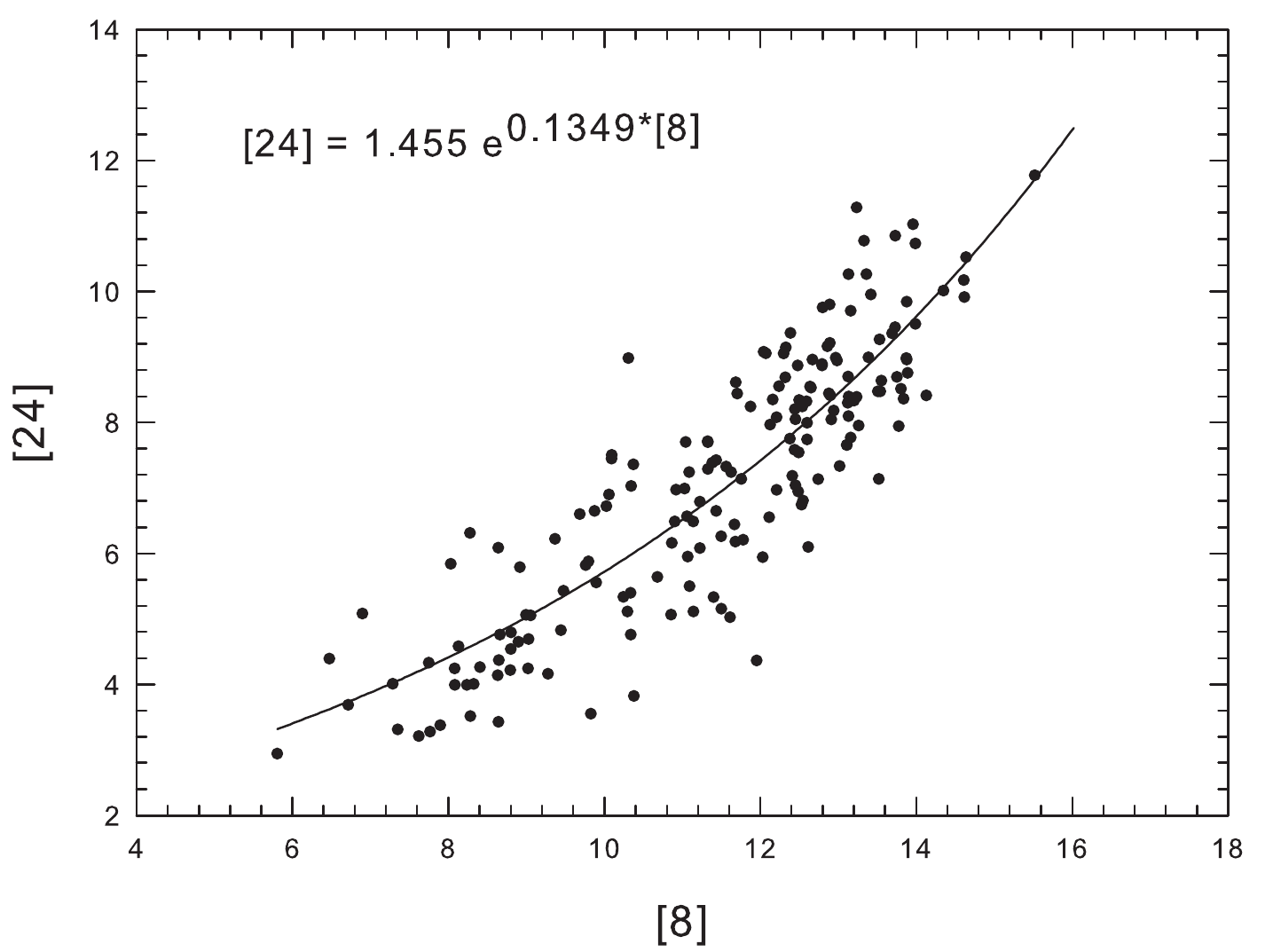}\\
  \caption{The [8] versus [24] magnitude--magnitude diagram. The equation for the exponential increase is shown on the plot. }
  \label{Figure13}
  \end{center}
  \end{figure}

   \begin{figure*}
\begin{center}
  \includegraphics[width=0.9\textwidth]{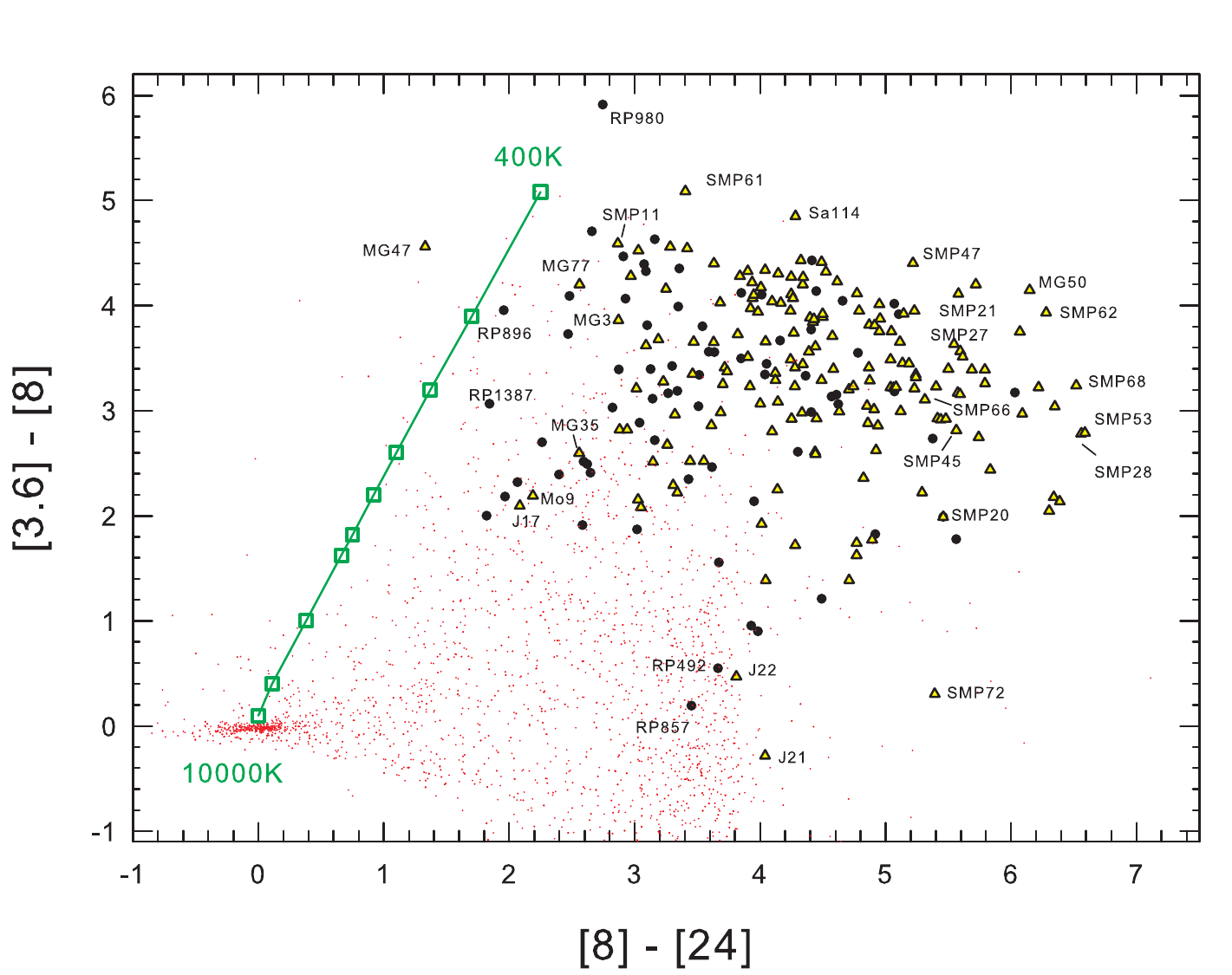}\\
  \caption{The [8] -- [24] versus [3.6] -- [8] colour-colour diagram. Blackbody temperatures are plotted as connected squares at 400, 500, 600, 700, 800, 900, 1000, 1500, 3000 and 10\,000\,K.}
  \label{Figure14}
  \end{center}
  \end{figure*}

  Like Fig.~\ref{Figure9}, Fig.~\ref{Figure10} shows the [3.6] -- [4.5] colour comparison on the $\textsl{y}$-axis but here changes to [4.5] -- [8] on the $\textsl{x}$-axis, thereby extending the colour range from --0.2 $<$ [4.5] -- [8] $<$ 4.4. The wider [4.5] to [8] range means we are comparing the fainter dust and low continuum emission in [4.5] with the stronger dust continuum, PAHs and other emission found in [8]. It is likely that PNe occupying positions less than [4.5] -- [8] $\sim$ 2.2 produce most of their emission from solid-state features rather than dust. A number of these PNe, such as MG55, MG12, RP1796 and SMP68, also have some of the highest [3.6] -- [4.5] levels in the sample, reaching in excess of 1.5. As a consequence, the SEDs for these PNe show a steep rise between [3.6] -- [4.5] but plateau considerably as they move toward [8].

  PNe at the right of the plot with [4.5] -- [8] values $>$ 3 decline in their maximum [3.6] -- [4.5] colour until (0 : 4.3). This suggests that the greater the rise in the SED from [4.5] -- [8] $\sim$2.8, the flatter it will be within [3.6] -- [4.5]. At the lower right of the plot, PNe such as RP1773, RP331, J32, SMP38, SMP75, SMP46, SMP36 and Sa114 all show large excess at [8] but no common features in the optical. Possible contributors to the flat [3.6] -- [4.5] levels for these PNe with [4.5] -- [8] $>$\,3.6 may also include a mixture of 3.3$\mu$m PAH emission and the Pf$\gamma$ H recombination line (see Hora et al. 2008).

  There is a small group of PNe centred at (-0.2 : 2.2) in the lower mid-section of the plot. For these objects the [4.5] -- [8] values are average but the negative [3.6] -- [4.5] values indicate that the central star dominates cool dust emission. Those PNe that share positions occupied by stars, close to (0 : 0), are strongly dominated by the central star in all MIR bands. Such objects are either highly evolved PNe, eclipsed by nearby intervening stars, or not PNe at all. Spectroscopy shows that they have a strong continuum component. If J21 is a PN, then it must lie to the immediate west of the object assumed to be its central star. RP445 has very strong \OIII\,5007 but a late-type stellar continuum. RP1676 also has very strong \OIII\,5007 but a B/A-type continuum. Since the data indicates that both of these objects are emission-line stars, their classification has been changed accordingly. RP491 has a low level continuum but SAGE resolution is not sensitive enough to separate any coeval stellar sources. RP1835 has a strong \NII/H$\alpha$ ratio and \HeII\,4686 is clearly seen but there is a high probability of intervening starlight due to coincidence with nearby sources. RP491, RP691, J22, SMP48, SMP72 and RP903 also suffer from intervening starlight, as discussed in previous plots. In these cases, the nebula emission can be seen extending away from one side of a bright stellar source which is partially eclipsing the possible nebula. Clearly, the highly luminous stellar source is erroneously being picked up as the central source of the PN in multi-wavelength data. This plot, more than any of the others, is able to clearly separate stellar sources from PNe.

\subsection{ [3.6] -- [8] versus [3.6]}

  In the [3.6] -- [8] versus [3.6] plot shown in Fig.~\ref{Figure11}, there is a concentration of PNe near position (16 : 3.5) despite the wide range of luminosities that are evident. This group shares an area of the plot with background galaxies (see Blum et al. 2006) and PNe blue-ward from this position occupy an area where YSOs are present (Whitney et al. 2008). Even so, PNe in this area of the plot are well separated from the main-sequence stars, which form a somewhat narrow line at the blue end where [3.6] -- [8]\,=\,0. We also see a number of extreme AGB stars which extend across the top of the plot towards (9 : 3). Here we find PNe such as J17, RP757, Mo9, and SMP64, all of which are extremely bright in the MIR. Of these, only RP757 shows evidence of low level continuum in the optical. This group generally has lower-than-average \OIII\,5007/\,H$\beta$ levels with SMP64 the most extreme, with \OIII\,5007/\,H$\beta$ = 0.3. This does not necessarily mean that these objects should be classified as extreme AGB stars. The optical spectra clearly show the characteristic PN emission lines, none of which would be detected through a dust-encased AGB star. It does indicate that these objects could be extremely young or proto-PNe which have not yet brightened to their maximum mass-determined optical luminosity.

  Evident in this plot is the sensitivity cut-off of the [8] band, which is close to magnitude 15. This is seen in the decreasing cut-off of objects from (15 : 0) to $\sim$(18.4 : 3).

  \subsection{[8] -- [24] versus [8]}

  Due to both the SAGE sensitivity limits and the increased strength of warm dust features in young PNe, plots that utilize the [24] band contain a larger proportion of the brighter, previously known LMC PNe. As in Fig.~\ref{Figure10}, Fig.~\ref{Figure12} shows another example where the sensitivity cut-off in the [24] band is clearly seen at the lower left, forming an approximate line from (10 : -1) to (15 : 4). The stars form a boomerang shape with those without emission shells gathering at 0 on the [8] -- [24] axis. Stars brighter than mag 10 on the [8] axis increasingly form a narrow line while the more highly evolved stars with dusty environments separate and extend from (9.5 : 0) to (13 : 3.5). The stars at (13 : 3.5) occupy a region of the plot where we could also find extragalactic background sources and YSOs (Meixner et al. 2006; Hora et al. 2008). The dusty stars did not separate in the previous plots due to the higher sensitivity of 2MASS and IRAC to stellar photospheres in preference to dust, consistent with the shorter IR wavelengths.

  The PNe are well separated from stars with a (0 : 0) position but share the area of the plot occupied by dusty evolved stars, AGNs and YSOs. A number of PNe, such as MG33, MG12, MG27 and MG55, are even fainter than the stellar population in [8] but still show a typically high increase in dust emission at [24]. PNe to the left of the main PN population ([8] -- [24] $<$\,3) will be weak in their carbonaceous dust features such as that at 30$\mu$m. The dust grains are thought to gradually evolve and evaporate over the lifetime of the PN (eg. Stanghellini et al. 2007) but dust levels are poor at conveying evolutionary information about any particular object in isolation. Where the carbonaceous dust is weak, spectroscopic observations show that the \NeV~and \OIV~emission lines in [24] are either missing or at insignificant levels (see Stanghellini et al. 2007). At the same time, such PNe are often dominated by strong PAH features at 8.6 and 11.3, which raises the flux level at [8]. Extreme examples can be seen at the top of the plot centred at (6 : 2.5).

  A direct comparison of [24] and [8] (Fig.~\ref{Figure13}) reveals a moderate correlation whereby the average difference between the two magnitude bands increases with decreasing luminosity. The line of best fit between the two bands is included in the plot and the resulting equation is shown. This modest correlation suggests that the slow decline in PAHs at 7.7 and 8.6$\mu$m and other components contributing to the magnitudes at [8] fade at an increasing rate with age compared to the solid-state features and dust at [24] whether or not the dust is carbon or oxygen rich.

\subsection{[8] -- [24] versus [3.6] -- [8]}

  Fig.~\ref{Figure14} shows the [8] -- [24] versus [3.6] -- [8] colour--colour plot including the blackbody temperatures from 400 to 10\,000\,K. This plot clearly separates the majority of PNe from the main sequence stars located at (0 : 0). The main body of PNe extend in a red-ward direction, away from the YSOs that would typically be centred at (3.3 : 2.5). The trend is for the maximum [3.6] -- [8] to decrease with increasing [8] -- [24] which results in an increasingly steep rise in the SED between [8] and [24]. This could be attributed to the relative strength of the \NeV~and \OIV~emission lines in featureless PNe such as SMP45, SMP66, SMP20 and SMP72, aided by a rise in the dust continuum in PNe with O-rich dust such as SMP21. On the other hand it could be purely due to the strengthening continuum in PNe with carbon-rich dust such as that found in SMP27 which rises towards the 30$\mu$m dust feature (assessed using MIR spectra in Stanghellini et al. 2007).

  At the top left of the plot MG47 stands alone. Optically this object is a very small, high-excitation PN (\HeII\,4686\,\AA~$\simeq$H$\beta$) but with extremely low \NII~and \SII~emission. The relatively low [8] -- [24] colour possibly indicates a lack of \NeV~and dust continuum around 24$\mu$m. As a cautionary note, the $Wide-field~Infrared~Explorer$ ($WISE$) channel-4 (22$\mu$m) band gives a magnitude of 5.49\,$\pm$0.016 for this PN, which is very different from 8.98\,$\pm$0.07 mag at [24] provided by the MIPS $Spitzer$ survey.

\begin{figure}
\begin{center}
  \includegraphics[width=0.49\textwidth]{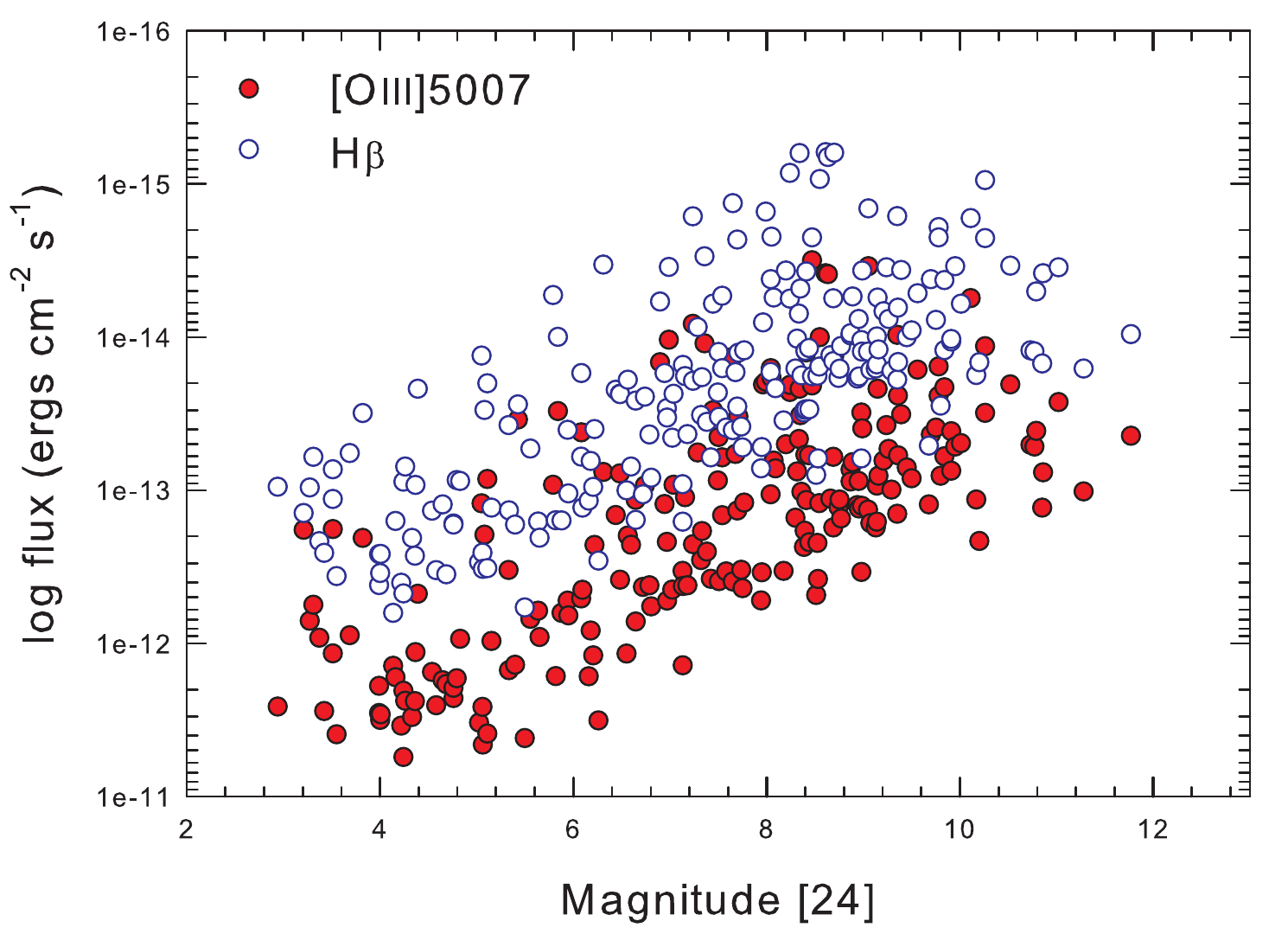}\\
  \caption{Magnitudes of the MIPS [24] band compared to the \OIII\,5007 and H$\beta$ fluxes. The correlation coefficient of --0.62 for \OIII\,5007 and [24] and --0.59 for H$\beta$ and [24] show that all three are correlated to the central star temperature. See the text for details.}
  \label{Figure15}
  \end{center}
  \end{figure}

  In the H$\alpha$/SR optical images PNe with [8] -- [24]\,$<$\,3 represent some of the smallest and faintest PNe in the sample. With the exception of SMP11, which is a very young PN, there are none of the brighter SMP objects found in this area. With [8] -- [24]$<$\,3 we do find PNe such as J17, Mo9, MG35, MG77 and MG3. All of these PNe also have very low levels of H$\alpha$ emission. A comparison of the MIPS [24] band magnitudes with published \OIII\,5007 and H$\beta$ fluxes (Reid \& Parker, 2010b) reveals a strong correlation (see Fig.~\ref{Figure15}). Therefore, it is reasonable to assume that the larger and denser the ionized zone, as reflected in the strength of H emission, the larger is the contribution from warm carbonaceous dust and probably \NeV~and \OIV~forbidden emission.

\section{Luminosity functions at MIR wavelengths}

\label{section4}

  Since the mechanisms leading to MIR emission in PNe are complex and varied, luminosity functions built on the SAGE bands will each be the result of different components such as PAHs, line radiation, optically thin thermal dust emission, Q-branches of various hydrocarbon molecules and emission from forbidden lines. Therefore it is not immediately apparent why they should correlate to the standard PNLFs built on optical emission lines such as \OIII\,5007 and H$\beta$, which loosely trace stellar temperature, excitation and metallicity. Nor is there any established reason why the standard truncated exponential curve (Ciardullo et al. 1989), based on the optical emission lines and used for a distance estimate should equally apply to MIR PNLFs. There are, nonetheless, MIR features such as forbidden emission lines and optically thin thermal dust emission that are linked to the UV luminosity of the central star (Hora et al. 2008). The MIR luminosity functions may therefore track the UV flux in bands where these components dominate.

\begin{figure}
\begin{center}
  \includegraphics[width=0.483\textwidth]{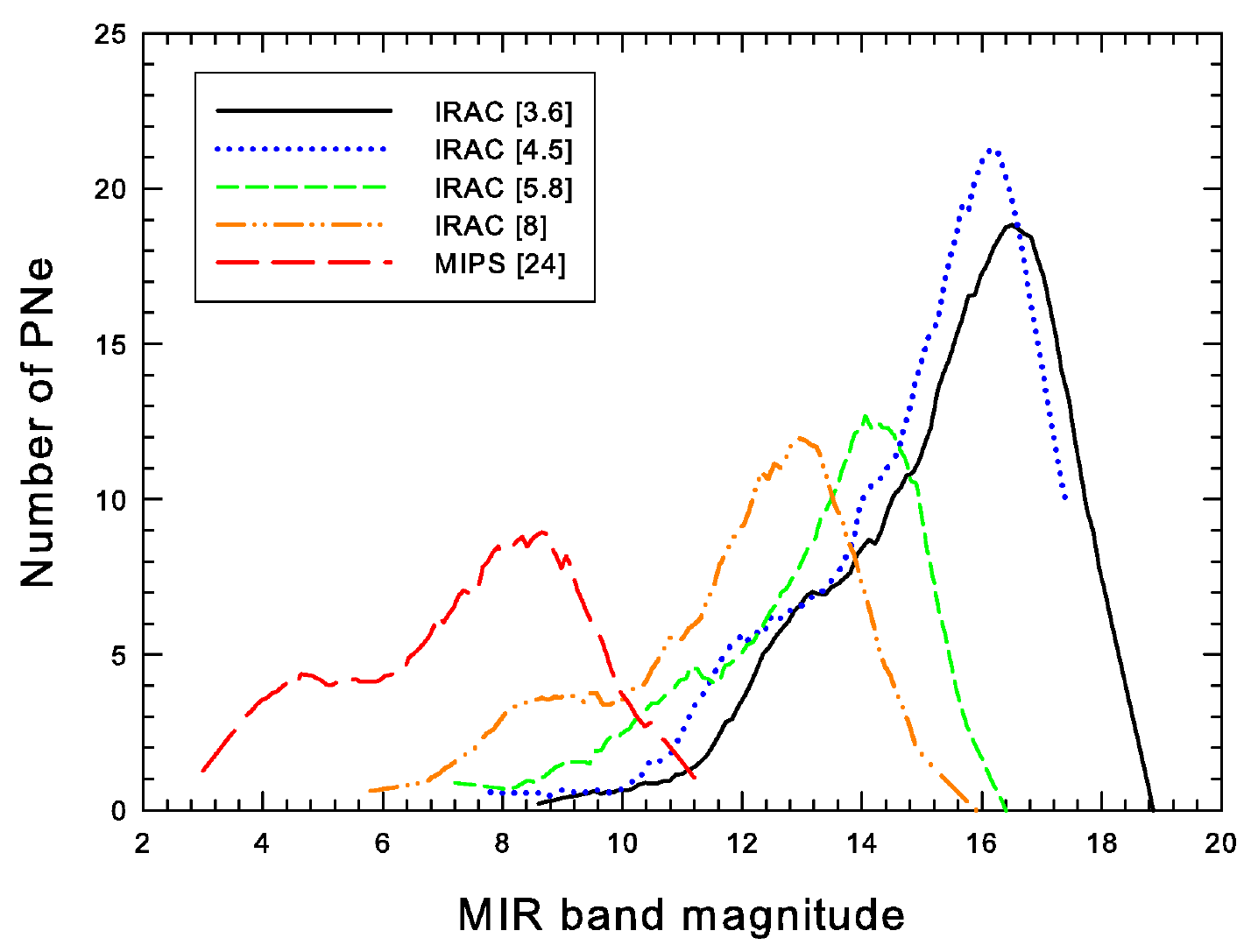}\\
  \caption{Luminosity functions for the [3.6], [4.5], [5.8] and [8] IRAC bands and the [24] MIPS band. The line plot for each band is smoothed over two 0.25mag bins. The number of detected objects decreases as a function of the cutoff in instrumental sensitivity. A noticeable `hump' and `dip' at the bright end becomes proportionally more prominent at longer wavelengths. }
  \label{Figure16}
  \end{center}
  \end{figure}

  Fig.~\ref{Figure16} shows a comparison of MIR PNLFs for the four IRAC and MIPS [24] bands. Although PNe were allotted to 0.25 mag bins, a line plot, smoothed over two bins, is presented here in order to facilitate the comparison. In each of the bands, the level of completeness corresponds closely to the peak, or highest point, of the function. The bright-end cut-offs also agree closely with the data presented by Hora et al. (2008), which is not surprising since the authors used some of the brightest PNe in the LMC. What is clear in Fig.~\ref{Figure16} is the changing shape of the bright end from band to band. Despite the larger number of PNe included in the IRAC [3.6] and [4.5] bands, the characteristic dip, well documented in optical wavelengths at the bright end (see Reid \& Parker, 2010b and references therein), grows proportionally more dominant with increasing wavelength. The dip also moves closer to the bright end of each function as the wavelength increases. At [3.6] it occurs $\sim$5 mag below the brightest, at [4.5]$\sim$5 below, at [5.8]$\sim$4.3 below, at [8]$\sim$4 below and at [24]$\sim$2.5 below. This dip in the function, also seen prominently in the \OIII\,5007 and H$\beta$ PNLFs, indicates a point of very rapid evolution of the central star (Jacoby \& de Marco 2002, Reid \& Parker 2010b).

  The prominent `hump' at the bright end prior to the dip in the [24] band suggests that PNe spend proportionally more time at the brightest 2 mag, close to the bright cut-off in the [24] band. It would then follow that the warm dust component, and possibly any emission features within that band, remains at proportionally high levels after other features at shorter wavelengths begin to evolve. Such a scenario would be in keeping with the direct action of the UV flux from the central star, which warms the dust at [24]. The evolutionary dip then reflects the excitation of the central star (Reid \& Parker, 2010a) and correlates to what we see in the optical. This also supports and partly explains the correlation found between the [24] band and the \OIII\,5007 and H$\beta$ fluxes shown in Fig.~\ref{Figure15}.

\section{Continuing object follow-up: classification results}
\label{section5}

In order to provide independent checks on the original 2dF data and create the most accurate and complete LMC PN sample possible, I have undertaken a deep multiwavelength study of PN candidates that exhibited unusual characteristics and/or were difficult to classify. A large proportion of PNe in the LMC sample, both previously known and new, have been re-analysed in order to identify and exclude any interlopers that may affect the PNLF. This is particularly critical at the bright end which is more important when the PNLF is used for distance determination. H$\alpha$ and off-band red images were examined alongside medium resolution optical spectra obtained from 6dF on the UKST. In addition SAGE false colour maps at 3.6, 4.5, 5.8 and 8$\mu$m were overlaid. Likewise I have done the same using $WISE$ (Wright et al. 2010) bands at 3.4, 4.6, 12 and 22$\mu$m to assist verification. Using mosaic radio maps of combined LMC data from ATCA and Parkes three ultra-bright `true' PNe were able to be re-classified as contaminants due to their strong radio fluxes $>$3\,mJy (see Filipovic et al. 2009). This is despite optical spectra which would otherwise strongly indicate a PN. These compact emission objects are now classified as \HII~regions. For a full list of re-classified objects, please see Table~\ref{table2} for objects previously given a probability of `possible' and Table~\ref{table3} for objects previously given a probability of `likely' or `true'.

Combining NIR images with MIR images further assisted the initial classification process. Prior to plotting the photometric data shown in Section~\ref{section3} the LMC PNe were cross-checked against the 2MASS (Skrutskie et al. 2006) NIR images at $\textsl{J}$\,(1.25 $\mu$m), $\textsl{H}$\,(1.65 $\mu$m) and $\textsl{K}$$_{s}$\,(2.16 $\mu$m) band-passes in order to identify and separate obvious bright stellar sources and large emission objects from PNe. Where a stellar continuum was stronger than expected for a PN central star the NIR flux was considerably higher towards the shorter wavelength. If the object was an extended emission source, the $\textsl{K}$$_{s}$ band was much stronger. Since a PN spectrum at each of these bands may be composed of various mixtures of emission lines and/or continuum, only objects with a large excess in either direction are unlikely to be PNe.

In particular, due to the possibility of minor positioning errors, coordinates were re-checked using both SuperCOSMOS and 2MASS. Once the coordinates were confirmed, the combination of images, photometric data, optical spectra and the plots shown in Section~\ref{section3} led to re-classification of the objects listed in Tables~\ref{table2} and~\ref{table3}. This is in addition to the 24 objects previously re-classified as other object types in Reid \& Parker (2010a). Since these changes were previously published they are not repeated with the changes listed in Tables~\ref{table2} and~\ref{table3}.

\subsection{Emission-line stars}
Where an object is re-classified as an emission-line star, I have provided the spectral classification and luminosity class where they were able to be determined. These were acquired through a method of cross-correlation which is described in Reid \& Parker (2006b, 2012). The emission-line stars that were previously published in Reid \& Parker (2012) have an asterisk after the ELS identification in column 4 of Table~\ref{table2}. Not included in this list are RP352, RP667, RP798, RP841, RP1079, RP1760 and RP1783, which were previously re-classified as emission-line stars in Reid \& Parker (2010b) and published as such in Reid \& Parker (2012). The only adjustment to the Reid \& Parker (2010b) list is object RP1691 which is likely to be a YSO due to its MIR colours.

\begin{table*}
\begin{center}
\caption{A list of RP objects originally published as possible PNe which have now been re-classified or had their status changed following new high resolution, higher S/N spectral observations and multiwavelength analysis.  ELS = emission-line star; LTS = late-type star; LPV = long period variable; YSO = young stellar object; EmO = emission object; PN = planetary nebula; SNR = supernova remnant; Prob. refers to classification probability where K = known; T = true; L = likely; P = possible. Under the heading `Reason', (1) refers to analysis using a 2.3m telescope optical spectrum, (2) refers to analysis using follow up observations on 2dF, `IR' refers to false infrared colours at 3.6, 4.5, 5.8 and 8\,$\mu$m indicating \HII~regions or hot stars, `Radio' strong radio source ($>$3\,mJy). Please see the text (Section~\ref{section5}) for more details. An * next to an ELS identification indicates that the object was re-classified as an ELS in Reid \& Parker (2012).}
\begin{tabular}{llllcl}
\hline\hline \noalign{\smallskip}
RP cat. & RA $^{(h~m~s)}$ &  DEC $^{(\circ~_{\prime}~_{\prime\prime})}$   & ID    & Prob. & Reason  \\
     No.   &  J2000   &  J2000                &  new   &   new  &       \\
  \hline\noalign{\smallskip}
46  &    05 39 52.25 & -71 09 02.42               &   YSO   & L   &  Strength of \SII4072 suggests a YSO \\
86  &   05 39 22.33  &  -70 31 24.07            &   \HII  &   L  &   Diffuse emission, weak \OIII~and \NII  \\
95	&	05 41 13.73	&	-70 23 24.73			&	ELS*	&	P	&	He absorption lines on B9V[e] continuum	\\
97  &   05 40 57.75  &  -70 18 38.25             &   \HII &   L   &  Low \OIII~and diffuse emission  \\
120  &  05 49 30.89  &  -70 09 52.15         &   Star + \HII &  L   &  Ambient emission creating most emission lines\\
187	&	05 42 36.06	&	-69 40 23.63			&	SNR	&	L	&	Diffuse patches, strong \SII/\NII	\\
188	&	05 42 32.75	&	-69 40 24.15			&	SNR	&	L	&	Compact emission, strong \SII/\NII, near a star	\\
189	&	05 44 48.10	&	-69 38 59.81			&	\HII	&	P	&	Low \OIII~continuum and diffuse emission	\\
198	&	05 44 19.12	&	-69 24 42.25			&	EmO	&	K	&	Low \OIII	\\
217 &   05 45 15.01 &   -69 38 37.34            &   \HII  &  L   &    Diffuse H$\alpha$ extending to the north \\
218	&	05 39 07.30	&	-69 35 14.79			&	\HII	&	P	&	Large diameter, low \OIII	\\
219	&	05 39 02.75	&	-69 35 09.34			&	Star + \HII	&	L	& Contribution from \HII~region dominates	\\
227	&	05 37 46.71	&	-69 31 53.99			&	Symbiotic	&	P	&  Late-type stellar continuum (M2III) + ambient emission. \\
 &  &    &   &   &  OGLE variability from 14.55 to 14.65 over $\sim$200 d. \\
240	&	05 40 55.46	&	-69 14 09.92			&	ELS*	&	L	&	Strong emission lines + absorption lines on a \\ & & & & &  B0.5V[e] spectrum	 \\
242	&	05 40 08.66	&	-68 58 26.46			&	Ecl binary	&	L	& High intrinsic H$\alpha$ but low \OIII~detection \\
  &  &   &  &  &     + only a stellar detection in MIR	\\
246	&	05 38 57.19	&	-69 33 56.53			&	ELS	&	T	&	Strong H$\alpha$ + forbidden lines on a B1.7V[e] spectrum	 \\
247	&	05 38 48.25	&	-69 34 07.95			&	Ecl binary	&	P	& Some intrinsic H$\alpha$ but ambient \\
 & & & & &  \HII~+ warm, extended dusty	environment in MIR\\
250	&	05 44 25.37	&	-69 16 42.96			&	Star	&	T	&	Two stars 4 arcsec apart encircled by H$\alpha$ \\
 & & & & &   emission. B-type continuum.	\\
251	&	05 44 15.81	&	-69 17 22.37			&	Star + \HII	&	P	&	Most of the diffuse H$\alpha$ emission is \\
 & & & & &  associated with the star despite some ambient \HII	\\
252 &   05 45 00.21 &   -69 18 19.21          &   Symbiotic & P &  Low \OIII~and very bright in all SAGE MIR colours \\
254 &   05 43 37.67 &   -69 20 09.86          &   Symbiotic   & P  &  Diffuse H$\alpha$ associated but bright and \\ & & & & &  extended in SAGE MOR colours \\
256	&	05 38 51.35	&	-69 44 51.17			&	ELS*	&	L	& Diffuse H$\alpha$ associated, low \OIII, no detection \\
 & & & & &  in SAGE MIR colours. B0.5Ve	\\
259	&	05 36 48.63	&	-69 26 44.67			&	ELS*	&	T	& Compact H$\alpha$ halo surrounding a B3V[e] star	\\
261	&	05 45 58.15	&	-69 08 57.52			&	ELS*	&	L	&	B0.5III[e] star	\\
263 &   05 45 11.01 &   -69 10 13.71           &  ELS  &  L   &  B0[e] star   \\
264	&	05 43 30.29	&	-69 24 46.55			&	Symbiotic	&	L	& Low \OIII~and extremely bright in all SAGE MIR colours	 \\
268	&	05 39 30.15	&	-68 58 57.75			&	Star	&	P	& Location is strongly contaminated	with H$\alpha$.\\
 & & & & &   Only a stellar detection in SAGE MIR colours\\
 277  &  05 41 26.73 &  -68 48 02.80           &     \HII   &   L  &  \OIII/H$\beta$=2, optically faint, flattish MIR SED  \\
283	&	05 37 48.30	&	-68 39 54.66			&	ELS*	&	P	& Weak in H$\alpha$ and \OIII, only stellar in MIR, B3V[e]	 \\
288	&	05 40 19.60	&	-68 29 19.94			&	ELS*	&	P	& Weak continuum and H$\alpha$, very low \OIII~present \\
295 & 05 48 22.26 &     -67 58 53.18          &   Symbiotic  &  L  &  Compact H$\alpha$, low \OIII,~no \HeII\,4686, very \\
 & & & & &  luminous in all SAGE MIR bands \\
 296 & 05 37 04.68   &   -68 14 43.76          &   YSO    &  L  &  Weak \OIII\,5007, very luminous MIR bands   \\
 303 &  05 36 02.40 &   -67 45 16.66         &   YSO$?$ & P  &  Emitting mainly in H$\alpha$ on a weak continuum \\
 &  & & & &  with weak forbidden lines. Not unlike a PN in SAGE \\
307	&	05 43 12.41	&	-67 50 53.39			&	ELS*	&	P	& Stellar colours in SAGE MIR, most of the emission\\
 &  & & & &  is ambient; B1.5V[e] star	with two more within 2.5 arcsec	\\
315	&	05 36 14.63	&	-68 56 23.81			&	ELS	&	L	&	Star with strong H$\alpha$ and \OIII~emission; LPV	\\
326	&	05 35 30.21	&	-67 38 05.94			&	ELS*	&	L	& Narrow intrinsic H$\alpha$ + forbidden lines on a \\
& & & & &  B1.7V[e] continuum	\\
328 &   05 36 44.47  &  -67 32 59.21            &  ELS    &   L   &   Weak colours in SAGE data; LTS continuum \\
445 &   05 31 19.68   &  -70 54 22.21          &   ELS &   T   &  Very strong \OIII\,5007 but has an M-type continuum  \\
449	&	05 32 28.35	&	-70 47 28.58			&	ELS*	&	T	&	Mainly Balmer series emission on a B2Ve spectrum \\
460	&	05 27 57.14	&	-70 51 31.62			&	YSO	&	T	& Faint and somewhat diffuse emission; large and bright in MIR \\
463 &   05 28 19.32 &   -70 52 59.21         &   \HII &  P  &  Low level continuum and H$\alpha$. Too faint to be sure\\
505	&	05 31 50.45	&	-71 09 55.88			&	LTS$?$	&	P	&	Very weak H$\alpha$ and \OIII\,5007; too faint to be sure	\\
\hline\noalign{\smallskip}
  \end{tabular}\label{table2}
 \end{center}
 \end{table*}

\begin{table*}
\begin{center}
\caption{Table 2 (cont.)}
\begin{tabular}{llllcl}
\hline\hline \noalign{\smallskip}
RP cat. & RA $^{(h~m~s)}$ &  DEC $^{(\circ~_{\prime}~_{\prime\prime})}$    & ID    & Prob. & Reason  \\
     No.   &  J2000   &  J2000                 &  new   &   new  &       \\
  \hline\noalign{\smallskip}
506  &  05 31 45.02 &    -71 10 08.15         &   EmO    &  L  &   No continuum but \OIII\,5007 too weak\\
613  &  05 41 24.23 &   -71 00 30.61        &   EmO  &   L   &   Too diffuse in H$\alpha$     \\
618  &  05 39 42.63 &  -71 10 44.08         &   YSO  & P  &  Strong H$\alpha$, weak \OIII, weak continuum \\
621  &  05 43 46.99  &  -70 58 04.13        &   \HII  &   L  &  \OIII/H$\beta$$\sim$2, optically faint, no MIR detection \\
682	&	05 32 41.08	&	-70 09 51.65			&	ELS*	&	T	& Emission too diffuse; B2V[e] spectrum	\\
711 &   05 23 32.76 &   -69 58 31.56         &   LPV  &   L  &   \NII/H$\alpha$ = 4.5 but strong continuum peaking at \NII\,6583. \\
   &                  &                    &          &     &       AGB star candidate                                                            \\
774 &   05 32 39.71 &   -69 30 49.48          &  YSO  &  P &  Low \OIII/H$\beta$ + low continuum, large MIR extension west \\
775	&	05 32 44.85	&	-69 30 05.53			&	ELS	&	L	& Strong intrinsic H$\alpha$ halo on a B3III[e] spectrum \\
803 &   05 24 09.90 &   -69 19 47.51           &   Symbiotic &   L   &  Very strong MIR colours, late-type M3III continuum plus \\
& & & & &  forbidden lines suggest a symbiotic\\
 & & & & &  or possibly an interacting LTS\\
828	&	05 33 40.24	&	-69 12 50.65			&	ELS*	&	L	&	Stellar H$\alpha$ line profile. \\
833	&	05 31 06.54	&	-69 10 42.62			&	YSO	&	P	&	Spectrum typical of a YSO	\\
883	&	05 35 56.82	&	-69 00 45.30			&	Symbiotic	&	P	&	Dense, compact nebula + very weak continuum + bright
\\ & & & & &  MIR SAGE colours	suggest a weakly interacting symbiotic\\
887	&	05 29 22.39	&	-69 00 11.87			&	ELS*	&	L	&	Strong H$\alpha$ and \OIII\,5007/H$\beta$ but no MIR emission. B3V[e]	\\
913	&	05 32 12.51	&	-68 39 24.45			&	LTS	&	P	&	No OGLE variability; may be a late-type star	\\
931 &   05 31 13.56 &    -68 27 42.20          &   YSO  &  L   &   Low level \OIII~plus low level MIR levels  \\
970	&	05 26 42.37	&	-67 45 05.92			&	\HII	&	L	&	Very faint and diffuse. 	\\
971	&	05 26 50.44	&	-67 45 54.55			&	\HII	&	L	& Optically faint, \OIII$\simeq$H$\beta$		\\
977	&	05 32 52.21	&	-67 41 08.81			&	Symbiotic	&	L	& Mainly H$\alpha$ in optical but massive in MIR \\
992	&	05 31 21.06	&	-68 31 34.37			&	\HII	&	L	&	Also known as an ELS, MIR shows it is more likely\\
 & & & & &  several stars within a dense \HII~region	\\
1018	&	05 40 55.07	&	-68 39 54.41			&	ELS	&	P	&	MIR points to an ELS	\\
1071    &   05 30 55.35  &   -67 20 05.92           &  YSO  &   L  &   Strong MIR colours, \OIII\,5007/\,H$\beta$ varies  \\
1281    &   05 12 50.68  &  -69 41 52.31            &   ELS & L     &   Weak MIR colours. Continuum peaks at 6563\AA \\
1283	&	05 11 57.68	&	-69 37 29.00			&	Star + \HII	&	L	&	B3Ve star + some \HII~emission mainly to west	 \\
1341	&	05 20 12.13	&	-69 40 30.06			&	LTS	&	L	&	Despite quite strong \OIII\,5007 it is a likely K5 giant	\\
1439	&	05 17 19.19	&	-68 46 40.45			&	LTS	&	L	&	M5III[e] 	\\
1540	&	05 21 20.68	&	-67 47 26.23			&	\HII	&	P	&	Low \OIII\,5007, bright, extended in SAGE MIR	\\
1601   &    05 13 54.17  &  -67 20 18.20         &   YSO  &  T   &  Low level continuum, \OIII\,5007$\simeq$H$\beta$, bright in SAGE MIR \\
1615  &  05 04 15.39  &  -71 07 17.02         &   Star + \HII   &  L & Weak in H$\alpha$, no dust in MIR  \\
1811	&	05 07 01.08	&	-68 46 60.08			&	ELS*	&	T	&	B3V[e]	\\
1923	&	05 04 40.40	&	-66 49 49.01			&	ELS*	&	P	&	Diffuse (associated) H$\alpha$ emission surrounding star. B1IIIe \\
1958   &    05 43 39.38  &  -70 34 03.69          &   ELS   & L   &   No \OIII, continuum, flat MIR SED  \\
1962    &    05 38 16.58 &  -70 42 03.29         &    YSO   &  P  &  Bright MIR SAGE detection extended to nearby stars\\
2180   &  05 28 04.93   &   -68 59 47.15          &   Symbiotic &  P  &  Bright in SAGE MIR \\
2182	&	05 28 58.82	&	-69 03 55.04			&	ELS	&	L	&	BVe	\\
\hline\noalign{\smallskip}
  \end{tabular}
   \end{center}
  \end{table*}


  \begin{table*}
\begin{center}
\caption{The same as for Table 2 but for RP objects originally given a probability of `likely' and `true' which have now been re-classified or had their status as PNe changed following new high resolution, higher S/N spectral observations and multiwavelength analysis. }
\begin{tabular}{lllclcl}
\hline\hline \noalign{\smallskip}
RP cat. & RA $^{(h~m~s)}$ &  DEC $^{(\circ~_{\prime}~_{\prime\prime})}$   & Prob. & ID    & Prob. & Reason  \\
     No.   &  J2000   &  J2000        &   2006          &  new   &   new  &       \\
  \hline\noalign{\smallskip}
1	&	05 40 38.00	&	-70 55 38.12		&	L	&	\HII	&	L	&	\HII~is too diffuse	\\
203	&	05 44 15.05	&	-69 10 50.03		&	T	&	\HII	&	T	&	Diffuse in H$\alpha$	\\
231	&	05 36 49.38	&	-69 23 55.20		&	L	&	ELS*	&	L	&	MIR colours show no dust. Stellar continuum B2V[e]	 \\
312	&	05 36 22.18	&	-68 55 43.51		&	L	&	ELS	&	L	&	\OIII\,5007/H$\beta$=4 however a continuum peaks near 5000\AA	\\
394	&	05 37 45.70	&	-67 11 53.29		&	T	&	LTS	&	T	& Evidence of TiO bands, low \OIII. Stellar colours in MIR	 \\
698	&	05 33 29.89	&	-69 52 28.79		&	L	&	ELS*	&	L	& Strong resolved nebula surrounding a B3III[e] star\\
727	&	05 25 44.66	&	-69 53 20.57		&	L	&	ELS*	&	L	&   B1V[e] spectrum	\\
790	&	05 32 33.61	&	-69 24 56.01		&	L	&	ELS*	&	L	& Subsequent spectra show a B3III[e] spectrum		\\
791	&	05 33 07.62	&	-69 29 46.15		&	L	&	ELS*	&	P	&	Weak continuum, strong \OIII/H$\beta$, no dust in SAGE MIR\\
793	&	05 34 41.50	&	-69 26 30.46		&	T	&	LPV/Mira	&	T	&	OGLE-III lightcurves show large amplitudes \\
& & & & & & (Soszy\~{n}ski et al., 2009), TiO bands are low but strong H$\alpha$\\
 & & & & & & and an \OIII\,5007~line are visible in all 4 spectra obtained.	\\
908	&	05 33 23.24	&	-68 39 34.85		&	L	&	ELS*	&	L	&	H$\alpha$ line profile suggest an ELS. B1V[e] likely \\
1078   &   05 24 21.33   &   -67 05 16.38       &   T   &   ELS  &   L  &  Contamination of original spectrum \\
1087	&	05 29 47.13	&	-66 41 30.74		&	L	&	\HII	&	L	&	Extremely faint H$\alpha$ emission but low \OIII\,5007	 \\
  1106	&	05 33 33.12	&	-67 24 54.05		&	L	&	ELS*	&	T	&	O7V[e] likely	\\
1111	&	05 25 58.12	&	-67 24 13.64		&	T	&	star + \HII	&	L	&	\HII~region extending to the SE	\\
1112	&	05 42 19.59	&	-67 18 58.04		&	L	&	ELS*	&	L	&	B1V[e]  likely	\\
1238	&	05 20 49.08	&	-70 12 40.84		&	L	&	ELS*	&	L	&	Between B9V[e] and A3V[e]	\\
1446    &   05 10 10.07  &  -68 42 56.12        &  T    &    PN  &  P    &  Faint and extended over 4.8arcsec radius   \\
1508	&	05 16 52.76	&	-68 10 01.78		&	L	&	LTS	&	L	&	M3III[e]	\\
1512	&	05 15 40.89	&	-68 06 28.41		&	L	&	LTS	&	L	&	M3III[e]	\\
1631  &    05 00 34.46  &   -70 52 00.09     &   T   &   AGB  &  L  &  Bright source in MIR,  ArCorB possible or PPN \\
1675	&	05 04 54.94	&	-70 43 33.73		&	L	&	ELS*	&	L	&	B3IIIe	\\
1676    &   05 04 54.67   &   -70 43 09.54      &   T   &   ELS*   &  L  &  Good PN lines but strong B$\sim$A-type continuum  \\
1822	&	05 06 38.01	&	-68 44 41.58		&	L	&	ELS*	&	L	&	B1.7V[e], star cluster	\\
1853   &    05 02 30.70 &   -68 28 48.34        &   T   &   ELS     &   T  &   Continuum, $J$ $>$ $K$, weak in MIR    \\
1862	&	05 07 47.09	&	-68 18 59.60		&	L	&	ELS*	&	L	&	B2V[e] + diffuse H$\alpha$ emission fading to the north	 \\
2171	&	05 22 16.18	&	-69 41 27.89		&	L	&	Star	&	T	&	Stellar continuum	\\
2202    &   05 45 15.80  &  -69 46 48.47       &  L    &   AGB   &  L  &   No continuum, strong \HeII\,4686 emission line.   \\
      &                  &                       &     &           &    &     MIR: bright but flattish SED. Semi-regular pulsations       \\
         &                  &                       &     &           &    & indicate a possible late AGB star or PPN.          \\
\hline\noalign{\smallskip}
  \end{tabular}\label{table3}
   \end{center}
  \end{table*}


\subsection{Notes on specific objects}


$\textbf{\textit{SMP11} (\textit{RP2241})}$ is one of the most luminous MIR PNe in the LMC sample (see Fig.~\ref{Figure1}). Its unusual dust qualities have been studied between 5 and 36$\mu$m by Bernard-Salas et al. (2006). The optical spectrum shows a typical young PNe, with \OIII\,5007/H$\beta$ = 2.08, \HeII\,4686/H$\beta$ = 0.22 and \NII\,6548+6583/H$\alpha$ = 0.82. The overall N abundance derived by Leisy \& Dennefeld (2006), however, is very low at [log (N/H) + 12 = 7.1], which is also low relative to carbon. Likewise, the MIR spectrum shows low abundances of nitrogen-based molecules such as HCN and NH3, a common finding in the low metallicity environment of the LMC (Matsuura et al. 2006). I find an electron temperature ($T$$_{e}$) of 12\,894K and electron density of 4100 cm$^{3}$, indicating a dense dusty envelope. The infrared spectrum shows a lack of emission features and peaks at 15.5$\mu$m, implying that most of the dust is emitting at 330K (Bernard-Salas et al. 2006), a common temperature for late AGB stars. The MIR spectrum reveals a cool dust continuum, absorption bands from hydrocarbon molecules and \NeII~(and possibly \NeIII) low ionization forbidden lines. The spectrum contains no PAH emission, which is probably due to its young age or possibly due to our line of sight towards a photo-dissociated edge-on torus.

$\textbf{\textit{SMP26} (\textit{RP1313})}$ has the lowest detectable \OIII\,5007/H\,$\beta$ line ratio, at only 0.018. Of more concern is that it also has a low \NII/H$\alpha$ ratio of 0.33 and no detectable \HeII\,4686. Although Ne, \OII~and S are present, there is not much else in the optical spectrum that provides confidence in its classification as a PN. In its defence however, the MIR SED is consistent with it being a bright young PN.

$\textbf{\textit{SMP31} (\textit{RP1397})}$ is rather like a small brother to SMP64. It has a low \OIII\,5007/H\,$\beta$ ratio of 1.29 and no detection of \HeII\,4686. MIR luminosity is high at 12.5, 11.1, 9.49, 7.61 and 3.2 for IRAC channels 1 to 4 and MIPS channel 1 at 24$\mu$m respectively. The steep gradient of this SED is further indication of a young PN where PAHs, silicate features and dust continuum are still very strong. Morphologically, both IRAC and H$\alpha$ images indicate a round nebula with some faint extensions (possibly a faint post-AGB halo) in H$\alpha$ which correlate to a faint extension detectable in IRAC [8].

$\textbf{\textit{SMP64} (\textit{RP645})}$ is likely to be a very young and massive PN given its position in Fig.~\ref{Figure10} and MIR luminosity of 9.95, 8.76, 7.78 and 6.56 for IRAC channels 1 to 4. The low \OIII\,5007/H\,$\beta$ ratio (0.3) is another indication that this PN is very young. Although no \HeII\,4686 has been detected, Ne, Ar and S lines are already present. The morphology appears to be very slightly elliptical, possibly more so in IRAC data than in the optical. It is also located very close to star cluster SL488.

$\textbf{\textit{SMP94} (\textit{LM2-44})}$ shows \OIII$\lambda$5007~emission at only 0.72 the strength of H$\beta$. It does however contain He$\lambda$4686 at half the strength of H$\beta$. Note that Belczy\'{n}ski et al. (2000) consider this object to be a strong symbiotic candidate. Based on this data and its position in Figs.~\ref{Figure4} and~\ref{Figure5}, I have removed this object from the catalog of known LMC PNe. For some time the location of this object has been confused with that of an ordinary star, located 10 arcsec to the south. The position provided by Leisy et al. (1997) is J2000 RA 05$^{h}$ 54$^{m}$ 10$^{s}$.73 Dec. --73$^{\circ}$ 02\,$^{\prime}$ 47\,$^{\prime\prime}$.4 and this is also the position recorded in SIMBAD. There is only an ordinary field star at this location but the intended object may be found, surrounded by an emission shell, 10 arcsec to the north at position J2000 RA 05$^{h}$ 54$^{m}$ 09$^{s}$.64 Dec. --73$^{\circ}$ 02\,$^{\prime}$ 35\,$^{\prime\prime}$.5. Once the position is corrected, rather than a flat SED through the NIR and MIR one finds a strongly reddened SED with 2MASS $J$ 14.493$\pm$\,0.024, $H$ 12.765 $\pm$\,0.025, $K$$_{s}$ 11.407 $\pm$\,0.023 and WISE Ch1 9.833 $\pm$\,0.021, Ch2 8.95 $\pm$\,0.017, Ch3 6.894 $\pm$\,0.010, Ch4 5.528 $\pm$\,0.027, creating an SED consistent with a symbiotic star (see Reid \& Parker 2013).

$\textbf{\textit{Mo4}}$ shows no H$\alpha$ excess in optical images and likewise none in spectroscopic observations. There is also no detection in MCELS images. Curiously there is a very PN-like ratio of \OIII\,5007,4959 and H$\beta$ emission lines in the blue but they are redshifted to about $z$=0.04966, clearly beyond the LMC if this is correct. The present data indicates that this object is not a PN.

$\textbf{\textit{Mo22}}$ is an object often excluded from PN followup analysis due in part to its far northerly location at J2000 RA 05$^{h}$ 21$^{m}$ 26$^{s}$.64, Dec. --62$^{\circ}$ 34$^{\prime}$ 12$^{\prime\prime}$.5. Spectroscopy confirms my previous suspicions that this object is an ordinary field star. The H$\alpha$/SR and SuperCOSMOS combined images suggest an ordinary star exhibiting no H$\alpha$ excess. Likewise, the 2MASS combined $J$, $H$ and $K_{s}$ images indicate an ordinary star. This is despite searching for any emission object in the immediate vicinity. Because the star occupies the precise published position for Mo22 it is hard to accept that there was an error in the original astrometry. This object has therefore been removed from the LMC PN catalogue.

$\textbf{\textit{Mo23 (RP660})}$ has a flat SED between J and H yet it steepens between H and K. Since this is unusual I checked the visual H-alpha/SR, 2MASS and SuperCOSMOS images for this PN. Due to coincidence with nearby stellar sources, the brightest star in the immediate 6 arcsec field was previously accepted as the central position of the PN (eg. Leisy et al. 1997). The H$\alpha$ emission, however, shows a larger extent to the SW which suggests that the current position catalogued as the centre may not in fact be associated with the PN. By shifting our central position to a faint Bmag =19.73 star firmly centred on the H$\alpha$ emission just 2.6 arcsec to the SW, the NIR colours for this object increase in magnitude with longer wavelengths, as expected for a PN. I therefore adopt a new position of J2000 RA 05$\textrm{$^{h}$}$ 21$\textrm{$^{m}$}$ 52$^{s}$.12 Dec -69$^{\circ}$ 43\,$^{\prime}$ 20\,$^{\prime\prime}$.00 for this PN.

$\textbf{\textit{Mo31}}$ at the far northern outskirts of the LMC, both imaging and spectroscopy confirm that this object does not have any H$\alpha$ emission. The 6dF spectroscopic observation of this object in blue shows some weak continuum but no significant emission lines. Since SAGE data also reveals an ordinary star, this object has been removed from the PN catalogue.

$\textbf{\textit{Mo53}}$ has a PN-like SED through the MIR but this is the only feature found that would suggest a PN. Optical H$\alpha$/SR and MCLES \OIII, \SII~and H$\alpha$ images show no emission at these wavelengths. Likewise, a lack of emission lines in the red and blue spectra suggest either an ordinary star or a background object.

$\textbf{\textit{MG57}}$ is located in a region far south of the LMC and has been overlooked in most studies. Spectroscopic observations confirm that it is not a PN, due to the lack of any \OIII~or H$\alpha$ emission in the H$\alpha$/SR deep stacked images and spectra.

$\textbf{\textit{MG79} (\textit{RP73})}$ has a very strong continuum and only faintly detected emission lines of \OIII\,5007 and \OIII4959 with almost no H$\alpha$ emission. Over the past nine years, through various programmes, I have made four separate observations of this object, each time attempting to stay clear of the other three stars in this very close cluster. Unfortunately I have not been able to positively detect a PN in this position. In three of the observations I see a late-type star resembling spectral classes between G2 and G5. Although the object shows more emission features than its immediate neighbours in multiwavelength imaging, its quasi-stellar appearance in all combined SAGE bands suggest that the object is unlikely to be a PN.

\begin{figure}
\begin{center}
  \includegraphics[width=0.50\textwidth]{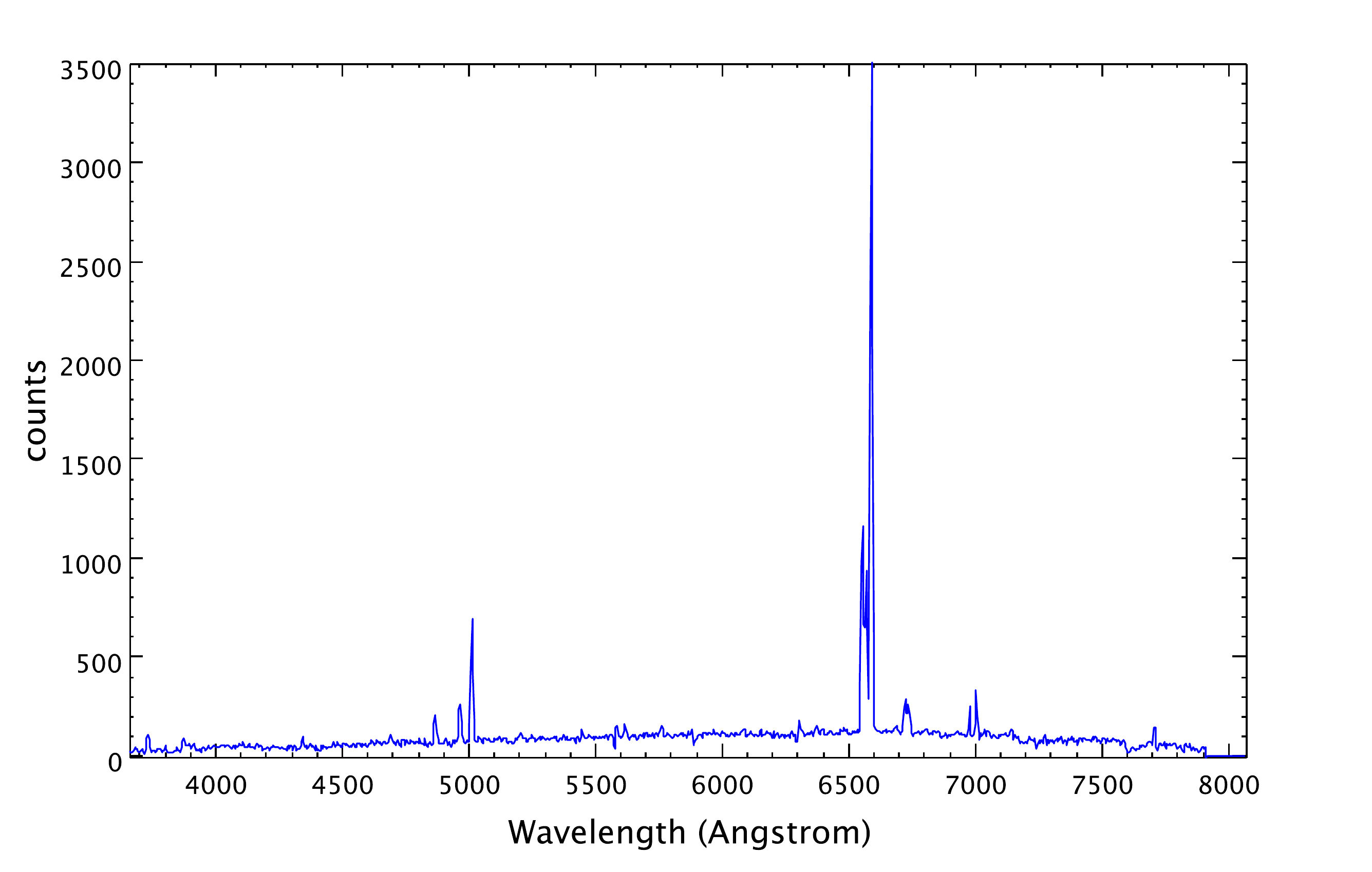}\\
  \caption{Optical spectrum of RP135}
  \label{Figure17}
  \end{center}
  \end{figure}

$\textbf{\textit{RP135}}$ was originally classified as a true PN due to its compact optical appearance and spectrum (see Fig.~\ref{Figure17}). Miszalski et al. (2011b) state that ``the NIR and MIR colours of this object suggest a non-PN classification, perhaps an emission-line star if there is indeed weak H$\alpha$ emission". This summary and reclassification by Miszalski et al. (2011b) is surprising, given that they do not have MIR (IRAC) or MIPS magnitudes for this object. However, the NIR $\textit{Y}$, $\textit{J}$ and $\textit{K$_{s}$}$ magnitudes that they do provide place it well in keeping with other LMC PNe.

\begin{figure}
\begin{center}
  \includegraphics[width=0.51\textwidth]{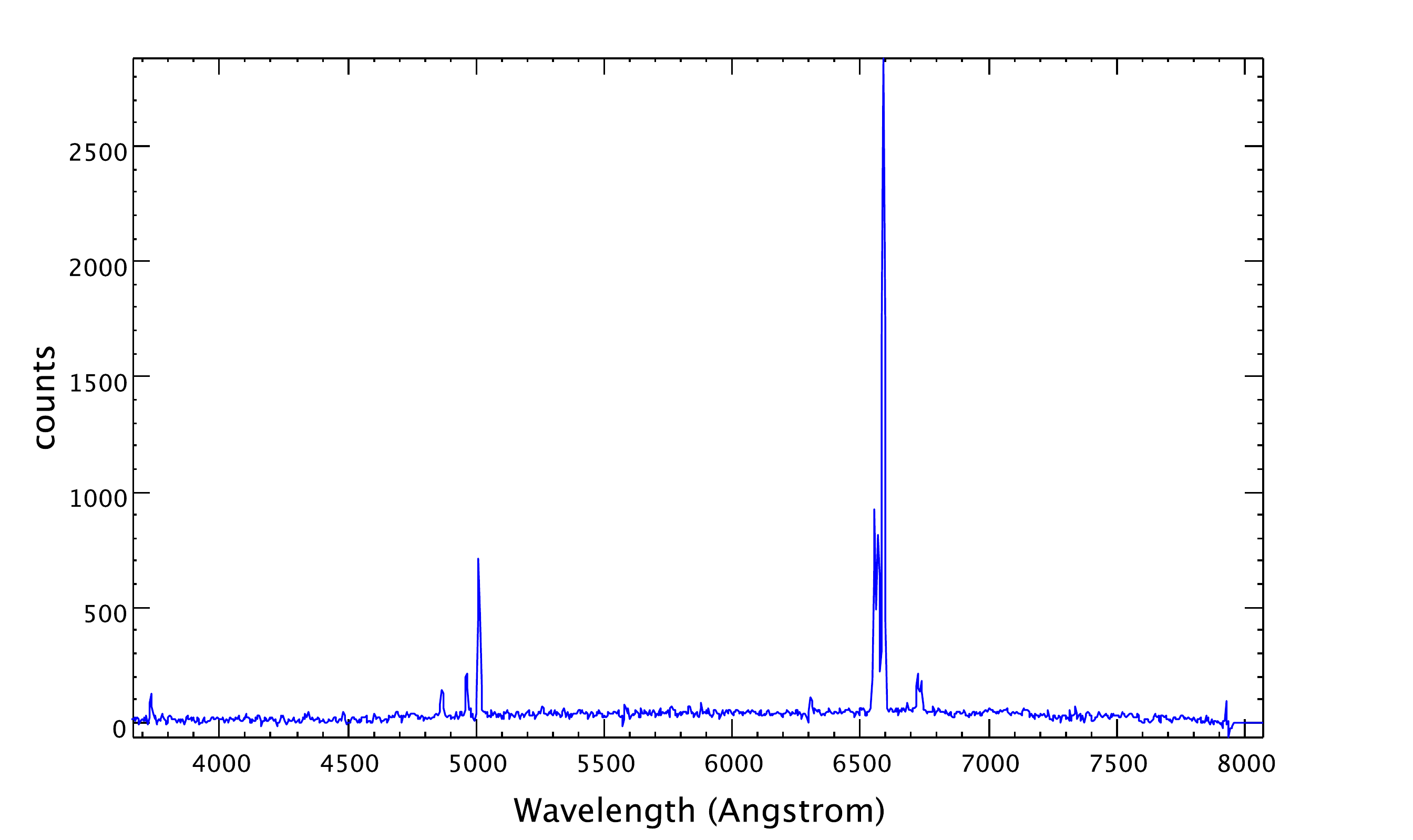}\\
  \caption{Optical spectrum of RP143}
  \label{Figure18}
  \end{center}
  \end{figure}

  $\textbf{\textit{RP143}}$ was originally classified as a true PN and I am not convinced that this should be changed. Miszalski et al. (2011b) have given the PN a classification of FD$?$,NL. The reason for this is not made clear as they have no optical magnitudes, NIR (VISTA) or MIR (SAGE) data on this object. The H$\alpha$/Red images show an extended nebula which almost fills a space between two bright stars. There is nothing detectable in the SAGE data but this may only indicate the low luminosity and/or evolutionary state of the nebula and the dust contained within it. The H$\alpha$ and \NII~fluxes remain a strong feature of this object (see Fig.~\ref{Figure18}).

$\textbf{\textit{RP187} and \textit{RP188}}$ were originally thought to be possible PNe but, judging by the \SII\,6716/\NII\,6583 ratios ($\sim$1.5), both these neighboring objects are likely to be part of a diffuse supernova remnant centred at $J$2000 RA 05$^{h}$ 42$^{m}$ 39$^{s}$.5, Dec 69$^{\circ}$ 40$^{\prime}$ 15$^{\prime\prime}$.1 where \SII\,6716/\NII\,6583 = 1.66. For clarity, this small SNR candidate has been allocated a new identification of RP186 at this central position.

$\textbf{\textit{RP328}}$ was originally classified as a possible PN but further spectroscopic follow-up shows that it is a very weak source with a combination of forbidden nebula lines on a rising red continuum. SAGE imaging primarily shows a stellar source.

$\textbf{\textit{RP774}}$ was identified as a possible symbiotic star by Miszalski et al (2011b) but I do not consider the MIR SED steep enough to support this classification. With \OIII\,5007/H\,$\beta$ = 1.65, a low continuum, very low N levels, no \HeII\,4686 or \OIII\,4363, and yet brightening from 14.77 $\pm$0.063 mag in $K$ to 12.181 $\pm$0.054 mag in [3.6], I classify it as a YSO in agreement with Gruendl \& Chu (2009). A tail, bright in [8], originating from the main object and extending almost 11 arcsec to the west could trace the formation history of this object.

$\textbf{\textit{RP1631}}$ was originally classified as a true PN, based on optical spectra. It has since been identified as several other object types including a He burning AGB star (Vijh et al. 2009), a variable star$?$ (Vijh et al. 2009), and a RCrB star (Woods et al. 2011). I agree that this object is He rich. A 2dF spectrum shows \HeII\,4686/H\,$\beta$ = 0.65 with typical nebulae \OIII~and Balmer lines as well as argon and neon. There is no continuum and there are no \SII\,6716/6731 lines. In addition, the optical lines are particularly faint as is the optical image in H$\alpha$ and short red. Since RCrB stars are variable stars that fade by several magnitudes over regular intervals, it is surprising that the stack of 12 digitised images, observed over a three-year period have not shown this variability. In the case of emission-line stars and other variable stars, the variability has been clearly detected (Reid \& Parker, 2012) in the stacked images. One possibility is that the exposures were made at times where the star was at its regular magnitude. RCrB stars usually fade suddenly (sometimes by factors of thousands) only to recover to their normal brightness over several months. RP1631 is indeed very bright in the SAGE bands placing it in the brightest bin for compact emission sources in the LMC. The current data supports the probability that it may be a He-rich late-stage AGB star which is undergoing semi-regular pulsations.

$\textbf{\textit{RP1796}}$ is a very small and faint object that shows a low 0.75 \OIII/H\,$\beta$ ratio plus \NII\,6583, \SII\,6716,6731 and \OII\,3727 nebula lines without any evidence of a strong continuum source, either spectroscopically or through continuum-subtracted imaging. The SED constantly brightens from $V$ through to 22$\mu$m ($WISE$ ch4), as expected for a PN. Although this object is published as an AGN by Kozlowski \& Kochanek (2009), repeated velocity measurements from five different observations endorse our previous heliocentric velocity measurement of 291.3 km\,s$^{-1}$ (Reid \& Parker, 2006b), placing the object well within the LMC.

$\textbf{\textit{RP4034} and \textit{RP4041}}$ are conspicuous in that their spectral ratios are almost identical. The measured \HeII\,4686/H\,$\beta$~ratios are 0.85 and 0.84 while the \OIII\,5007/H\,$\beta$ ratios are 2.15 and 2.55 respectively. After questioning this similarity I took further spectroscopic observations of these PNe only to find results where the ratios are the same to within 6\%. Since both of these PNe are linearly only 11\,min 24\,arcsec apart, it is possible that the similar chemical tagging is the result of a common stellar nursery for the progenitors.

$\textbf{\textit{MNC1} and \textit{MNC2} (\textit{RP4332} and \textit{RP4333})}$ were first published by Miszalski et al. (2011a) who, without spectroscopic confirmation, assigned a PN status of `true' to both of them based on colour composite H$\alpha$, $V$, \OIII~and $B$ band and VMC images. With the advantage of spectroscopic observations, I prefer to classify them as `possible' PNe. MNC1 has a weak blue continuum and weak emission lines. With a low \OIII\,5007/H\,$\beta$ ratio of only $\sim$3 and no \HeII\,4686 emission, such a low excitation PN should have a high \NII\,6583/H$\alpha$ ratio but in the case of MNC1 it is only 0.14. MNC2 is somewhat weaker still with \OIII\,5007/H\,$\beta$ =2.37 and \NII/H$\alpha$ =0.14. Since both objects are in a location with considerable ambient emission and other compact \HII~regions with similar emission-line ratios, I give these objects the lowest PN probability of `possible'.

$\textbf{\textit{MNC3} and \textit{MNC4}}$ were also published by Miszalski et al. (2011a). MNC3 was classified as a possible PN at J2000 RA 05$^{h}$ 41$^{m}$ 28$^{s}$.35, Dec --69$^{\circ}$ 43\,$^{\prime}$ 53\,$^{\prime\prime}$.0, and MNC4, a true PN at J2000 RA 06$^{h}$ 00$^{m}$ 59$^{s}$.20, Dec -66$^{\circ}$ 36\,$^{\prime}$ 15\,$^{\prime\prime}$.3. Both of these objects have now been observed and, based on the spectroscopic confirmations, I would classify MNC3 as a likely emission-line star with some possible weak symbiotic interaction and MNC4 as an \HII~region. MNC3 shows low and weak \OIII\,5007 emission (\OIII\,5007/H\,$\beta$\,$<$\,1.5) but has Balmer lines and some \NII~which is probably due to ambient emission which we see superimposed on a strong reddening stellar continuum. MNC4 has no \OIII\,5007 emission but it does have a strong blue continuum. I suspect that the weak H$\alpha$ and \NII~lines may also be due to ambient emission.

\label{subsection5.2}

\subsection{Objects in the 30\,Doradus region}

\label{subsection5.3}

The 30 Doradus (NGC\,2070) nebula is an extremely complex area, making it difficult to separate discrete emission objects such as PNe from strong ambient emission. For this reason, as Reid \& Parker (2006a,b) discovered and observed discrete emission objects in this area, they classified them as `possible' or `likely' PNe. This allowed them to be tagged for a full multiwavelength follow-up without a full commitment to their classification as PNe. The follow-up work has been on-going. Multiwavelength imaging and deep spectroscopic observations have allowed me to re-analyse and re-classify many of these discrete emission sources. This work was initially conducted without reference to the follow-up results from other studies in order to make a totally independent assessment.

Using the high resolution and optical depth of the Vista Magellanic Cloud (VMC) survey, Miszalski et al. (2011b) have examined and re-classified PN candidates within a couple of strategic areas of the LMC, such as 30 Doradus. The VMC survey has gone only part of the way to secure object classification since, in my opinion, the heavy reliance on absent or very faint NIR and MIR detections does not provide a sufficient basis for re-classification. For example, changing an object from `PN possible' or `PN likely' to ND (non-detection), FD (faint detection), NL (neutral) or Em$?$ (emission object of some type), points more to the faintness of the objects and limits of the surveys that were employed than to the classification of the objects.


\section{Conclusion}
\label{section6}

An optical spectrum and a high resolution, deep optical image are a good starting point for the identification of a PN. A follow-up multiwavelength analysis of PNe is the most complete and robust method of assessing the veracity of these objects.  SAGE observations in the MIR provide much needed information regarding the PAHs, dust content and other thermal emissions. It is additionally helpful to examine these bands in association with 2MASS $J$, $H$ and $K$$_{s}$ colours.

I have examined and compared optical, NIR and MIR images and plotted photometric data for all previously known and newly identified PNe in the LMC. The presented data includes the additional 61 PNe recently discovered in the outer regions of the LMC (Reid \& Paker 2013). Beginning with a total of 813 catalogued LMC PNe, 739 objects (previously known and new) with at least one detection in IRAC, MIPS, 2MASS or MCPS were re-analysed with additional spectroscopic data where previous detections were ambiguous. Detailed multiwavelength plots supported by spectroscopy have enabled me to reduce the number of previously published `likely' and `possible' PNe, thus improving the purity of the set for evolutionary studies. Altogether I have re-classified 76 possible, 19 likely and 8 true PNe as other object types. This work has reduced the total number of LMC PNe to 715, a number that is still short of the maximum $\sim$900 predicted in the LMC based on its mass and luminosity (see Reid 2012).

In terms of completeness, 466 PNe were detected at the $\textsl{J}$ magnitude. This number reduces to 408 for \textsl{H} and 354 for $\textsl{K}_{s}$. Data for all three bands are available down to a $\textsl{J}$ magnitude of $\sim$18.5, after which the remaining 47 detections in $\textsl{H}$ and 33 detections in $\textsl{K}_{s}$ become sporadic to the limiting magnitude of the survey. There are 303 detections in all three 2MASS bands, 376 in at least two bands and 360 in at least two consecutive bands. All magnitudes used including errors where available are provided in the appendix.

Due to the detection limits imposed on LMC objects by both IRAC and MIPS, I was able to use a maximum 440 PNe in [3.6], 444 in [4.5], 267 in [5.8], 294 in [8] and 209 in [24]. There were 444 PNe in at least two bands, 267 in three consecutive bands, 242 in all four IRAC bands and 170 PNe in all four IRAC bands plus the MIPS [24] band. The position of individual PNe in each plot relies on various emission, absorption and continuum features which depend on the age, mass and abundance ratios of the PNe. Although the spread of positions is wide in each plot, most of the PNe separate well from main-sequence stars. However, their position with respect to main-sequence stars is also dependent on the wavelengths that are being compared. Certain combinations such as $\textsl{H -- K}_{s}$ versus $\textsl{J -- H}$ permit almost half the PN population to share colour-colour space with stars. These PNe are shown to have mean \NII\,6563/H$\alpha$ ratios of 1.26$\pm$1.59 whereas the brighter and presumably younger PNe away from the stellar population have mean \NII\,6563/H$\alpha$ ratios at 0.57$\pm$1.35.

Strong evidence is found for the evaporation of dust grains and PAHs under the continuous bombardment of UV radiation from the central star. Dust levels will therefore be extremely low in the later stages of PN evolution. I also find a strong correlation between the dust type, the slope of the MIR SED and the optical emission-line ratios of \NII/H$\alpha$ and \OIII/H\,$\beta$, implying the level of carbon abundance in the nebula.


\label{section 9}

\section*{Acknowledgements}

I wish to thank Robert Gruendl and You-Hua Chu for their generous supply of SAGE magnitude data. I also thank Macquarie University, Sydney, for research and travel grants, Suzanne Reid from Kalidus Systems for designing a data base to facilitate the analysis and Prof. Quentin Parker for his help with spectroscopic observations.

\label{section 10}

\bsp

\label{lastpage}

\end{document}



\pagerange{\pageref{firstpage}--\pageref{lastpage}} \pubyear{2011}


\label{firstpage}
\setcounter{table}{1}
\begin{landscape}
\begin{table*}
\begin{minipage}{240mm}
\section{Appendix: Table 1}
Magnitude data for each LMC PNe. Column 1 gives the common name and reference number of the PN. Columns 2 to 5 give the optical $\textsl{U,B,V}$ and $\textsl{I}$ band magnitudes from the MCPS survey. Columns 6 to 8 give the NIR band magnitudes where the $\textsl{J,H}$ and $\textsl{K}_{s}$ data from the 2MASS survey is marked with an astrict after the $\textsl{J}$ magnitude. All other NIR $\textsl{J,H}$ and $\textsl{K}_{s}$ magnitudes are from the IRSF survey. Columns 9 to 13 give the MIR [3.6], [4.5], [5.8], [8] IRAC and [24] MIPS band magnitudes. The lack of an estimated error in 2MASS data depicts a low confidence result. Where one MIR magnitude may be missing in one catalogue, we substitute a magnitude from the other catalogue only as long as general agreement between both catalogues is within 0.2 magnitudes. The $\dagger$ symbol indicates a magnitude for a point source taken from the SAGE catalogue. The $\ddagger$ symbol indicates a magnitude taken from Miszalski et al (2011). Magnitudes with a $\dagger$ or $\ddagger$ symbol were not used in the plots presented in this paper.
\centering
\scriptsize
\begin{tabular}{|l|c|c|c|c|c|c|c|c|c|c|c|c|}
\hline
LMC    &   $\textsl{U}$ &  $\textsl{B}$  &  $\textsl{V}$   &  $\textsl{I}$   &   $\textsl{J}$   &  $\textsl{H}$  &  $\textsl{K}$  &   [3.6] & [4.5] & [5.8] & [8] & [24]   \\
PN  &  band  &   band  &   band  &   band  &  band     &   band  &  band   &     band     &    band   &   band    &  band   &   band    \\
\hline\hline
J80	4		&	20.145\,$\pm$0.166	&	20.356\,$\pm$0.073	&	19.91\,$\pm$0.076	&				&					&				&				&				&				&				&				 &				 \\
J80	7		&	19.922\,$\pm$0.182	&	20.179\,$\pm$0.087	&	19.727\,$\pm$0.111	&	19.6\,$\pm$0.152	&	19.75\,$\pm$0.24		&				&				&	16.964\,$\pm$0.101	&	16.47	 $\pm$	0.182	&	 14.875\,$\pm$0.162	&				&				\\
J80	10		&	20.173\,$\pm$0.231	&	20.729\,$\pm$0.1	&	19.934\,$\pm$0.077	&	19.831\,$\pm$0.172	&	19.05\,$\pm$0.11		&	18.75\,$\pm$0.2	&				&	18.41\,$\pm$0.37	&	 17.463\,$\pm$0.149	&				 &				&	9.78\,$\pm$0.12	\\
J80	14		&	17.975\,$\pm$0.063	&	18.217\,$\pm$0.072	&	17.853\,$\pm$0.12	&	18.029\,$\pm$0.078	&	17.93\,$\pm$0.05		&	18.11\,$\pm$0.13	&				&	17.074\,$\pm$0.086	 &				&	 14.859	 $\pm$	0.105	&	13.419\,$\pm$0.108	&	9.95\,$\pm$0.16	\\
J80	16		&	19.068\,$\pm$0.115	&	19.332\,$\pm$0.098	&	19.295\,$\pm$0.126	&	19.58\,$\pm$0.167	&	18.55\,$\pm$0.09		&				&				&				&				&				 &				 &				\\
J80	17		&	18.896\,$\pm$0.127	&	19.566\,$\pm$0.056	&	19.039\,$\pm$0.249	&	18.705\,$\pm$0.09	&	15.03\,$\pm$0.01		&	12.75\,$\pm$0.01	&	10.89\,$\pm$0.01	&	8.568	 $\pm$	0.025	&	7.827	 $\pm$	0.021	&	7.19\,$\pm$0.016	&	6.474\,$\pm$0.021	&	4.39\,$\pm$0.01	\\
\hline
\end{tabular}
\end{minipage}
\end{table*}
\end{landscape}
\label{secondpage}
\setcounter{table}{2}
\begin{landscape}
\begin{table*}
\begin{minipage}{240mm}
\centering
\scriptsize
\begin{tabular}{|l|c|c|c|c|c|c|c|c|c|c|c|c|}
\hline
LMC    &   $\textsl{U}$ &  $\textsl{B}$  &  $\textsl{V}$   &  $\textsl{I}$   &   $\textsl{J}$   &  $\textsl{H}$  &  $\textsl{K}$  &   [3.6] & [4.5] & [5.8] & [8] & [24]   \\
PN  &  band  &   band  &   band  &   band  &  band     &   band  &  band   &     band     &    band   &   band    &  band   &   band    \\
\hline\hline
MG	41		&				&				&				&				&					&				&				&	16.321\,$\pm$0.03	&	16.12\,$\pm$0.045	&	13.854\,$\pm$0.036	&	12.043	 $\pm$	0.014	 &	9.074\,$\pm$0.154	\\
MG	42		&				&	19.922\,$\pm$0.061	&	19.102\,$\pm$0.046	&	18.13\,$\pm$0.057	&	17.66\,$\pm$0.05		&	17.27\,$\pm$0.07	&	17.04\,$\pm$0.12	&	17.07\,$\pm$0.034	 &	16.96\,$\pm$0.049	&				 &				&				\\
MG	43		&				&	20.651\,$\pm$0.091	&	19.866\,$\pm$0.127	&				&	19.7\,$\pm$0.21		&				&				&				&	16.46\,$\pm$0.1	&				&	 14.13\,$\pm$0.09	&	 8.41\,$\pm$0.06	\\
MG	44		&				&	18.37	$\dagger$		&				&				&					&				&				&				&				&				&				&				\\
MG	45		&				&				&				&				&	15.923\,$\pm$0.034	$\ast$	&	15.099\,$\pm$0.039	&	13.468\,$\pm$0.026	&				&				&				&				 &				 \\
MG	46		&	18.034\,$\pm$0.06	&	18.238\,$\pm$0.041	&	17.312\,$\pm$0.061	&	16.477\,$\pm$0.048	&	15.556\,$\pm$0.066	$\ast$	&	15.295\,$\pm$0.111	&	14.794\,$\pm$0.136	&	14.369	 $\pm$	0.01	&	 14.028	 $\pm$	0.012	&	13.669\,$\pm$0.037	&	12.984\,$\pm$0.034	&	8.94\,$\pm$0.144	\\
\hline
\end{tabular}
\end{minipage}
\end{table*}
\end{landscape}
\label{secondpage}
\setcounter{table}{2}
\begin{landscape}
\begin{table*}
\begin{minipage}{240mm}
\centering
\scriptsize
\begin{tabular}{|l|c|c|c|c|c|c|c|c|c|c|c|c|}
\hline
LMC    &   $\textsl{U}$ &  $\textsl{B}$  &  $\textsl{V}$   &  $\textsl{I}$   &   $\textsl{J}$   &  $\textsl{H}$  &  $\textsl{K}$  &   [3.6] & [4.5] & [5.8] & [8] & [24]   \\
PN  &  band  &   band  &   band  &   band  &  band     &   band  &  band   &     band     &    band   &   band    &  band   &   band    \\
\hline\hline
Mo	21		&				&				&				&				&					&				&				&	17.708\,$\pm$0.106	&	16.41\,$\pm$0.09	&				&	13.981\,$\pm$0.153	 &				 \\
Mo	22		&				&				&				&				&	17.363\,$\pm$0.116	$\ast$	&	16.74\,$\pm$0.164	&	15.644			&				&				&				&				&				 \\
Mo	23		&	19.406\,$\pm$0.154	&	19.73\,$\pm$107	&	18.827\,$\pm$0.133	&	17.855\,$\pm$0.111	&	17.52\,$\pm$0.04		&	16.42\,$\pm$0.07	&	15.94\,$\pm$0.12	&	15.477\,$\pm$ 0.153	&	14.761\,$\pm$ 0.073	&	14.628\,$\pm$0.136	&	13.139\,$\pm$0.17	&				\\
Mo	24		&	18.594\,$\pm$0.071	&	19.927\,$\pm$0.113	&	19.012\,$\pm$0.092	&	20.094\,$\pm$0.119	&	18.22\,$\pm$0.24		&	17.68\,$\pm$0.12	&	17.45\,$\pm$0.21	&	16.921	 $\pm$	0.053	&	15.995	 $\pm$	0.052	&	14.802\,$\pm$0.114	&	12.892\,$\pm$0.029	&	9.21\,$\pm$0.164	\\
Mo	25		&				&				&				&				&					&				&				&	17.086\,$\pm$0.074	&	16.287\,$\pm$0.059	&	15.148\,$\pm$0.171	&	13.595			 &				 \\
Mo	26		&	19.238\,$\pm$0.142	&	20.196\,$\pm$0.079	&	18.964\,$\pm$0.074	&	20.57\,$\pm$0.229	&	19\,$\pm$0.27		&				&				&	16.72\,$\pm$0.12	&	16.054	 $\pm$	0.104	&	 14.427	 $\pm$	0.181	&	12.321\,$\pm$0.059	&	8.689\,$\pm$0.131	\\
\hline
\end{tabular}
\end{minipage}
\end{table*}
\end{landscape}
\label{secondpage}
\setcounter{table}{2}
\begin{landscape}
\begin{table*}
\begin{minipage}{240mm}
\centering
\scriptsize
\begin{tabular}{|l|c|c|c|c|c|c|c|c|c|c|c|c|}
\hline
LMC    &   $\textsl{U}$ &  $\textsl{B}$  &  $\textsl{V}$   &  $\textsl{I}$   &   $\textsl{J}$   &  $\textsl{H}$  &  $\textsl{K}$  &   [3.6] & [4.5] & [5.8] & [8] & [24]   \\
PN  &  band  &   band  &   band  &   band  &  band     &   band  &  band   &     band     &    band   &   band    &  band   &   band    \\
\hline\hline
SMP	9		&	17.219\,$\pm$0.05	&	18.412\,$\pm$0.036	&	17.305\,$\pm$0.023	&	18.73\,$\pm$0.122	&	18.17\,$\pm$0.09		&	18.36\,$\pm$0.21	&	17.1\,$\pm$0.2	&	14.745\,$\pm$ 0.012	&	14.443\,$\pm$ 0.015	&	12.666\,$\pm$0.017	&	10.903\,$\pm$0.01	&	6.482\,$\pm$0.046	\\
SMP	10		&	18.195\,$\pm$0.088	&	17.209\,$\pm$0.056	&	16.178\,$\pm$0.048	&				&	18.05\,$\pm$0.07		&	18.4\,$\pm$0.15	&				&	16.026\,$\pm$0.034	&	15.13	 $\pm$	0.032	&	 15.029	 $\pm$	0.125	&	13.804\,$\pm$0.107	&	8.512\,$\pm$0.118	\\
SMP	11		&	19.156\,$\pm$0.093	&	20.303\,$\pm$0.053	&	20.142\,$\pm$0.06	&	19.582\,$\pm$0.092	&	18.36\,$\pm$0.1		&	17.07\,$\pm$0.06	&	15.04\,$\pm$0.04	&	10.393\,$\pm$ 0.008	&	8.803\,$\pm$ 0.004	 &	7.299\,$\pm$0.005	&	5.805\,$\pm$0.002	&	2.941\,$\pm$0.009	\\
SMP	12		&	18.371\,$\pm$0.104	&	19.156\,$\pm$0.044	&	18.183\,$\pm$0.044	&	19.479\,$\pm$0.108	&	18.83\,$\pm$0.14		&	18.91\,$\pm$0.24	&				&	16.408\,$\pm$0.043	 &	15.501\,$\pm$0.037	&	 14.455\,$\pm$0.106	&	13.018\,$\pm$0.036	&	7.331\,$\pm$0.068	\\
SMP	13		&	16.692\,$\pm$0.079	&	16.809\,$\pm$0.057	&	15.662\,$\pm$0.043	&	17.75\,$\pm$0.063	&	17.105\,$\pm$0.091	$\ast$	&	16.788\,$\pm$0.149	&	15.574\,$\pm$0.094	&	13.83	 $\pm$	0.02	&	13.04	 $\pm$	0.02	&	11.76\,$\pm$0.01	&	9.76\,$\pm$0.01	&	5.82\,$\pm$0.03	\\
SMP	14		&	18.103\,$\pm$0.201	&	19.574\,$\pm$0.069	&	18.107\,$\pm$0.07	&	19.887\,$\pm$0.12	&					&				&				&	17.05\,$\pm$0.077	&	16.131\,$\pm$ 0.051	&				&	 13.826\,$\pm$0.13	&	8.743\,$\pm$0.132	\\
\hline
\end{tabular}
\end{minipage}
\end{table*}
\end{landscape}
\label{secondpage}
\setcounter{table}{2}
\begin{landscape}
\begin{table*}
\begin{minipage}{240mm}
\centering
\scriptsize
\begin{tabular}{|l|c|c|c|c|c|c|c|c|c|c|c|c|}
\hline
LMC    &   $\textsl{U}$ &  $\textsl{B}$  &  $\textsl{V}$   &  $\textsl{I}$   &   $\textsl{J}$   &  $\textsl{H}$  &  $\textsl{K}$  &   [3.6] & [4.5] & [5.8] & [8] & [24]   \\
PN  &  band  &   band  &   band  &   band  &  band     &   band  &  band   &     band     &    band   &   band    &  band   &   band    \\
\hline\hline
SMP	70		&				&				&				&				&					&				&				&	16.199\,$\pm$0.036	&	15.153\,$\pm$0.025	&	14.089\,$\pm$0.051	&	12.913	 $\pm$	0.03	 &	8.04\,$\pm$0.095	\\
SMP	71		&	16.348\,$\pm$0.038	&	16.781\,$\pm$0.031	&	15.661\,$\pm$0.021	&	17.112\,$\pm$0.035	&	16.44\,$\pm$0.01		&	16.11\,$\pm$0.02	&	15.11\,$\pm$0.03	&	12.87	 $\pm$	0.005	&	12.376	 $\pm$	0.005	&	10.615\,$\pm$0.006	&	8.802\,$\pm$0.011	&	4.539\,$\pm$0.019	\\
SMP	72		&	18.987\,$\pm$0.09	&	18.234\,$\pm$0.064	&	16.669\,$\pm$0.1	&	15.469\,$\pm$0.028	&	14.45\,$\pm$0.02		&	13.76\,$\pm$0.01	&	13.57\,$\pm$0.01	&	13.462\,$\pm$ 0.008	&	13.443\,$\pm$ 0.009	&	13.396\,$\pm$0.03	&	13.158\,$\pm$0.066	&	7.766\,$\pm$0.083	\\
SMP	73		&	15.439\,$\pm$0.039	&	15.842\,$\pm$0.029	&	14.676\,$\pm$0.022	&	16.53\,$\pm$0.038	&	15.94\,$\pm$0.02		&	15.62\,$\pm$0.02	&	14.58\,$\pm$0.02	&	12.359	 $\pm$	0.005	&	11.702	 $\pm$	0.004	&	10.11\,$\pm$0.004	&	8.081\,$\pm$0.006	&	4.242\,$\pm$0.013	\\
SMP	74		&	15.961\,$\pm$0.054	&	16.327\,$\pm$0.032	&	15.199\,$\pm$0.113	&	17.082\,$\pm$0.033	&	16.226\,$\pm$0.11	$\ast$	&	16.129\,$\pm$0.24	&	15.14			&	13.216\,$\pm$ 0.007	&	12.556\,$\pm$ 0.006	&	11.018\,$\pm$0.007	&	8.995\,$\pm$0.011	&	5.06\,$\pm$0.024	\\
SMP	75		&	15.753\,$\pm$0.034	&	16.181\,$\pm$0.02	&	14.927\,$\pm$0.026	&	16.631\,$\pm$0.041	&	15.63\,$\pm$0.02		&	15.23\,$\pm$0.01	&	14.03\,$\pm$0.02	&	11.847	 $\pm$	0.003	&	11.262	 $\pm$	0.003	&	9.292\,$\pm$0.003	&	7.288\,$\pm$0.003	&	4.006\,$\pm$0.015	\\
\hline
\end{tabular}
\end{minipage}
\end{table*}
\end{landscape}
\label{secondpage}
\setcounter{table}{2}
\begin{landscape}
\begin{table*}
\begin{minipage}{240mm}
\centering
\scriptsize
\begin{tabular}{|l|c|c|c|c|c|c|c|c|c|c|c|c|}
\hline
LMC    &   $\textsl{U}$ &  $\textsl{B}$  &  $\textsl{V}$   &  $\textsl{I}$   &   $\textsl{J}$   &  $\textsl{H}$  &  $\textsl{K}$  &   [3.6] & [4.5] & [5.8] & [8] & [24]   \\
PN  &  band  &   band  &   band  &   band  &  band     &   band  &  band   &     band     &    band   &   band    &  band   &   band    \\
\hline\hline
		RP145	&				&				&				&				&	17.92\,$\pm$0.05		&	17.36\,$\pm$0.08	&	17.08\,$\pm$0.19	&	15.989\,$\pm$0.122	&	16.278\,$\pm$0.159	&				 &				 &				\\
		RP147	&	20.549\,$\pm$0.246	&	19.664\,$\pm$0.129	&	18.338\,$\pm$0.077	&	18.02\,$\pm$0.054	&	16.339\,$\pm$0.074	$\ast$	&	15.727\,$\pm$0.097	&	15.589\,$\pm$0.131	&	 15.306\,$\pm$0.077	&	 15.504	 $\pm$	0.093	&				&				&				\\
		RP162	&				&				&				&				&	17.72\,$\pm$0.05		&	17.14\,$\pm$0.05	&	16.68\,$\pm$0.18	&	16.42\,$\pm$0.25	&	15.99\,$\pm$0.25	&	 15.23\,$\pm$ 0.3	 &	14.44\,$\pm$0.35	&				\\
		RP163	&	19.805\,$\pm$0.136	&	20.671\,$\pm$0.068	&	20.459\,$\pm$0.072	&	20.406\,$\pm$0.145	&					&				&				&				&				&				 &				 &				\\
		RP171	&	18.846\,$\pm$0.094	&	19.987\,$\pm$0.074	&	19.855\,$\pm$0.076	&				&	19.75\,$\pm$0.27		&				&				&	17.443\,$\pm$0.117	&	16.264\,$\pm$ 0.097	&				 &	13.804\,$\pm$0.134	&				\\
		RP178	&	19.747\,$\pm$0.132	&	19.532\,$\pm$0.085	&	18.797\,$\pm$0.077	&	17.976\,$\pm$0.089	&	17.45\,$\pm$0.05		&	17.13\,$\pm$0.08	&	17.1\,$\pm$0.2	&	15.672	 $\pm$	0.036	&	15.442	 $\pm$	0.054	&				&				&	9.147\,$\pm$0.236	\\
\hline
\end{tabular}
\end{minipage}
\end{table*}
\end{landscape}
\label{secondpage}
\setcounter{table}{2}
\begin{landscape}
\begin{table*}
\begin{minipage}{240mm}
\centering
\scriptsize
\begin{tabular}{|l|c|c|c|c|c|c|c|c|c|c|c|c|}
\hline
LMC    &   $\textsl{U}$ &  $\textsl{B}$  &  $\textsl{V}$   &  $\textsl{I}$   &   $\textsl{J}$   &  $\textsl{H}$  &  $\textsl{K}$  &   [3.6] & [4.5] & [5.8] & [8] & [24]   \\
PN  &  band  &   band  &   band  &   band  &  band     &   band  &  band   &     band     &    band   &   band    &  band   &   band    \\
\hline\hline
		RP666	&				&	18.721\,$\pm$0.06	&	17.621\,$\pm$0.07	&	16.971\,$\pm$0.051	&	16.32\,$\pm$0.03		&	15.85\,$\pm$0.02	&	15.76\,$\pm$0.05	&	15.571\,$\pm$ 0.044	&	15.494\,$\pm$ 0.039	&				&				&				\\
		RP671	&	19.623\,$\pm$0.144	&	19.764\,$\pm$0.138	&	18.899\,$\pm$0.148	&	19.531\,$\pm$0.129	&	18.05\,$\pm$0.07		&	17.37\,$\pm$0.07	&	16.91\,$\pm$0.11	&	15.674	 $\pm$	0.085	&	 15.248	 $\pm$	0.067	&	14.991\,$\pm$0.151	&	14.241\,$\pm$0.152	&				\\
		RP676	&	20.211\,$\pm$0.184	&	19.954\,$\pm$0.104	&	18.976\,$\pm$0.114	&	18.153\,$\pm$0.127	&	17.42\,$\pm$0.04		&	17.01\,$\pm$0.05	&	16.77\,$\pm$0.15	&	16.803	 $\pm$	0.069	&	 16.612	 $\pm$	0.127	&				&				&				\\
		RP678	&	19.301\,$\pm$0.155	&	19.305\,$\pm$0.092	&	18.163\,$\pm$0.07	&	17.585\,$\pm$0.064	&	17.06\,$\pm$0.06		&	16.93\,$\pm$0.11	&	16.52\,$\pm$0.08	&	16.089	 $\pm$	0.091	&	 15.713	 $\pm$	0.081	&	14.867\,$\pm$0.149	&	14.311\,$\pm$0.21	&				\\
		RP679	&	19.81\,$\pm$0.165	&	19.345\,$\pm$0.162	&	18.372\,$\pm$0.167	&	17.695\,$\pm$0.058	&	17.58\,$\pm$0.03		&	17.01\,$\pm$0.05	&	16.99\,$\pm$0.14	&	16.194	 $\pm$	0.061	&	 15.823	 $\pm$	0.175	&				&				&				\\
		RP680	&				&	20.458\,$\pm$0.143	&	19.915\,$\pm$0.176	&	19.598\,$\pm$0.354	&	18.84\,$\pm$0.13		&	17.05\,$\pm$0.05	&	15.29\,$\pm$0.04	&	12.612\,$\pm$ 0.006	&	11.776\,$\pm$ 0.005	&	10.946\,$\pm$0.007	&	10.096\,$\pm$0.005	&	7.502\,$\pm$0.06	\\
\hline
\end{tabular}
\end{minipage}
\end{table*}
\end{landscape}
\label{secondpage}
\setcounter{table}{2}
\begin{landscape}
\begin{table*}
\begin{minipage}{240mm}
\centering
\scriptsize
\begin{tabular}{|l|c|c|c|c|c|c|c|c|c|c|c|c|}
\hline
LMC    &   $\textsl{U}$ &  $\textsl{B}$  &  $\textsl{V}$   &  $\textsl{I}$   &   $\textsl{J}$   &  $\textsl{H}$  &  $\textsl{K}$  &   [3.6] & [4.5] & [5.8] & [8] & [24]   \\
PN  &  band  &   band  &   band  &   band  &  band     &   band  &  band   &     band     &    band   &   band    &  band   &   band    \\
\hline\hline
		RP1037	&				&				&				&				&	20.28				&				&	18.5			&	17.42\,$\pm$0.25	&	16.49\,$\pm$0.25	&	16.11\,$\pm$0.3	&	 14.93\,$\pm$ 0.35	&				\\
		RP1038	&	21.543\,$\pm$0.433	&	21.074\,$\pm$0.128	&	20.327\,$\pm$0.123	&	19.727\,$\pm$0.142	&	19.59\,$\pm$0.25		&				&				&				&				&				 &				 &				\\
		RP1040	&	20.914\,$\pm$0.392	&	21.053\,$\pm$0.157	&	20.128\,$\pm$0.151	&	20.222\,$\pm$0.183	&					&				&				&				&				&				 &				 &				\\
		RP1046	&				&	20.625\,$\pm$0.074	&	20.578\,$\pm$0.069	&	20.434\,$\pm$0.122	&					&				&				&				&				&				&				 &				 \\
		RP1065	&	20.753\,$\pm$0.295	&	21.331\,$\pm$0.126	&	21.067\,$\pm$0.197	&	20.121\,$\pm$0.159	&					&				&				&				&				&				 &				 &				\\
		RP1067	&				&				&				&				&	19.28\,$\pm$0.16		&				&				&				&				&				&				&				 \\
\hline
\end{tabular}
\end{minipage}
\end{table*}
\end{landscape}
\label{secondpage}
\setcounter{table}{2}
\begin{landscape}
\begin{table*}
\begin{minipage}{240mm}
\centering
\scriptsize
\begin{tabular}{|l|c|c|c|c|c|c|c|c|c|c|c|c|}
\hline
LMC    &   $\textsl{U}$ &  $\textsl{B}$  &  $\textsl{V}$   &  $\textsl{I}$   &   $\textsl{J}$   &  $\textsl{H}$  &  $\textsl{K}$  &   [3.6] & [4.5] & [5.8] & [8] & [24]   \\
PN  &  band  &   band  &   band  &   band  &  band     &   band  &  band   &     band     &    band   &   band    &  band   &   band    \\
\hline\hline
		RP1378	&				&				&				&				&	15.19\,$\pm$0.069	$\ast$	&	14.55\,$\pm$0.074	&	14.391\,$\pm$0.101	&	14.025\,$\pm$0.013	&	14.14\,$\pm$0.017	 &	14.506\,$\pm$ 0.128	&				&				\\
		RP1387	&	18.421\,$\pm$0.082	&	17.763\,$\pm$0.066	&	16.269\,$\pm$0.058	&	15.173\,$\pm$0.064	&	14.406\,$\pm$0.032	$\ast$	&	13.748\,$\pm$0.043	&	13.582\,$\pm$0.04	&	 13.305\,$\pm$0.008	&	 13.384	 $\pm$	0.01	&	13.339\,$\pm$0.059	&				&				\\
		RP1415	&	20.159\,$\pm$0.181	&	18.879\,$\pm$0.056	&	17.59\,$\pm$0.04	&	16.448\,$\pm$0.07	&	15.49\,$\pm$0.02		&	14.82\,$\pm$0.02	&	14.65\,$\pm$0.02	&	14.457	 $\pm$	0.018	&	 14.574	 $\pm$	0.024	&	14.458\,$\pm$0.097	&				&				\\
		RP1416	&				&	21.048\,$\pm$0.182	&	20.472\,$\pm$0.188	&	20.446\,$\pm$0.251	&					&				&				&	18.41\,$\pm$0.37	&	17.38\,$\pm$0.22	&				 &				 &	9.78\,$\pm$0.12	\\
		RP1426	&	19.431\,$\pm$0.114	&	19.747\,$\pm$0.085	&	19.401\,$\pm$0.09	&	19.738\,$\pm$0.126	&	17.08\,$\pm$0.04	$\ast$	&	16.52\,$\pm$0.04	&	16.32\,$\pm$0.1	&	16.221	 $\pm$	0.062	&	 15.385	 $\pm$	0.036	&	14.185\,$\pm$0.058	&	12.451\,$\pm$0.036	&	8.043\,$\pm$0.078	\\
		RP1429	&				&	21.137\,$\pm$0.101	&	19.663\,$\pm$0.058	&	19.109\,$\pm$0.089	&	18.19\,$\pm$0.06		&	17.69\,$\pm$0.09	&	17.16\,$\pm$0.19	&				&				 &				 &				&				\\
\hline
\end{tabular}
\end{minipage}
\end{table*}
\end{landscape}
\label{secondpage}
\setcounter{table}{2}
\begin{landscape}
\begin{table*}
\begin{minipage}{240mm}
\centering
\scriptsize
\begin{tabular}{|l|c|c|c|c|c|c|c|c|c|c|c|c|}
\hline
LMC    &   $\textsl{U}$ &  $\textsl{B}$  &  $\textsl{V}$   &  $\textsl{I}$   &   $\textsl{J}$   &  $\textsl{H}$  &  $\textsl{K}$  &   [3.6] & [4.5] & [5.8] & [8] & [24]   \\
PN  &  band  &   band  &   band  &   band  &  band     &   band  &  band   &     band     &    band   &   band    &  band   &   band    \\
\hline\hline
		RP1773	&	18.509\,$\pm$0.074	&	18.868\,$\pm$0.094	&	18.853\,$\pm$0.176	&	17.697\,$\pm$0.11	&	17.65\,$\pm$0.06	$\ast$	&	17.55\,$\pm$0.23	&	17.14\,$\pm$0.26	&	 14.685\,$\pm$0.032	&	 14.466	 $\pm$	0.027	&	11.94\,$\pm$0.025	&	10.057\,$\pm$0.018	&	6.897\,$\pm$0.046	\\
		RP1791	&				&				&				&				&	17.12\,$\pm$0.02		&	17.11\,$\pm$0.04	&	17.17\,$\pm$0.13	&	16.153\,$\pm$0.066	&	16.14\,$\pm$0.077	&				 &				 &				\\
		RP1796	&	19.783\,$\pm$0.184	&	20.509\,$\pm$0.141	&	20.284\,$\pm$0.11	&	20.925\,$\pm$0.47	&	19.22\,$\pm$0.13		&	18.1\,$\pm$0.08	&	16.69\,$\pm$0.1	&	14.28	 $\pm$	0.047	&	13.412\,$\pm$ 0.034	&	12.599\,$\pm$0.051	&	11.528\,$\pm$0.054	&				\\
		RP1803	&	20.583\,$\pm$0.214	&	19.971\,$\pm$0.046	&	19.413\,$\pm$0.059	&	18.653\,$\pm$0.062	&	18.09\,$\pm$0.06		&	17.55\,$\pm$0.09	&	17.33\,$\pm$0.25	&	17.332	 $\pm$	0.103	&	 17.199	 $\pm$	0.099	&				&				&				\\
		RP1805	&	18.258\,$\pm$0.071	&	18.537\,$\pm$0.047	&	18.726\,$\pm$0.187	&	18.425\,$\pm$0.072	&					&	18.19\,$\pm$0.16	&				&	16.815\,$\pm$0.137	&	 16.713\,$\pm$0.144	&				 &				&				\\
		RP1807	&				&	23.313\,$\pm$0.529	&	21.992\,$\pm$0.424	&				&					&				&				&				&				&				&				&				 \\
\hline
\end{tabular}
\end{minipage}
\end{table*}
\end{landscape}
\label{secondpage}
\setcounter{table}{2}
\begin{landscape}
\begin{table*}
\begin{minipage}{240mm}
\centering
\scriptsize
\begin{tabular}{|l|c|c|c|c|c|c|c|c|c|c|c|c|}
\hline
LMC    &   $\textsl{U}$ &  $\textsl{B}$  &  $\textsl{V}$   &  $\textsl{I}$   &   $\textsl{J}$   &  $\textsl{H}$  &  $\textsl{K}$  &   [3.6] & [4.5] & [5.8] & [8] & [24]   \\
PN  &  band  &   band  &   band  &   band  &  band     &   band  &  band   &     band     &    band   &   band    &  band   &   band    \\
\hline\hline
		RP3923	&				&	23.024\,$\pm$0.532	&	22.138\,$\pm$0.221	&				&					&				&				&				&				&				&				&				 \\
		RP4029	&	21.281\,$\pm$0.366	&	19.925\,$\pm$0.154	&	20.709\,$\pm$0.103	&	20.645\,$\pm$0.229	&	20.18\,$\pm$0.21		&				&				&				&				&				 &				 &				\\
		RP4034	&	19.713\,$\pm$0.122	&	20.379\,$\pm$0.06	&	20.388\,$\pm$0.067	&	21.114\,$\pm$0.201	&					&				&				&	18.414\,$\pm$0.177	&	16.988\,$\pm$ 0.105	&				 &	14.233\,$\pm$0.148	&				\\
		RP4041	&				&				&				&				&					&				&				&	17.527\,$\pm$0.139	&	16.898\,$\pm$0.145	&				&				&				 \\
		RP4062	&				&	21.527\,$\pm$0.245	&	21.346\,$\pm$0.308	&	20.361\,$\pm$0.131	&					&	18.73\,$\pm$0.23	&				&	16.87\,$\pm$0.064	&	16.929\,$\pm$ 0.076	&				 &				&				\\
		RP4065	&				&				&				&				&	18.284\,$\pm$0.221	$\ast$	&	17.73			&	16.125\,$\pm$0.139	&				&				&				&				 &				 \\
\hline\hline
\end{tabular}
\end{minipage}
\end{table*}
\end{landscape}
\label{secondpage}

\newpage
\newpage

\bsp

\label{lastpage}